\documentclass[review]{elsarticle}

\usepackage[breaklinks=true, pdftex, 
  pdfborder={0 0 0}, colorlinks=true, linkcolor=blue, citecolor=blue, urlcolor=blue, 
  bookmarks=false, pdfpagelabels=false, pdftitle={Theory of Nuclear Fission}]{hyperref}
\usepackage{fullpage}
\usepackage{times}
\usepackage{graphicx,graphics,graphbox}
\usepackage{revsymb,amssymb,amsmath,amsthm}
\usepackage{color}
\usepackage{subfigure,wrapfig}
\usepackage{multirow}
\usepackage{pdfpages}
\usepackage{datetime2}
\usepackage{enumitem} 
\usepackage{refcheck}           
\usepackage[it,small]{caption}
\usepackage[top=1in,bottom=1in,left=1in,right=1in]{geometry} 
\usepackage{doi}
\usepackage[numbers]{natbib}
\usepackage{lineno}
\usepackage{braket}
\usepackage{textgreek}
\usepackage[utf8]{inputenc}
\usepackage[french,russian,english]{babel}
\usepackage{fancyhdr}

\newcommand{\gras}[1]{\boldsymbol{#1}}
\newcommand{\newtensor}[1]{\mathsf{#1}}

\newcommand{\definition}[1]{{\bf #1}}

\modulolinenumbers[5]
\journal{Journal of \LaTeX\ Templates}

\graphicspath{{figs/}}

\begin{document}

\begin{frontmatter}

\title{Theory of Nuclear Fission}

\author{Nicolas Schunck\corref{mycorrespondingauthor}}
\address{Nuclear and Chemical Sciences Division, Lawrence Livermore National Laboratory, Livermore, California 94551, USA}
\ead{schunck1@llnl.gov}
\author{David Regnier}
\address{CEA, DAM, DIF, 91297 Arpajon, France}
\address{Universit\'e Paris-Saclay, CEA, LMCE, 91680 Bruy\`eres-le-Ch\^atel, France}
\ead{david.regnier@cea.fr}

\cortext[mycorrespondingauthor]{Corresponding author}

\begin{abstract}
Atomic nuclei are quantum many-body systems of protons and neutrons held 
together by strong nuclear forces. Under the proper conditions, nuclei can 
break into two (sometimes three) fragments which will subsequently decay by 
emitting particles. This phenomenon is called nuclear fission. Since different 
fission events may produce different fragmentations, the end-products of all 
fissions that occurred in a small chemical sample of matter comprise hundreds 
of different isotopes, including $\alpha$ particles, together with a large 
number of emitted neutrons, photons, electrons and antineutrinos. 
The extraordinary complexity of this process, which happens at length scales of 
the order of a femtometer, mostly takes less than a femtosecond but is not 
completely over until all the lingering $\beta$ decays have completed -- which 
can take years -- is a fascinating window into the physics of atomic nuclei. 
While fission may be more naturally known in the context of its technological 
applications, it also plays a pivotal role in the synthesis of heavy elements 
in astrophysical environments. In both cases, experimental measurements are not 
sufficient to provide complete data. Simulations are needed, yet at 
levels of accuracy and precision that pose formidable challenges to nuclear 
theory. The goal of this article is to provide a comprehensive overview of the 
theoretical methods employed in the description of nuclear fission.
\end{abstract}

\begin{keyword}
Fission\sep 
Fission fragment yields
\sep Cross sections
\sep Prompt fission spectrum
\sep Delayed spectrum
\sep Large-amplitude collective motion
\sep Energy density functional theory
\sep Time-dependent density functional theory
\MSC[2010] 00-01\sep  99-00
\end{keyword}

\end{frontmatter}

\renewcommand{\headrulewidth}{0pt}
\thispagestyle{fancy}
\fancyhf{}
\rhead{LLNL-JRNL-830603}
\lhead{}


\tableofcontents


\section{Introduction}
\label{sec:intro}

The first experimental evidence of nuclear fission was obtained in late 1938 by 
German scientists Hahn and Strassman \cite{hahn1938ueber,hahn1939ueber}, who 
observed that ``{\it isotopes of Barium ($Z=56$) are formed as a consequence of 
the bombardment of Uranium ($Z=92$) with neutrons}'' [excerpt from 
\cite{meitner1939disintegration}]. The first theoretical interpretation of this 
phenomenon was given shortly thereafter by Lise Meitner and was based on the 
liquid drop model of the atomic nucleus \cite{meitner1939disintegration}. Only 
a few months later, N. Bohr and J.A. Wheeler published the first comprehensive 
theoretical study of the process \cite{bohr1939mechanism}. These initial 
discoveries pertained to what we now call neutron-induced fission; G.N. Flerov 
and K.A. Petrzhak separately discovered the spontaneous fission of Uranium 
isotopes in 1941 \cite{flerov1940discovery,scharff-goldhaber1946spontaneous}.

\begin{figure}[!ht]
\centering
\includegraphics[width=0.70\textwidth]{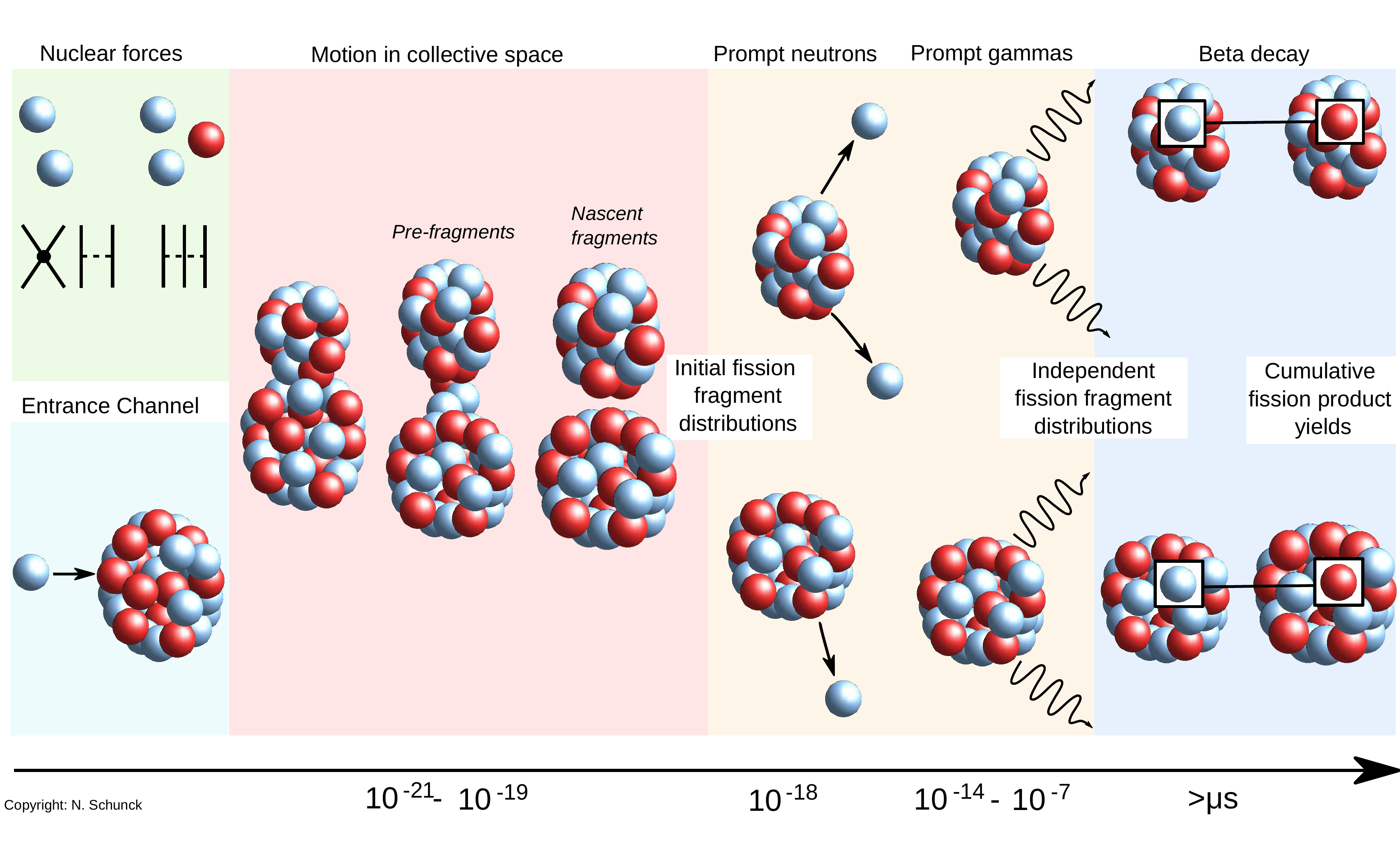}
\caption{Schematic representation of the fission process. A heavy atomic 
nucleus, possibly formed in a nuclear reaction, deforms until it divides into 
two (sometimes three) fragments, thereby releasing a large amount of energy of 
the order of about 200 MeV for actinide nuclei. The excited fission fragments 
rapidly emit neutrons followed by photons (prompt fission spectrum). About a 
few picoseconds after the scission of the original nucleus, many of the 
resulting fission fragments undergo a sequence of $\beta$ decays in competition 
with additional photon and or neutron emission (delayed emission).}
\label{fig:schematic}
\end{figure}

In simple terms, nuclear fission is the process by which a heavy atomic nucleus 
divides into two (binary) or three (ternary) fragments. Although there 
is no formal definition for it, the usual consensus is that any object bigger 
than an $\alpha$ particle $(Z=2, N=2$) formed by the breaking up of a heavy 
nucleus is considered a fission fragment; anything smaller would be considered 
as particle emission or radioactivity. For a given sample of fissioning 
material, not all nuclei fission identically: the relative proportion of each 
isotope is called the {\bf primary fission fragments distribution}. Fission 
fragments are always in an excited state. Therefore, as soon as they are 
formed, they undergo a sequence of decays. Initially, this decay proceeds 
mainly through neutron emission. When the energy of the fragment becomes 
smaller than the separation energy $S_n$ of a single neutron, photon emission 
becomes the dominant mode of deexcitation. This initial phase of the decay 
process is very rapid, of the order of $\tau \lesssim 10^{-13}$s and defines what is 
called the {\bf prompt particle emission}. Since neutron emission changes the isotopic 
composition of the sample, it is customary to refer to the distribution of 
fission fragments after prompt emission as the {\bf independent fission yields}. 
After neutron and photon emission, many fragments may be unstable against 
$\beta$ decay -- the conversion of a neutron into a proton with the emission of 
an electron and antineutrino. Beta decay takes place over time scales ranging 
from a few picoseconds up to years. Since the daughter nuclei produced in the 
decay may also be excited, $\beta$ decay is often followed by photon 
and sometimes neutron emission. This phase of the decay of the fission 
fragments is called {\bf delayed 
emission}. At the end of all emissions, both the isotopic and elemental 
composition of the sample have changed. The relative proportion of each isotope 
is encoded in what is called the {\bf cumulative fission product yields} \cite{mills1995fission}. Figure 
\ref{fig:schematic} gives a qualitative representation of the entire process. 

From a theoretical perspective, nuclear fission is a striking example of 
large-amplitude collective motion for a quantum many-body system. In other 
words, it results, or emerges, from the quantum many-body effects induced by 
in-medium nuclear forces. The characteristic features of atomic nuclei, e.g., 
the competition between individual and collective behavior, their self-bound 
nature and the fact that they are open quantum systems, all play a role in our 
understanding of the various phases of the fission process. Finally, fission is 
intrinsically a time-dependent process that eventually populates a huge number
of output channels. 
This short enumeration 
should already suggest to the informed reader that the theoretical description 
of nuclear fission poses enormous challenges. In addition, fission plays a 
major role in understanding the limits of nuclear stability, especially 
concerning superheavy elements \cite{smolanczuk1995spontaneousfission,
hofmann2000discovery,berger2001superheavy,burvenich2004systematics,
schindzielorz2009fission,pei2009fission,sheikh2009systematic,warda2012fission,
staszczak2013spontaneous,baran2015fission,warda2018cluster,
giuliani2019colloquium}. Partly for this reason, fission is one of the main 
mechanisms that terminates the cycle of nuclear reactions responsible for the 
formation of elements in astrophysical environments, especially in the 
rapid-neutron capture process (r process) \cite{erler2012fission,
panov2013influence,giuliani2018fission,mumpower2018beta}. Last but not least, 
the fission mechanism is at the heart of nuclear technology. Apart from the 
standard example of nuclear energy \cite{aliberti2006nuclear,
sartori2013nuclear}, details of the fission process are also important for 
nuclear forensics \cite{kristo2020chapter} and international non-proliferation 
efforts and safeguards by the international community. Knowledge about the 
physics of the fission process is summarized in evaluated nuclear
data libraries such as JEFF~\cite{plompen2020joint} and ENDF~\cite{brown2018endf}, which 
list the cross sections of a large
range of nuclear reactions along with the fission yields, or ENSDF \cite{evaluated}, which is
more oriented toward nuclear structure data.

Whether one is interested in the fundamental description of fission or its 
applications in other scientific or technological disciplines, the recent years 
have brought many interesting developments. The rapid and sustained increase in 
computational capabilities has rejuvenated fission theory by enabling 
simultaneously realistic large-scale simulations of fission properties 
\cite{giuliani2018fission,mumpower2020primary} and more fundamental 
descriptions of the process \cite{schunck2016microscopic,bender2020future}. 
Many of these most recent results and methods have yet to find their way in the 
nuclear data libraries, which -- as far as fission is concerned -- often rely 
on very phenomenological models and concepts from the 1950ies or 1960ies. At 
the same time, the quality and robustness of these libraries keeps improving 
partly thanks to the introduction of novel techniques from artificial 
intelligence and machine learning to evaluate and verify nuclear data 
\cite{radaideh2019combining,neudecker2020enhancing,whewell2020evaluating}.

Surprisingly there are only a few textbooks providing a comprehensive 
introduction to nuclear fission, especially fission theory, and they are just a 
few years old \cite{krappe2012theory,younes2019microscopic,younes2021introduction}. In the past five 
years, three review articles have been published covering either microscopic 
methods \cite{schunck2016microscopic}, experimental results 
\cite{andreyev2017nuclear} or phenomenology \cite{schmidt2018review}. While 
these papers have provided a welcome update of recent results, their focus is 
mostly on the chain of events leading to the divide of the nucleus and not so 
much on the probability that this process takes place nor on the decay of the 
fragments, even though nearly all experimental information on fission is 
collected, directly or indirectly, during this decay phase. We also feel that 
there is a disconnect between the basic nuclear science community, the goal of 
which is understand the basic mechanisms at play in fission, and the 
specialized user communities were fission data is employed. For example, the 
producers and users of global nuclear data libraries may not always be aware of 
the most recent developments in nuclear theory that could impact the process of 
evaluation; conversely, nuclear scientists may not always be aware of the 
possible impact of their work nor of the constraints (especially in terms of 
precision) imposed by applications. The goal of this review is to try to 
provide an overview of fission theory that could be beneficial to both types of 
communities.

Our presentation is organized to follow the chronology of the fission process. 
Following Bohr \cite{bohr1939mechanism,nix1965studies,nix1969further}, one 
still thinks of fission as a deformation process. In Section 
\ref{subsec:deformation}, we will therefore discuss in details how the 
qualitative concept of deformation emerges from the nuclear mean field theory, 
and how fission fragments are mapped to specific, deformed configurations of 
the fissioning nucleus at the point of scission. This section has some 
overlap with \cite{schunck2016microscopic}, but it is necessary to introduce 
some very basic concepts that will be used throughout this review. Section 
\ref{sec:proba} deals with the probability that fission occurs. In the case of 
spontaneous fission, this probability is in fact the main quantity that must be 
computed to determine the fission half-life; in induced fission, it is related 
to the very challenging problem of computing the fission cross section, namely 
the probability that fission takes place relatively to other possible mode 
of decay such as neutron scattering, evaporation, photoemission, etc. The 
mechanisms by which the nucleus will change deformation and reach the point of 
scission are described in Section \ref{sec:lacm}. In the past two or three 
decades, this is perhaps the area where the most progress has been made. The 
static models of fission en vogue from the 1950ies to the 1990ies have been 
replaced by explicit time-dependent description, either semi-classical or full 
quantum-mechanical. As mentioned earlier, the decay of fission fragments has 
been seldom addressed in a review. It will be presented in Section 
\ref{sec:deexcitation}. This decay leads to the emission of neutrons and 
photons and to a series of $\beta$ decays. Some fundamental problems in basic 
science such as the role of fission in nucleosynthesis mechanisms 
\cite{beun2008fission,goriely2013new,mumpower2016impact,mumpower2018beta,
misch2020astromers,cowan2021origin} or what has been dubbed the antineutrino 
reactor anomaly \cite{mention2011reactor,mueller2011improved,
sonzogni2015nuclear,sonzogni2017dissecting,sonzogni2018revealing,
prospectcollaboration2021,ashtariesfahani2021bayesian,
prospectcollaboration2021a} cannot be properly understood without invoking 
fission fragment deexcitation. Finally, Section \ref{sec:initial_fragments} 
will review some of the most recent attempts at predicting the properties
of the primary fission fragments after the point of scission with microscopic 
methods. By connecting the physics of large amplitude collective motion 
with the statistical decay of 
the fragments, these efforts offer the prospect of building a truly consistent 
theory of fission in the near future.


\section{Nuclear Deformation}
\label{subsec:deformation}

The concept of nuclear deformation is a phenomenological one. The nuclear 
Hamiltonian $\hat{H}$ that determines, in principle, all nuclear properties, is 
rotationally invariant \cite{bohr1998nuclear,ring2004nuclear}. Theories that 
attempt to solve directly the many-body Schrödinger equation, such as the 
no-core shell model \cite{barrett2013initio} or the nuclear shell model 
\cite{stroberg2019nonempirical} never need consider deformation: the 
eigenvectors of $\hat{H}$ are always rotationally-invariant. This observation 
also applies to the other types of many-body methods that give approximations 
to the true eigenvectors of $\hat{H}$ \cite{hagen2014coupledcluster,carlson2015quantum,
hergert2016inmedium}. 

However, it was discovered long ago that one could describe, qualitatively 
and quantitatively, many important nuclear properties such as, e.g., the value 
of nuclear quadrupole moments, the strength of electromagnetic transitions 
between excited states or the particular distribution of discrete energy 
levels, by invoking collective behaviors associated with a deformed average 
nuclear potential in which nucleons move independently 
\cite{rainwater1950nuclear,bohr1951nuclear,rainwater1976background}. While 
originally introduced in a phenomenological manner, the concept of deformation 
can be formalized as a spontaneous symmetry breaking of the nuclear mean field: 
the average potential is not invariant under rotation \cite{schunck2019energy}. Deformation is a very pervasive concept 
in fission theory. Potential energy surfaces, i.e., the potential energy as a 
function of nuclear deformation, determine fission paths and the 
large-amplitude dynamics of the nucleus. The number of particles in fission 
fragments is typically mapped onto the deformation of the fissioning nucleus at 
the point of scission. The overlap, in the quantum mechanical sense, between 
the nuclear wave functions at different deformations is directly related to the 
calculation of fission probabilities. For these reasons, models to compute 
deformed nuclei are the cornerstones of any fission theory. 

Historically, such models were based on phenomenological mean-field approaches, 
where the average nuclear potential is parametrized directly with an 
appropriate mathematical function. Examples of such phenomenological 
potentials include the Nilsson, Woods-Saxon, and Folded-Yukawa potentials. 
Within this framework, some of the most comprehensive studies of fission were 
reported already in the 1970ies \cite{brack1972funny,bolsterli1972new}. In 
recent years, this phenomenology has also been used to provide large-scale, 
systematic surveys of nuclear properties \cite{moller1995nuclear,
moller1997nuclear,moller2006global,bonneau2007global,moller2016nuclear}. We 
briefly review this approach in Section \ref{subsec:micmac}. 
Nuclear energy density functional theory provides a more fundamental 
description of nuclear deformation, since it basically relates the average 
potential to an energy density functional or an effective pseudo-potential that 
describe effective interactions between nucleons \cite{schunck2019energy}. The 
theory can be viewed as a (considerable) extension of the self-consistent 
Hartree-Fock theory \cite{lowdin1955quantuma,lowdin1955quantumb,
lowdin1955quantumc} and is inspired in part by the density functional theory of 
electrons in atoms \cite{kohn1965selfconsistent}. As mentioned above, nuclear 
deformation emerges when the density of particles breaks some of the symmetries 
of the nuclear Hamiltonian \cite{nazarewicz1994microscopic}. While 
computationally more involved, the advantage of EDF methods is that they 
provide (effective) Hamiltonians and an arsenal of many-body techniques to 
compute quantum numbers. While there were pioneering applications of EDF to 
fission in the 1970ies \cite{flocard1974selfconsistent,negele1978dynamics}, it 
is fair to say that self-consistent calculations of fission properties only 
became relevant in the late 1980ies and 1990ies. In the past two decades, these 
methods have now become competitive with more phenomenological ones thanks to 
advances in computing capabilities \cite{schunck2016microscopic}. We will 
summarize the most relevant notions of the EDF approach in Section 
\ref{subsec:EDF}. Finally, we will discuss in Section 
\ref{subsec:scission} how to infer from potential energy surfaces the points at 
which the nucleus breaks into two fragments.


\subsection{Phenomenological Nuclear Mean Field}
\label{subsec:micmac}

What is known as the macroscopic-microscopic approach to nuclear 
structure is at the heart of nuclear phenomenology. It relies on modeling the 
atomic nucleus as a quantum, charged liquid drop with additional corrective 
terms that account for quantum many-body effects. In the language of modern 
many-body theory, one would call such an approach a 0-body theory with 
corrective terms of 1- and 2-body origin. In this picture, fission is the 
extreme distortion of the droplet that leads to its breakup. In more details, 
the total energy of the atomic nucleus is written as the sum of several 
contributions: (i) the macroscopic energy $E_{\rm mac}(\gras{q})$, typically 
represented by the liquid drop or droplet model, (ii) a shell correction 
$\delta E_{\rm shell}(\gras{q})$ originating from the single-particle degrees 
of freedom and (iii) a pairing correction $\delta E_{\rm pair}(\gras{q})$ to 
account for nuclear superfluidity. Each of these three terms parametrically 
depends on the underlying deformation $\gras{q} = (q_{1},\cdots,q_{N})$ of the 
nuclear shape, 
\begin{equation}
E(\gras{q}) = E_{\rm mac}(\gras{q}) + \delta E_{\rm shell}(\gras{q}) + \delta E_{\rm pair}(\gras{q})
\label{eq:micmac}
\end{equation}

The leading term in this expression is the macroscopic energy which is computed 
as the energy of a charge drop of nuclear matter in the liquid drop model 
(LDM). This finite piece of nuclear matter can also be expressed as a series of 
powers of $Z$ and $A$ -- the lepdotermous expansion \cite{myers1969average,
myers1974nuclear} -- and be related to predictions from more microscopic 
theories \cite{treiner1986semiclassical,reinhard2006finite,nikolov2011surface}. 
Initial parameterizations of the liquid drop were relatively simple 
\cite{myers1966nuclear}; more recent models have included additional terms to 
mock up various effects (charge asymmetry, odd-even effect of pairing, proton 
form factors, etc.) that vary smoothly with $Z$ and $N$ \cite{moller1995nuclear,
moller1997nuclear,moller2016nuclear}. Historically, the LDM provided the 
theoretical basis for the first explanation of fission by Meitner, Bohr and 
Wheeler \cite{meitner1939disintegration,bohr1939mechanism}. The stability of a 
liquid drop against fission is captured in the dimensionless fissility 
parameter $x$, which is the ratio of the Coulomb to the surface energy of the 
drop \cite{bohr1939mechanism,nix1965studies}. Empirically, the fissility 
parameter can be approximated as
\begin{equation}
x \approx \frac{Z^2}{47A(1 - \eta I^2)},
\label{eq:fissility}
\end{equation}
with $I = (N-Z)/A$ and $\eta = - a_{\rm ssym}/a_{\rm surf}$ is the ratio of the 
surface-symmetry to surface coefficients of the leptodermous expansion of the 
nuclear energy. These two terms are proportional to $I^2 A^{2/3}$ and 
$A^{2/3}$, respectively. They can be either fitted to reproduce average 
properties of nuclei \cite{myers1969average}, or computed from a model of 
nuclear matter.

Even if the macroscopic energy represents nearly 99\% of the total binding of a 
heavy atomic nucleus such as an actinide, corrective terms must be added to 
obtain a more precise estimate of the total energy. The shell correction 
effectively simulates the fact that neutrons and protons are quantum-mechanical 
particles occupying single-particle levels, the distribution of which impacts 
the stability of the nucleus, hence its binding energy \cite{strutinsky1967shell,
strutinsky1968shells}. Generically, it reads
\begin{equation}
\delta E = \sum_{k\in\mathcal{S}} e_{k} - \left\langle \sum_{k\in\mathcal{S}} e_{k} \right\rangle
\end{equation}
where $\mathcal{E} = \{ e_{k} \}_{k=1,\dots, p}$ refers to a full set of 
single-particle (s.p.) energies and $\mathcal{S}$ is a subset of $\mathcal{E}$. 
To extract the energy for the lowest configuration at deformation $\gras{q}$, 
$\mathcal{S}$ would simply contain the $Z$ and $N$ lowest energies; any other 
combination of s.p. levels would yield the energy of an excited state. The 
symbol $\langle \cdots \rangle$ refers to the Strutinsky averaging procedure to 
produce the ``smooth'' energy. Full details about how to compute this term are 
given in \cite{bolsterli1972new}; variants of the standard Strutinsky procedure 
to remove the spurious effect of continuum or pairing effects are discussed in 
\cite{vertse1998shell,kruppa1998calculation,vertse2000shell,
pomorski2004particle}. The introduction of the shell correction was key to 
explain two major features of the fission process: (i) the energy threshold at which fission takes place is nearly constant 
for actinides, contrary to LDM predictions; (ii) the LDM cannot account for the 
existence of fission isomers. These meta-stable states are more deformed than 
the ground state thanks to the stabilizing effect of the shell correction. 

The shell correction can be interpreted as having a one-body origin: it is 
associated with an independent-particle model of the nucleus, where each 
nucleon moves, independently of one another, in the average mean field. In the 
language of second quantization, the resulting Hamiltonian operator only 
contains one-body operators. To describe the different energy spectra between 
even-even and odd nuclei, pairing correlations must be taken into account 
explicitly; see \cite{brink2005nuclear} for a comprehensive introduction to 
nuclear pairing. Historically, the effect of pairing on the total energy was 
quantified by introducing a pairing correction $\delta E_{\rm pair}$ akin to 
the shell correction. In most applications of the macroscopic-microscopic 
model, this pairing correction is computed from the BCS solution and can thus be 
interpreted as a corrective term of two-body origin. All three terms of 
\eqref{eq:micmac} can be modified to describe nuclei at finite angular momentum 
or finite temperature \cite{moretto1972statistical,jensen1973shell,
diebel1981microscopic,devoigt1983highspin,bengtsson1985rotational,
dudek1988pairing,werner1995shape,ivanyuk2018temperature}; see also the textbook 
\cite{nilsson1995shapes}. 

\begin{figure}[!ht]
\centering
\includegraphics[width=0.45\textwidth]{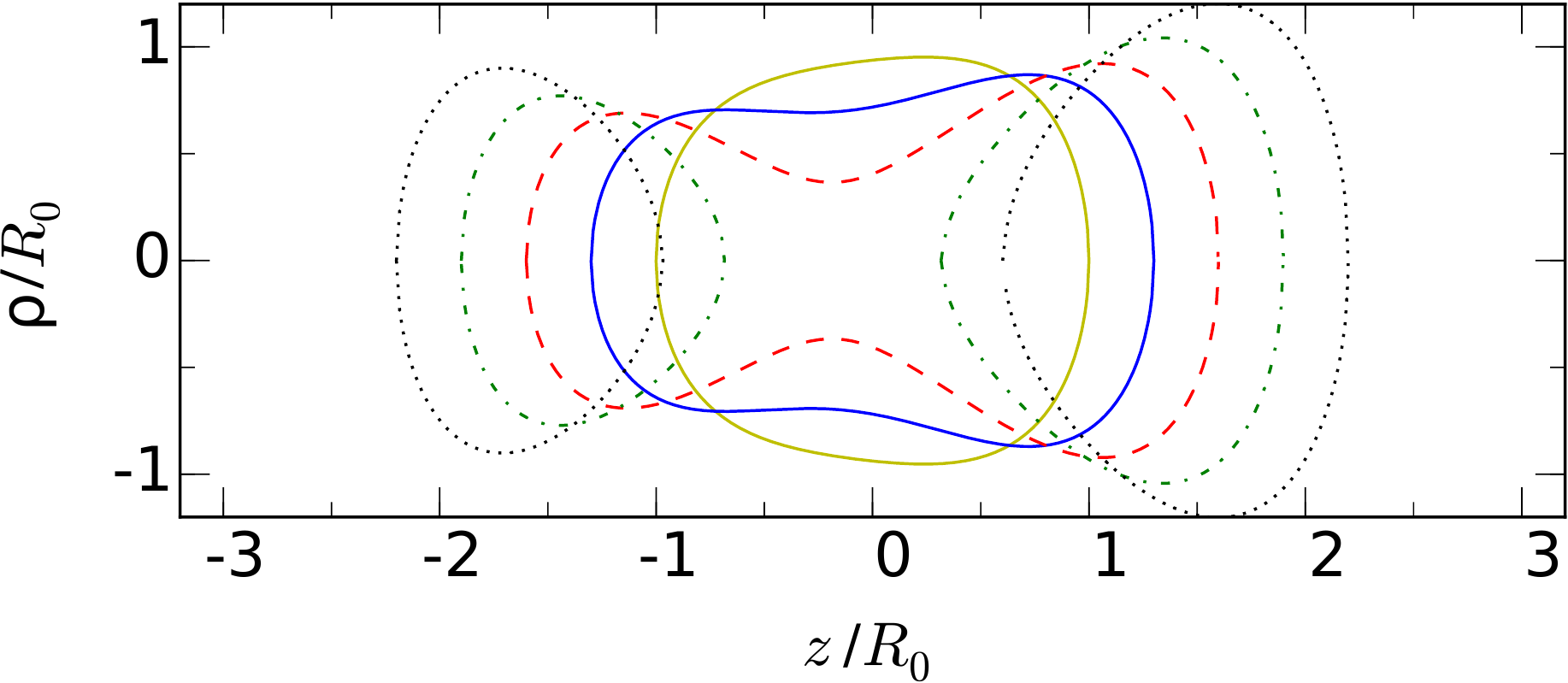}
\includegraphics[width=0.45\textwidth]{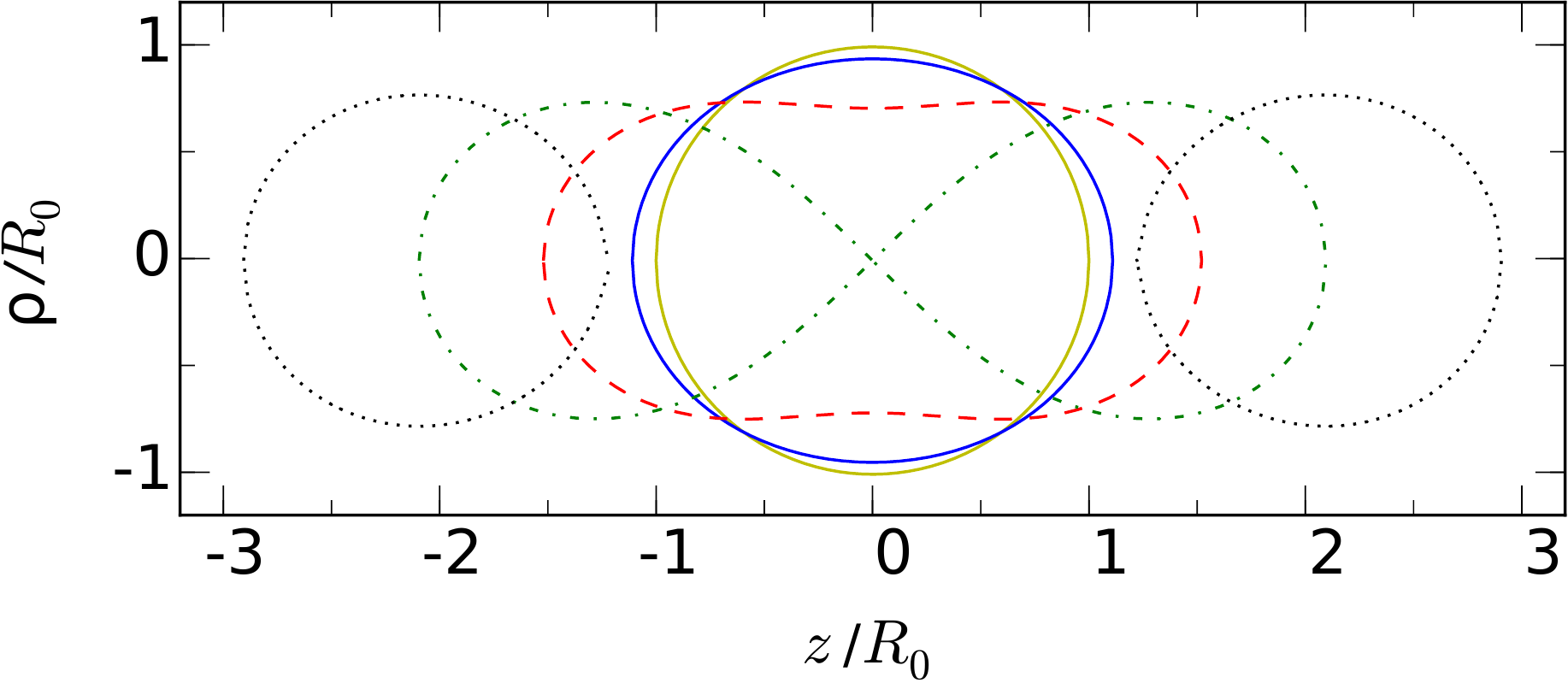}
\caption{Left: Funny Hills parameterization \cite{brack1972funny} of the nuclear 
shape for $(h,\alpha) = (0.2,0.3)$ with $c$ varying from 1.0 to 2.2 by steps of 
0.3. Right: Axial shapes with the Cassini oval parameterization 
\cite{pashkevich1971asymmetric} for $u = 0.0, 0.4, 0.8, 1.0, 1.2$. The value 
$u=0$ corresponds to the sphere, the value $u = 1.0$ to the point of scission 
and the value $u=1.2$ to separated fragments.
Figures reproduced with permission from \cite{schunck2016microscopic} 
courtesy of Schunck; copyright 2016 by Institute of Physics.
}
\label{fig:shapes}
\end{figure}

All terms in the decomposition \eqref{eq:micmac} of the total energy depend on 
the deformation of the nuclear shape which, in this approach, is an {\it input} 
to the calculation. The determination of a suitable parameterization of the 
nuclear shape is in fact a challenging problem. Given a parameterization 
$\mathcal{P}$ and a set of shape parameters $\gras{q} = (q_{1},\cdots,q_{N})$, 
the following criteria should be met: (i) there should be a one-to-one 
correspondence between any vector $\gras{q}$ and the actual shape; (ii) since 
the shape parameterization is an input to any calculation, the number $N$ of 
components of $\gras{q}$, i.e., the number of shape parameters, should be as 
small as possible (iii) the parameterization must be capable of describing 
weakly-deformed, compact shapes, extremely elongated shapes and even disjoint 
shapes representing the separated fragments. The standard multipole expansion 
of the nuclear radius \cite{bohr1998nucleara}, for example, does not meet any 
of these criteria: it is not bijective \cite{rohozinski1981hexadecapole,
rohozinski1997parametrization}, the number of parameters needed to describe 
well-deformed shapes increases with deformation, and it cannot characterize 
separated fragments. Parametrizations more adapted to describing fission 
include the Funny-Hills parameterization \cite{brack1972funny}, the Los Alamos 
parameterization \cite{nix1969further}, Cassini ovals 
\cite{pashkevich1971asymmetric} or Legendre polynomials 
\cite{dobrowolski2016solving}; see also \cite{schunck2016microscopic} for an 
explicit summary of some of these parameterizations. Figure \ref{fig:shapes}
shows an example of the Funny Hills and Cassini ovals parameterization.

\begin{figure}[!ht]
\centering
\includegraphics[width=0.6\textwidth]{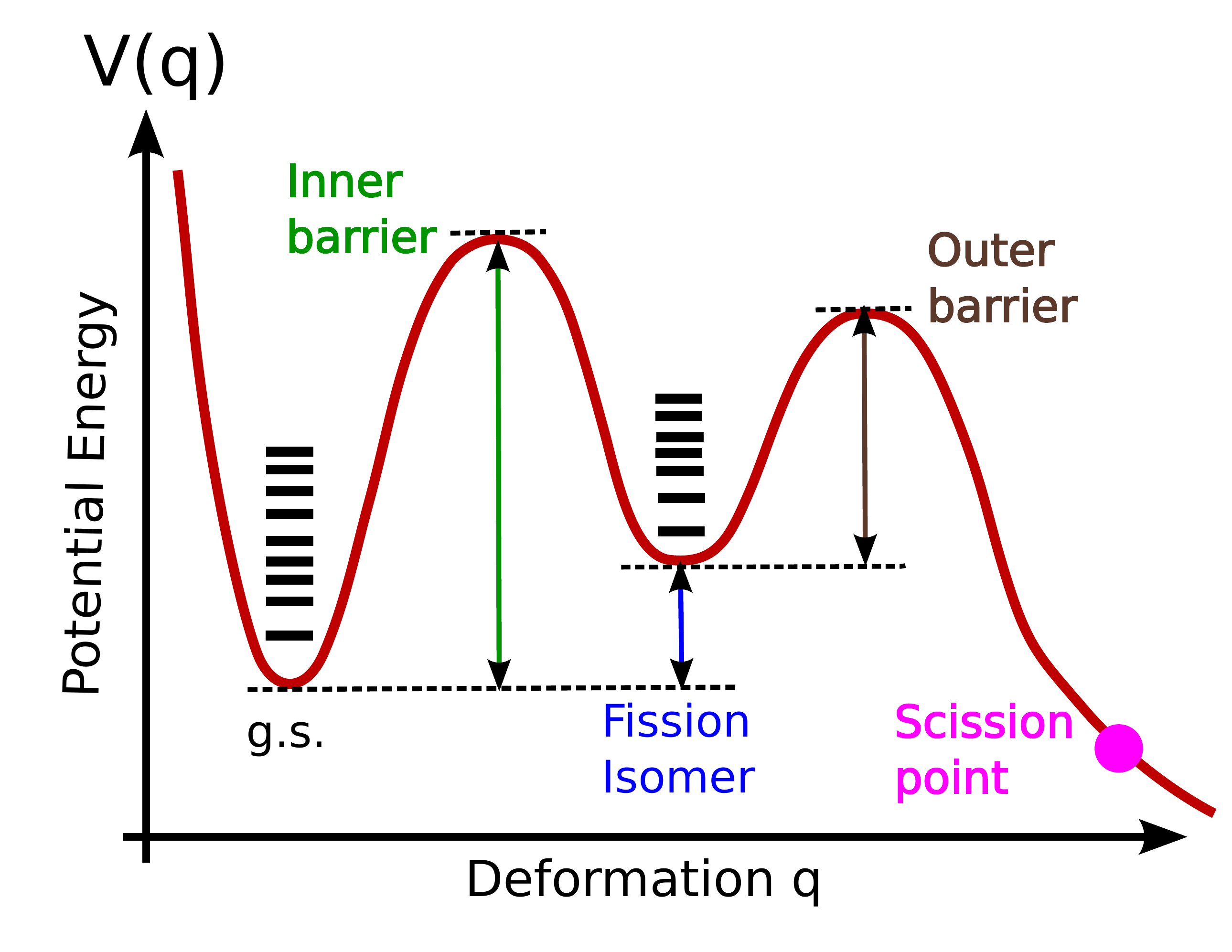}
\caption{One-dimensional representation of fission barriers in actinide 
nuclei. Note that the scission point is not represented at scale: in practice, 
it is ``far'' from the outer barrier. The distance in the deformation space 
between the location of the scission point and that of the fission isomer may 
be four to five times larger than the distance between the fission isomer and 
the ground state.}
\label{fig:barriers_actinides}
\end{figure}

This nuclear phenomenology based on the macroscopic-microscopic model has been 
extremely successful in predicting many facets of nuclear properties \cite{moller1995nuclear,moller1997nuclear,moller2006global,moller2016nuclear,
werner1995shape,jachimowicz2021properties}. In the context of fission, this 
approach provides a computationally efficient, realistic method to perform 
large-scale calculations of deformation properties, namely \definition{potential energy 
surfaces (PES)}. In many actinides, there is empirical evidence that the PES 
require several deformation degrees of freedom to adequately capture the full 
dynamics of fission: elongation of the whole nucleus, reflection asymmetry of 
the shape, triaxiality parameters, but also deformation parameters 
characterizing the fragments themselves \cite{moller2001nuclear}. Recent 
investigations proved that fragments can also be reflection asymmetric, which 
would require at least another two parameters to describe the shape asymmetry for each 
fragment \cite{scamps2019effect,scamps2018impact}. Such multi-dimensional PES 
cannot be visualized, but a lot of important information can be extracted by 
projecting them onto a one-dimensional subspace of lowest energy: the \definition{fission path}. In actinide 
nuclei, this fission path is typically characterized by a deformed potential minimum 
(i.e. the classical ground 
state) and an additional, meta-stable minimum at larger deformations: the \definition{fission isomer}. 
Both minima are separated 
by potential energy \definition{barriers}: the inner barrier, between the ground state 
and the fission isomer, and the outer barrier, between the fission isomer and 
regions corresponding to ever larger deformations. Precise calculations of these fission barriers 
are especially important for estimating spontaneous fission half-lives and 
fission cross sections, as will be presented in Section \ref{sec:proba}. Figure
\ref{fig:barriers_actinides} shows a typical, schematic representation of the 
one-dimensional fission path in actinide nuclei. In such one-dimensional 
representations, the  \definition{saddle point} is simply the point on the fission 
path that has the highest energy (here: the top of the inner barrier). 

\begin{figure}[!ht]
\centering
\includegraphics[width=0.6\textwidth]{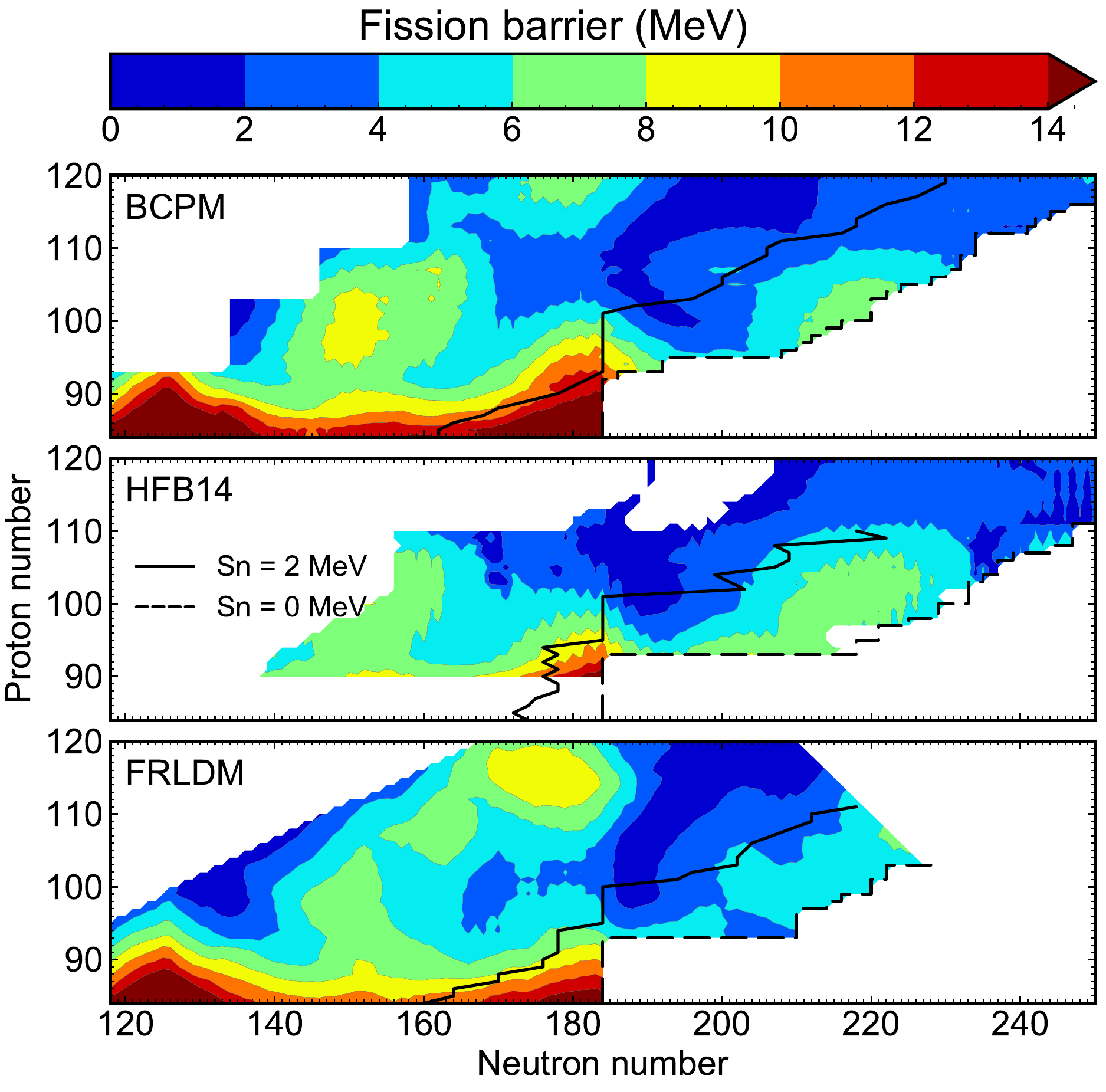}
\caption{Highest fission barrier heights for heavy and superheavy elements with 
$84 \geq Z \geq 120$ and $118 \geq N \geq 250$ for three different models of 
nuclear deformation: FRLDM corresponds to the macroscopic-microscopic 
calculation of \cite{moller2009heavyelement} while the BPCM 
\cite{baldo2017barcelonacataniaparismadrid} and HFB14 
\cite{goriely2009prediction} are two different version of energy density 
functionals; see discussion in Section \ref{subsec:EDF}.
Figures reproduced with permission from \cite{giuliani2018fission} 
courtesy of Giuliani; copyright 2018 by The American Physical Society.
}
\label{fig:barriers}
\end{figure}

Historically, the predictions of fission isomeric states and ``double-humped'' 
fission barriers was a major success of the macroscopic-microscopic approach 
\cite{brack1972funny,bolsterli1972new,nix1972calculation,larsson1972fission}, 
as was the explanation of asymmetric fission by reflection-asymmetric shapes 
stabilized by the shell correction \cite{pashkevich1971asymmetric}. Refinements 
in the models combined with modern computational capabilities have led to 
global predictions of static fission barriers \cite{moller1989new,
moller2009heavyelement,kowal2010fission,jachimowicz2012secondary,
jachimowicz2013eightdimensional,jachimowicz2017adiabatic,
jachimowicz2020static}. Such calculations are necessary, since as we go away 
from the valley of stability, the topology of the potential energy surfaces may 
change quite dramatically and the traditional features of fission barriers in 
actinides may not apply anymore. Figure \ref{fig:barriers} shows a systematics 
of fission barrier heights (more precisely, the highest fission barrier) in 
heavy and superheavy nuclei taken from \cite{giuliani2018fission}. It is clear 
from these results that uncertainties from nuclear models alone can reach a few 
MeV, which corresponds to dozens of orders of magnitude in spontaneous fission 
half-lives; see Section \ref{subsec:sf}.

\subsection{Energy Density Functional Theory}
\label{subsec:EDF}

The EDF approach is the modern name for what was known for a long time as the 
self-consistent mean field theory. Its root can be found both in the 
generalized Hartree-Fock theory, which was formalized in a few seminal papers 
in the early 1960ies \cite{thouless1960stability,valatin1961generalized}, as 
well as in the density functional theory for electrons 
\cite{kohn1965selfconsistent,hohenberg1964inhomogeneous}. In a nutshell, the 
basic idea of the EDF approach is to compute the nuclear mean field from a 
suitable model of nuclear forces among nucleons rather than parameterizing it as 
in the macroscopic-microscopic approach. In practice, this calculation is often 
based on a model for in-medium, two-body effective nuclear interactions (or 
pseudo-potentials) such as the Skyrme \cite{skyrme1959effective} or Gogny 
\cite{decharge1980hartreefockbogolyubov} force. When computing the expectation 
value of such pseudo-potentials on a many-body product state of single-particle 
wave functions (Slater determinant), both the resulting energy and the nuclear 
mean field become functionals of the density of particles 
\cite{bender2003selfconsistent}. Direct parameterizations of the energy density 
have also been proposed, with varying success \cite{fayans1994isotope,
baldo2008kohnsham,baldo2013new,baldo2017barcelonacataniaparismadrid,
bulgac2018minimal}. Note that the effective potentials need not be restricted 
to two-body terms only: three- and higher-order many-body interactions could be 
included as well \cite{sadoudi2013skyrme,sadoudi2013skyrmea}. The EDF approach 
comprises a full hierarchy of approximations and methods yielding increasing 
accuracy and precision in the determination of nuclear properties. One may 
consider two main categories: single-reference energy density functional 
(SR-EDF) and multi-reference density functional (MR-EDF). They are most of the time
used to compute static properties of the nucleus but could also be implemented
in a time-dependent flavor.
The EDF approach has a decade-old rich history and it 
would go far beyond the scope of this review to describe it in too many 
details. We refer the interested reader to \cite{schunck2019energy,
bender2003selfconsistent,mang1975selfconsistent,lacroix2010quantum,
bulgac2013timedependent,duguet2014nuclear,nakatsukasa2016timedependent}. Below, 
we will only summarize the essential concepts and formulas most relevant to 
fission theory.


\subsubsection{Single-Reference Energy Density Functional Theory: Self-Consistent Mean Field}
\label{subsubsec:SR-EDF}

Today, the single-reference EDF approach is largely built on the 
Hartree-Fock-Bogoliubov (HFB) theory, where the nuclear many-body wave function 
is approximated by a generalized product of quasiparticle wave functions
\cite{bender2003selfconsistent,ring2004nuclear,schunck2019energy}. In the 
formalism of second quantization, these quasiparticles (q.p.) are associated 
with annihilation and creation operators $(\beta_{\mu}, 
\beta_{\mu}^{\dagger})$ and the many-body wave function reads
\begin{equation}
\ket{\Phi} = \prod_{\mu} \beta_{\mu}\ket{0},
\label{eq:HFB_ansatz}
\end{equation}
where $\ket{0}$ is the particle vacuum. The q.p. operators are themselves 
related to the annihilation and creation operators $(c_{k}, c_{k}^{\dagger})$ of an 
arbitrary single-particle basis by the Bogoliubov 
transformation
\cite{valatin1961generalized}
\begin{subequations}
\begin{align}
\beta_{\mu} & = \sum_{m} \left[ U^{\dagger}_{\mu m}\,c_{m} + V^{\dagger}_{\mu m}\,c_{m}^{\dagger} \right] ,
\medskip\\
\beta_{\mu}^{\dagger} & = \sum_{m} \left[ V^{T}_{\mu m}\,c_{m} + U^{T}_{\mu m}\,c_{m}^{\dagger} \right].
\end{align}
\end{subequations}
Let us consider an effective two-body pseudo-potential $\hat{V}$ characterized 
by the matrix elements $v_{ijkl} \equiv \braket{ij|\hat{V}|kl}$ in the 
single-particle basis, with $\bar{v}_{ijkl}$ the anti-symmetrized version of 
these matrix elements ($\bar{v}_{ijkl} = v_{ijkl} - v_{ijlk}$). The HFB energy 
reads
\begin{equation}
E[\rho,\kappa,\kappa^{*}] = 
\sum_{ij} t_{ij} \rho_{ji}
+
\frac{1}{2} \sum_{ijkl} \bar{v}_{ijkl}\rho_{lj}\rho_{ki}
+
\frac{1}{4}\sum_{ijkl} \bar{v}_{ijkl} \kappa^{*}_{ij}\kappa_{kl}.
\label{eq:HFB_energy}
\end{equation}
where $t_{ij}$ are the matrix elements of the kinetic energy operator, $\rho$ 
is the one-body density matrix and $\kappa$ the pairing tensor. These two 
quantities encode the basic degrees of freedom of the HFB theory. They can be 
expressed in terms of the matrices of the Bogoliubov transformation as 
$\rho = V^{*}V^{T}$ and $\kappa = V^{*}U^{T}$. The main idea of the HFB theory 
is to minimize the energy \eqref{eq:HFB_energy} with respect to the matrix 
elements of $U$ and $V$. This results in the HFB equation
\begin{equation}
\big[ \mathcal{H}, \mathcal{R} \big] = 0 ,
\label{eq:HFB_equ}
\end{equation}
where $\mathcal{H}$ is the HFB matrix and $\mathcal{R}$ the generalized 
density, respectively given by
\begin{equation}
\mathcal{H} = \left( \begin{array}{cc}
h   & \Delta  \\
-\Delta^{*} & -h^{*}
\end{array}\right),
\qquad
\mathcal{R} =
\left( \begin{array}{cc}
\rho   & \kappa \\
-\kappa^{*} & 1 - \rho^{*}
\end{array}\right).
\label{eq:HFB_matrices}
\end{equation}
In these expressions, $h = t + \Gamma$ is the mean field, $t$ is the 
single-particle kinetic energy potential and $\Gamma \equiv \Gamma_{ik} = \sum_{jl} 
\bar{v}_{ijkl}\rho_{lj}$ the mean field potential, and $\Delta \equiv 
\Delta_{ij} = \tfrac{1}{2}\sum_{kl}\bar{v}_{ijkl}\kappa_{kl}$ is the pairing 
field. Both objects are functionals of $\rho$ and $\kappa$. 

Applications of the formalism rely on effective two-body potentials $\hat{V}$ 
such as the Skyrme or Gogny interactions; see 
\cite{bender2003selfconsistent,stone2007skyrme,robledo2019mean} for recent 
reviews. There is also a relativistic version of the EDF approach based on 
solving a self-consistent Dirac equation \cite{ring1996relativistic,
niksic2011relativistic,schunck2019energy}. Both non-relativistic and 
relativistic EDFs depend on a small set of adjustable parameters that must be 
calibrated on data. The most commonly used parameterizations in fission studies 
are the SkM* \cite{bartel1982better} and UNEDF1 \cite{kortelainen2012nuclear} 
Skyrme EDFs, the D1S parameterization of the Gogny force 
\cite{berger1991timedependent}, and the DD-PC1 \cite{niksic2008finite} and 
DD-ME2 \cite{lalazissis2005new} parameterizations of relativistic Lagrangians. 
Pairing correlations are central to the HFB theory. As can be seen from 
\eqref{eq:HFB_energy}, the same effective pseudo-potential $\hat{V}$ should be 
at the origin of all components of the energy: the particle-hole part 
proportional to $\rho$ and the particle-particle one proportional to $\kappa$. 
This requirement is enforced only in the case of the Gogny force. In practice, 
users of the Skyrme EDF adopt a different effective potential in the pairing 
channel based, e.g., on simple density-dependent forces 
\cite{chasman1976densitydependent}, separable finite-range interactions 
\cite{tian2009separable}, or even non-empirical pairing forces built out of 
chiral effective field theory \cite{duguet2008nonempirical,
lesinski2009nonempirical,hebeler2009nonempirical}. Such choices are largely 
motivated by pragmatic reasons such as the divergences induced by zero-range 
effective potentials, which require a regularization procedure 
\cite{bulgac2002renormalization,bulgac2002local} or/and the difficulty in 
probing the pairing channel when fitting the parameters of the EDF 
\cite{schunck2019energy}.

The HFB approach works best when the symmetries of the nuclear Hamiltonian are 
broken. For example, the minimization of the HFB energy may lead to solutions 
where both densities $\rho$ and $\kappa$  are not rotationally-invariant. These 
symmetries are not restricted to spatial ones such as translational and 
rotational invariance, axial symmetry or reflection symmetry: they also include 
time-reversal or isospin symmetry \cite{dobaczewski2000point,
dobaczewski2000pointa,rohozinski2010selfconsistent}. In fact the ansatz 
\eqref{eq:HFB_ansatz} by itself breaks a fundamental symmetry of the nuclear 
Hamiltonian: particle number. Therefore, HFB solutions, strictly speaking, 
correspond to wave packets of Slater determinants with different particle 
numbers \cite{ring2004nuclear}.

\begin{figure}[!ht]
\centering
\includegraphics[width=0.70\textwidth]{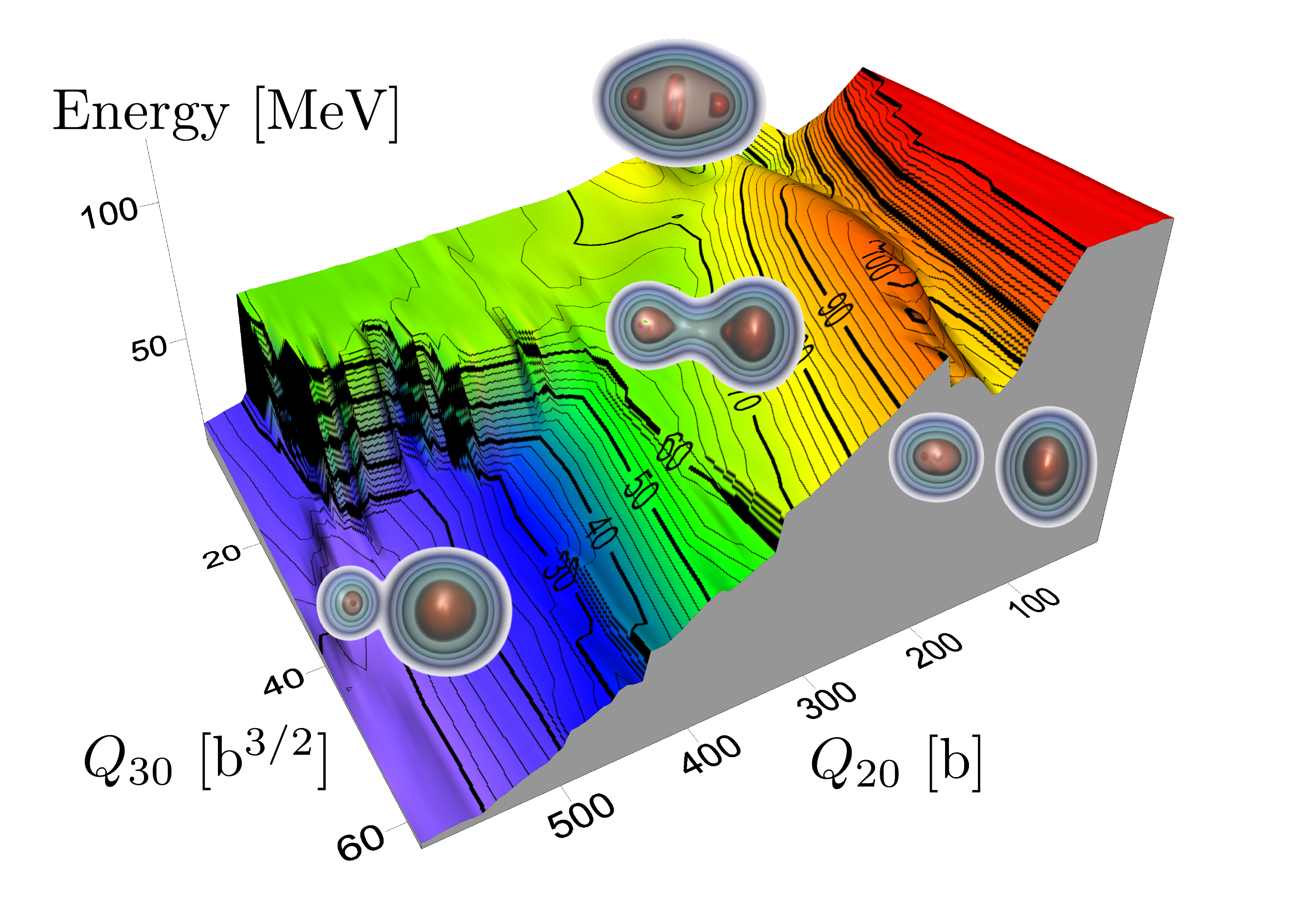}
\caption{Potential energy surface of $^{240}$Pu in the $(Q_{20},Q_{30})$ 
collective space. The four volume renderings give a visual representation of 
the nuclear shape at various locations on the PES. The calculation was 
performed for the SkM* parameterization of the Skyrme functional and corresponds 
to Fig. 3 of \cite{schunck2014description}.}
\label{fig:PES_HFB}
\end{figure}

The SR-EDF framework based on {\it deformed} HFB solutions is thus the direct, 
``microscopic'' equivalent of the macroscopic-microscopic approach and can in 
fact be viewed as its theoretical justification \cite{brack1981strutinsky}. 
Qualitatively, the main difference between the two is how the average nuclear 
potential is determined: it is approximated by a suitable mathematical function 
in the macroscopic-microscopic approach and computed self-consistently from a 
model of nuclear forces in the HFB theory. Potential energy surfaces can be 
constructed by imposing various constraints on the HFB solutions. For example, 
the expectation value of the multipole moment operators 
$\hat{Q}_{\lambda\mu}(\gras{r}) = r^2 Y_{\lambda\mu}(\theta,\varphi)$, where 
$Y_{\lambda\mu}(\theta,\varphi)$ are the spherical harmonics 
\cite{varshalovich1988quantum}, is given by 
\begin{equation}
\braket{\hat{Q}_{\lambda\mu}}
=
\int d^{3}\gras{r}\, \hat{Q}_{\lambda\mu}(\gras{r})\rho(\gras{r}),
\label{eq:multipoles}
\end{equation}
where $\rho(\gras{r})$ is the total density of particles. These moments have a 
straightforward physics interpretation: $\hat{Q}_{20}$ quantifies the degree of 
elongation of the shape, $\hat{Q}_{30}$ the degree of left-right asymmetry, $\hat{Q}_{40}$ 
the bulging around the mid-point of the shape, etc.. By solving the HFB 
equation \eqref{eq:HFB_equ} with constraints on the values 
$\braket{\hat{Q}_{\lambda\mu}}$ as given by \eqref{eq:multipoles}, we can 
generate a manifold of HFB solutions representing many different shapes. Figure 
\ref{fig:PES_HFB} shows a simple example of such a potential energy surface in 
the case where only two constraints were imposed on $\braket{\hat{Q}_{20}}$ and 
$\braket{\hat{Q}_{30}}$.

\subsubsection{Multi-Reference Energy Density Functional Theory: Configuration Mixing}
\label{subsubsec:MR-EDF}

While the SR-EDF approach introduces symmetry-breaking as a means to build more 
correlations in what would be otherwise a very simplified nuclear wave 
function, the multi-reference EDF ansatz takes these deformed solutions and 
combines them \cite{wawong1975generatorcoordinate}. Schematically, this mixing 
takes the form
\begin{equation}
\ket{\Psi} = \int dq\, f(q) \ket{\Phi(q)} ,
\label{eq:MREDF_ansatz}
\end{equation}
where $q$ is a generic label for the HFB solutions $\ket{\Phi(q)}$ given by 
\eqref{eq:HFB_ansatz} and $f(q)$ are some weight functions to be determined 
\cite{griffin1957collective}. This quantum configuration mixing can be used, in 
particular, to restore all the broken symmetries and therefore recover the 
quantum numbers that were lost during the symmetry breaking 
\cite{sheikh2021symmetry}. In this case, the labels $q$ correspond to the 
parameters of the symmetry group, the basis states $\ket{\Phi(q)}$ are 
typically given by $\ket{\Phi(q)} \equiv \hat{R}\ket{\Phi}$ where $\hat{R}$ is 
the generator of the relevant symmetry group, and $f(q)$ is an appropriate 
weight function. In the simplest case of particle-number projection, which is 
associated with the $U(1)$ symmetry group, the ansatz \eqref{eq:MREDF_ansatz}
thus becomes
\begin{equation}
\ket{\Psi} 
= \int d\varphi\, \left(\frac{1}{2\pi}e^{-i\varphi N}\right) e^{i\varphi\hat{N}}\ket{\Phi}
= \frac{1}{2\pi}\int_{0}^{2\pi} d\varphi\, e^{i\varphi(\hat{N}-N)}\ket{\Phi} .
\end{equation}
The most relevant symmetries for fission include parity 
\cite{egido1991parityprojected}, particle number \cite{anguiano2001particle,
dobaczewski2007particlenumber} and angular momentum \cite{bally2021projection}.
MR-EDF techniques also provide a 
powerful way to improve the ansatz for the nuclear wave function by adding yet 
another layer of correlations that could not be captured at the SR-EDF level 
\cite{egido2016stateoftheart}. One of the most standard techniques of MR-EDF is 
the generator coordinate method (GCM) \cite{brink1968generator,
reinhard1987generator}. The GCM is especially well adapted to describing 
collective correlations that involve all nucleonic degrees of freedom and play 
such a central role in understanding fission. In such a case, the labels $q$ 
correspond to different values of some constraint operators $\hat{Q}$ and the 
basis states are the constrained HFB solutions. In fact, GCM-based techniques 
can be used to extract a purely collective Hamiltonian called the Bohr 
Hamiltonian that is somewhat reminiscent of a quantized version of the deformed 
liquid drop, in the sense that it depends on global, collective coordinates 
that typically represent nuclear deformation \cite{rohozinski2012gaussian,
rohozinski2015gaussian,matsuyanagi2016microscopic}. 


\subsection{Scission Configurations and Fission Fragments}
\label{subsec:scission}

Qualitatively, the moment where the two fragments just separate and the whole 
nucleus breaks apart is called the \definition{scission point}. This picture results from 
the analogy of the fission process with the deformation of a charged liquid 
drop that overcomes surface tension \cite{meitner1939disintegration}. Within 
the macroscopic-microscopic model of Section \ref{subsec:micmac}, such a naive 
concept of scission can still make sense: in all the parameterizations of the 
nuclear shape specifically designed to describe very elongated shapes, there 
always exists a set of parameter values for which the neck between the two 
prefragments reduces to a single point -- the scission point 
\cite{nix1969further,pashkevich1971asymmetric,trentalange1980shape}. 

One may thus be tempted to simply define scission as the locations on the 
potential energy surfaces where the two fragments are fully formed and well 
separated. However, when determining the solutions to the nuclear mean field, 
whether semi-phenomenologically as in Section \ref{subsec:micmac} or 
self-consistently as in Section \ref{subsec:EDF}, the obtained wave functions are 
for the whole nucleus: if only because of the Pauli principle, some of the 
single-particle wave functions spread over each of the two prefragments. In 
other words, the system is (heavily) entangled \cite{younes2011nuclear}. In 
fact, the system remains so even when the two fragments are far apart from one 
another. These difficulties are compounded by the use of a variational 
principle (in the EDF approach) or by the naive filling of orbits (in the 
macroscopic-microscopic approach). For nuclear shapes corresponding to 
well-separated fragments -- such configurations are sometimes called the 
asymptotic conditions -- the total energy computed in both frameworks is, in 
practice, the one that minimizes the energy of each fragment. As a result,  
calculations of fission fragments properties for such configurations are highly 
non-physical: against all experimental evidence, fragments have no excitation 
energy.

To mitigate these problems, one is thus forced to resort to phenomenological 
criteria to define scission {\it before} the nuclear shape truly corresponds to 
two fragments. The simplest of these criteria involve setting a critical value 
for the density of particles in the neck region between the prefragments 
\cite{dubray2008structure} or for the number of particles in that same region 
\cite{schunck2014description}. Random neck rupture models provide an 
additional layer of refinement to such approaches \cite{brosa1990nuclear,
rizea2013dynamical,warda2015fission}. Since fission in general, and scission in 
particular, is the result of a competition between attractive nuclear forces 
and repulsive Coulomb forces, one may define scission as the point where the 
nuclear interaction energy between the two prefragments is smaller than the 
Coulomb repulsion between them \cite{davies1977rupture}. However, as mentioned above, 
the two prefragments are entangled and the very notion of a nuclear and Coulomb 
energy for a prefragment is ambiguous at best. In fact, it was shown that such 
quantities are not invariant under unitary transformations of the whole 
fissioning nucleus \cite{younes2011nuclear}. It is possible to design specific 
transformations that maximize the localizations of quasiparticles in each 
prefragment and minimize their interaction energy and, as a consequence, 
change the definition of scission \cite{schunck2014description}. More recently, 
there have been a few attempts to better understand the physics of scission by 
borrowing from configuration-interaction and reaction theory techniques, the 
goal being to establish a continuous pathway between pre- and post-scission 
configurations \cite{bertsch2018scission,bertsch2019decay,bertsch2019diabatic}.

From these remarks, it should be clear that there is no unique definition of 
scission and that predictions of fission fragment properties, including their 
initial conditions in terms of numbers of particle, excitation energy, spin and 
parity, are highly model-dependent. This has a significant impact on the 
modeling of the deexcitation of the fission 
fragments, which is largely determined by these initial conditions as will be 
discussed in Section~\ref{sec:deexcitation}. Furthermore, the class of fission 
models called the scission-point models are based on using the hypothesis of 
statistical equilibrium at scission to determine fission fragment properties 
\cite{fong1964fission,wilkins1976scissionpoint}. The predictions of such models 
are contingent on what definitions of scission they adopt.


\section{Probabilities, Rates and Cross Sections}
\label{sec:proba}

The previous section summarized the various theoretical methods used to 
compute the properties of deformed atomic nuclei -- including the extreme 
deformations characteristic of the point of scission. This prologue was 
necessary to introduce the basic concepts of fission theory: deformation (or 
symmetry breaking), collective space, potential energy surfaces, fission 
barriers, scission point, etc. The next three sections will build upon these 
notions to address the three most fundamental problems in fission theory: 
computing the probability that fission takes place -- possibly in competition 
with other decay modes, characterizing the evolution of the nucleus from a 
near ground-state configuration to the point of scission and predicting the 
outcome of the overall process, namely all the particles emitted and the 
fission products. This section is focused on fission probabilities. 

In nuclear physics jargon, fission probabilities are related to what is called 
the entrance channel, which we can define here as the mechanism that produces 
the proper conditions for an atomic nucleus $(Z,N)$ to undergo fission. Many of 
the elements heavier than Lead have in fact a significant probability of 
undergoing fission simply as a consequence of the large number of protons they 
contain. The 
spontaneous fission half-life $\tau_{\rm sf}$ is the key quantity that 
theorists must compute and compare with experimental measurements. This is in 
fact very challenging since the experimental half-lives cover about 30 orders 
of magnitude, from about $\tau_{\rm sf} \approx 4.1 \mu$s for $^{250}$No to 
$\tau_{\rm sf} >$ 14 billions years for $^{232}$Th \cite{krappe2012theory}.

The case of induced fission is even more complex. By definition, the 
fissioning nucleus is formed in a nuclear reaction involving a projectile, a 
target and a set of initial conditions: type of incoming particle, beam energy, 
impact parameter. While neutron-induced and photofission (fission induced by 
photons) are the most common and most important cases, other scenarii such as 
$\beta$-delayed fission (where fission follows the $\beta$ decay of an 
isotope), proton-induced fission or fission in heavy-ion collisions are also 
important, especially in nucleosynthesis or superheavy science. It is often 
assumed that fission takes place after the equilibration of all degrees of 
freedom that follows the reaction, resulting in what is called the compound 
nucleus \cite{bohr1939mechanism,hodgson1987compound}. The validity of this 
approximation depends on how typical fission times compare with the relaxation 
time of the various degrees of freedom including single-particle, collective 
vibrations such as the Giant Dipole Resonance (GDR) or neutron emission 
\cite{brink1990compound}. Estimates of these various timescales suggest that 
the notion of a compound nucleus is valid for up to several dozens of MeV of 
excitation energy \cite{brink1990compound,egido1993decay,simenel2020timescales}. 
In such cases, the main challenge for nuclear theory becomes describing the 
exact structure of this compound nucleus. There are also instances where 
fission takes place before full equilibration. This is the quasifission, that 
is especially important in heavy ion collisions \cite{toke1985quasifission}.


\subsection{Spontaneous Fission}
\label{subsec:sf}

As discussed earlier in Section \ref{subsec:micmac}, fission results from 
the competition between Coulomb repulsion and surface tension. This mechanism 
is encoded either directly, in the values of the liquid drop parameters, or 
indirectly in the parameters of the energy functional, and is best understood 
as a deformation process. This process itself depends on the variations of the 
potential energy as a function of the deformation of the nuclear shape -- the 
potential energy surface (PES). If first-year students of quantum mechanics 
were to look at an example of a one-dimensional PES 
such as Fig.~\ref{fig:barriers_actinides}, they would be immediately 
reminisced of quantum tunneling. Indeed, spontaneous fission is most 
conveniently described as quantum tunnelling through a (multi-dimensional) 
potential energy barrier. Qualitatively, the problem thus posed is quite 
simple: initially, the nucleus is in its ground state. The probability of 
fission is simply related to the probability of tunneling from this initial 
condition to a deformed shape outside the barrier. Such a probability is often 
computed semi-classically with the WKB approximation 
\cite{merzbacher1961quantum} and it is given by the following formula
\begin{equation}
\tau_{1/2}^{\rm SF} = \frac{1}{\nu}\exp\left( 
\frac{2}{\hbar}\int_{a}^{b} ds\sqrt{2B(s)(V(s) - E_0)}
\right) ,
\label{eq:t_sf}
\end{equation}
where $s$ is the curvilinear abscissa along the most likely fission path, i.e., 
the trajectory in the $N$-dimensional collective space spanned by the 
collective variables $\gras{q} = (q_1,\dots,q_{N})$ that minimizes the classical 
action \cite{krappe2012theory}. The nuclear potential energy along this path is 
noted $V(s)$ and $B(s)$ is the collective inertia. $\nu$ is the number of 
assaults to the barrier per unit time \cite{brack1972funny}. The ground-state 
energy is noted $E_0$ and the points $a$ and $b$ are referred to as the inner 
and outer turning points, respectively.

While qualitatively simple, the devil hides in the details. Firstly, in 
addition to involving the introduction of a set of collective variables 
$\gras{q}$, the derivation leading to \eqref{eq:t_sf} also relies on an 
explicit decoupling between these collective variables and the other nuclear 
degrees of freedom, which expresses the fact that the collective motion is 
confined within the space spanned by the variables $\gras{q}$ 
\cite{klein1991classical,hofmann1997quantal,dang2000selfconsistent,
nakatsukasa2012density,matsuyanagi2016microscopic,nakatsukasa2016timedependent}.
As a result, the potential energy cannot be taken directly as either the 
macroscopic-microscopic potential energy \eqref{eq:micmac} or the HFB energy 
\eqref{eq:HFB_energy}. Instead, these values are modified by zero-point energy 
corrections, the specifics of which depend on the nature of the collective 
variables and of the theoretical framework used to express the decoupling. Most 
of the time, these zero-point energy corrections are computed with the GCM 
\cite{reinhard1987generator}. Importantly, their values do not only affect the 
overall potential energy surface $V(\gras{q})$, but also the initial energy of 
the nucleus $E_0$: in principle, the latter should correspond to the lowest 
eigenstate of the collective Hamiltonian. In practice, most calculations treat 
the value of $E_0$ as an adjustable parameter that can be used to quantify 
uncertainties of the predictions.

\begin{figure}[!ht]
\centering
\includegraphics[width=0.6\textwidth]{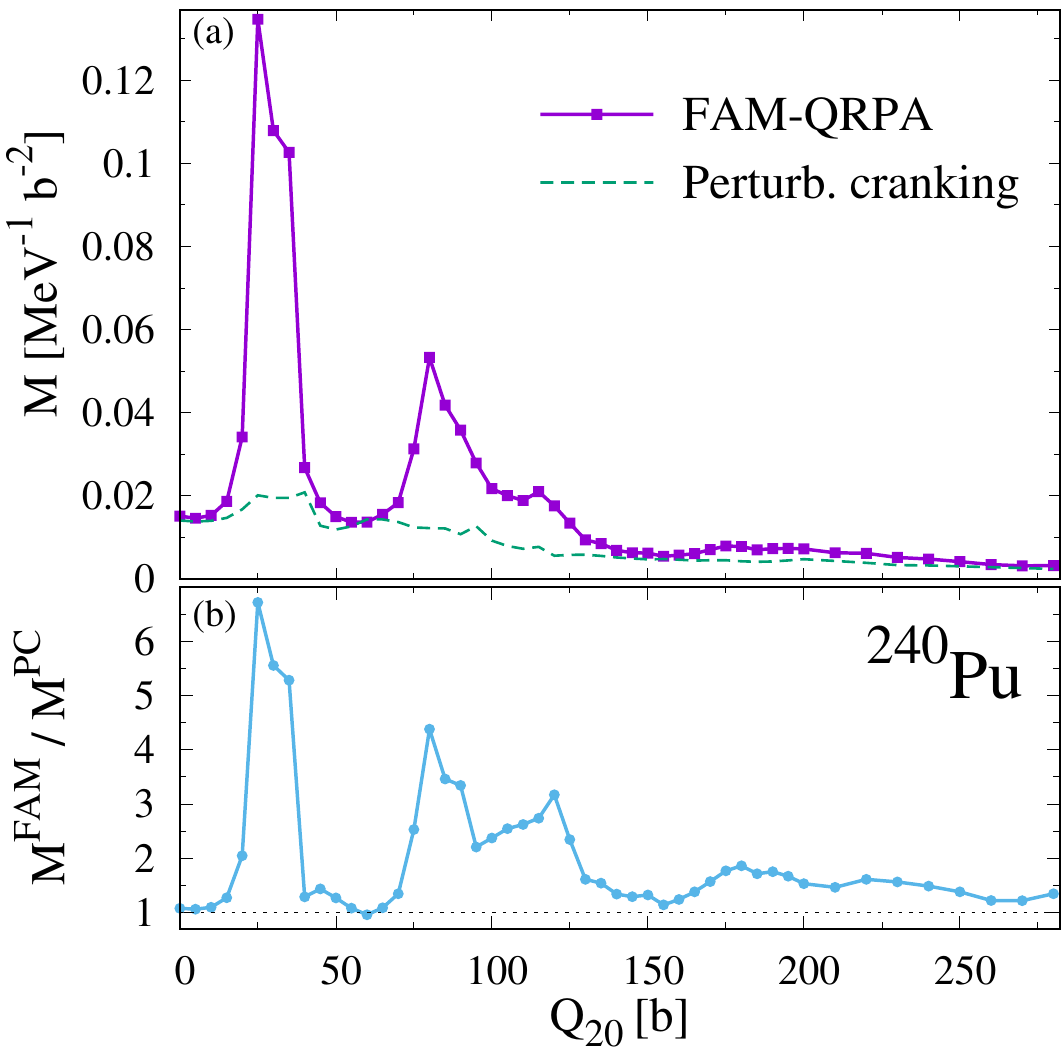}
\caption{Top panel: Comparison between the perturbative cranking approximation 
to the collective mass tensor $\newtensor{M} = \newtensor{B}^{-1}$ and the exact calculation along the symmetric 
fission path of $^{240}$Pu. Bottom panel: ratio between the exact and 
perturbative cranking inertia.
Figures reproduced with permission from \cite{washiyama2021finiteamplitude} 
courtesy of Washiyama; copyright 2021 by The American Physical Society.
}
\label{fig:coll_mass}
\end{figure}

Secondly, in contrast to textbook examples of tunnelling, the potential energy 
surface is defined in the abstract space that represents deformation. In 
practice the set of collective coordinates $\gras{q}$ can be either the 
deformations of the nuclear surface, see Section~\ref{subsec:micmac}, or the
expectation value of relevant observables, see Section~\ref{subsec:EDF}. This 
means that the ter $B$ in the WKB formula is not related to the mass of the 
nucleus, but is a more abstract object called the collective 
inertia tensor. It is in fact a rank-2 tensor $\newtensor{B}(\gras{q})\equiv 
B_{ij}(\gras{q})$ with $i,j = 1,\dots,N$ where $N$ is the number of collective 
variables. The calculation of this object is far from simple. Traditional 
formulas rely on the GCM or the adiabatic time-dependent HFB (ATDHFB) 
approximation; see \cite{schunck2016microscopic} for a summary of the relevant 
formulas. Most practical implementations of these methods neglect the effect of 
velocity, time-odd terms induced by the generalized momenta $\gras{p}$ 
associated with the collective variables $\gras{q}$, both in the GCM where 
these momenta should be introduced explicitly, which would double the number of 
collective variables \cite{goeke1980generatorcoordinatemethod}, or in the 
ATDHFB formulas where their effect is simply neglected 
\cite{dobaczewski1981quadrupole,hinohara2012effect}. In addition, both the GCM 
and ATDHFB collective inertia involve the derivative operator 
$\partial/\partial q_{i}$. This operator is often reduced to a local 
approximation: this perturbative approximation has been shown to have a large 
impact on half-lives \cite{baran2011quadrupole,sadhukhan2013spontaneous,
giuliani2018nonperturbative}. Because of these limitations, other methods have 
been proposed to compute the collective inertia, including ``dynamic'' path 
calculations \cite{yuldashbaeva1999mass} and the adiabatic self-consistent 
collective coordinate method \cite{matsuo2000adiabatic,
hinohara2007gaugeinvariant}. Recently, there have been very promising attempts 
to perform ``exact'' ATDHFB calculation of the collective inertia tensor by 
inverting the full QRPA inertia \cite{washiyama2021finiteamplitude}. Results 
suggest rather large differences with both local and non-local cranking 
approximations as seen in Fig.~\ref{fig:coll_mass}.

\begin{figure}[!ht]
\centering
\includegraphics[width=0.5\textwidth]{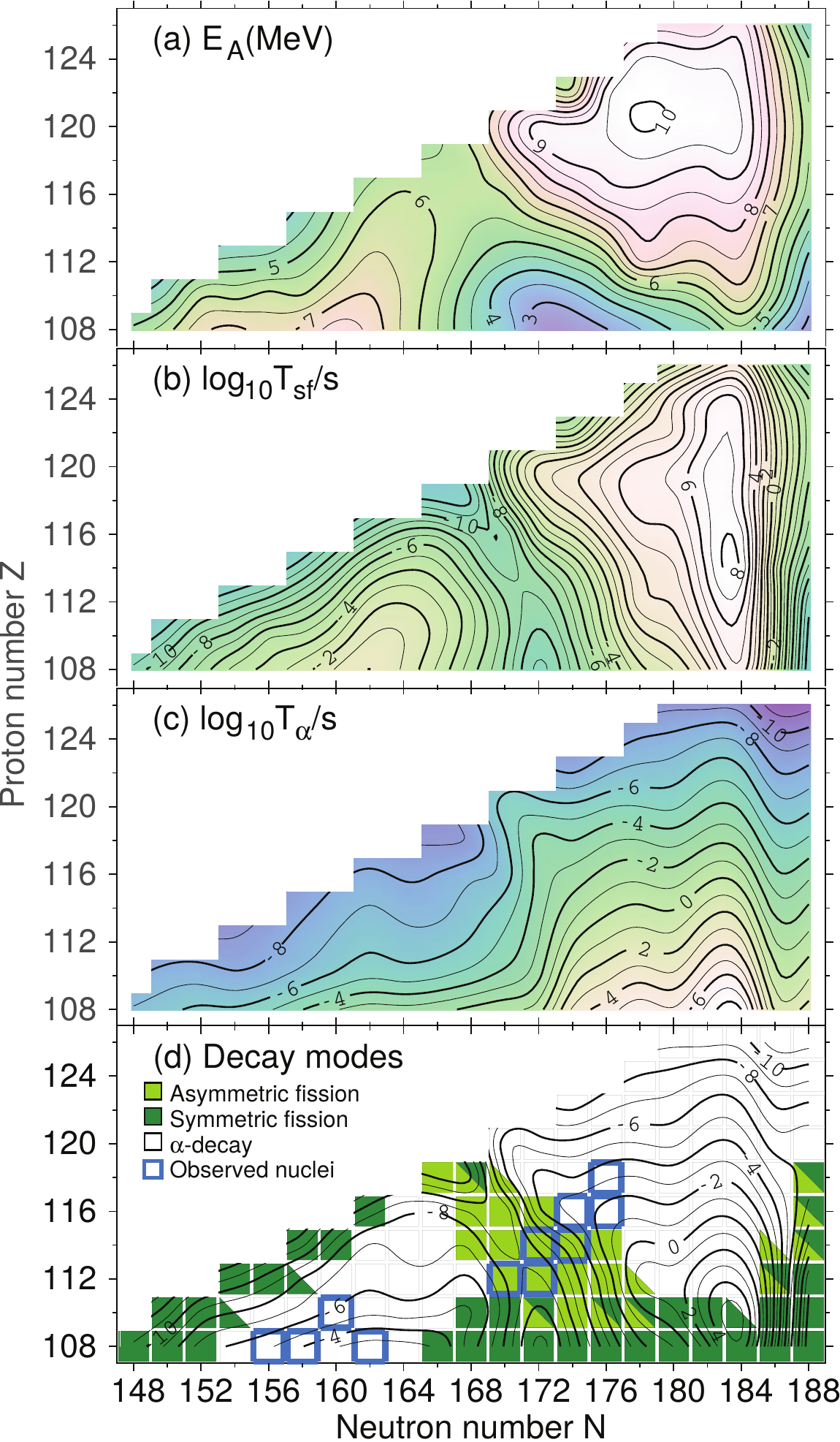}
\caption{Summary of spontaneous fission calculations with the SkM* EDF for 
even-even SH nuclei. (a) Inner fission barrier heights $E_A$ (in MeV); (b) SF
half-lives log$_{10}$ $\tau_{\rm SF}$ (in seconds); (c) $\alpha$-decay half-lives 
log$_{10}$ $\tau_{\alpha}$ (in seconds); (d) Dominant decay modes. If two modes 
compete, this is marked by coexisting triangles.
Figures reproduced with permission from \cite{staszczak2013spontaneous} 
courtesy of Staszczak; copyright 2013 by The American Physical Society.
}
\label{fig:t_sf}
\end{figure}

Spontaneous fission half-lives depend very sensitively on each of these 
ingredients. Because of the exponential factor, small variations in the value 
of the potential energy, zero-point energy or collective inertia can change the 
half life by orders of magnitude \cite{schindzielorz2009fission,
giuliani2013fission,sadhukhan2013spontaneous}. Among the quantities that can 
impact the final result are the number and type of collective variables 
$\gras{q}$ that define the PES. It has been known for a long time, for example, 
that pairing correlations can have a dramatic impact on $\tau_{\rm sf}$, 
largely by reducing the collective inertia which, in a simple solvable 
one-dimensional model scales like $B \propto 1/\Delta^{2}$ where $\Delta$ is the 
pairing gap \cite{urin1966spontaneous,moretto1972influence,moretto1974large,
gozdz1985mass,lazarev1987influence}. This observation was an incentive to take 
collective variables associated with pairing correlations, such as the 
constraint on the number fluctuations $\braket{\Delta\hat{N}^{2}} = 
\braket{\hat{N}^{2}} - \braket{\hat{N}}^{2}$ or the pairing gap $\Delta$ 
\cite{staszczak1989influence,sadhukhan2013spontaneous,
sadhukhan2014pairinginduced,giuliani2014dynamic,
zhao2016multidimensionallyconstrained,bernard2019role}.

\begin{figure}[!ht]
\centering
\includegraphics[width=0.7\textwidth]{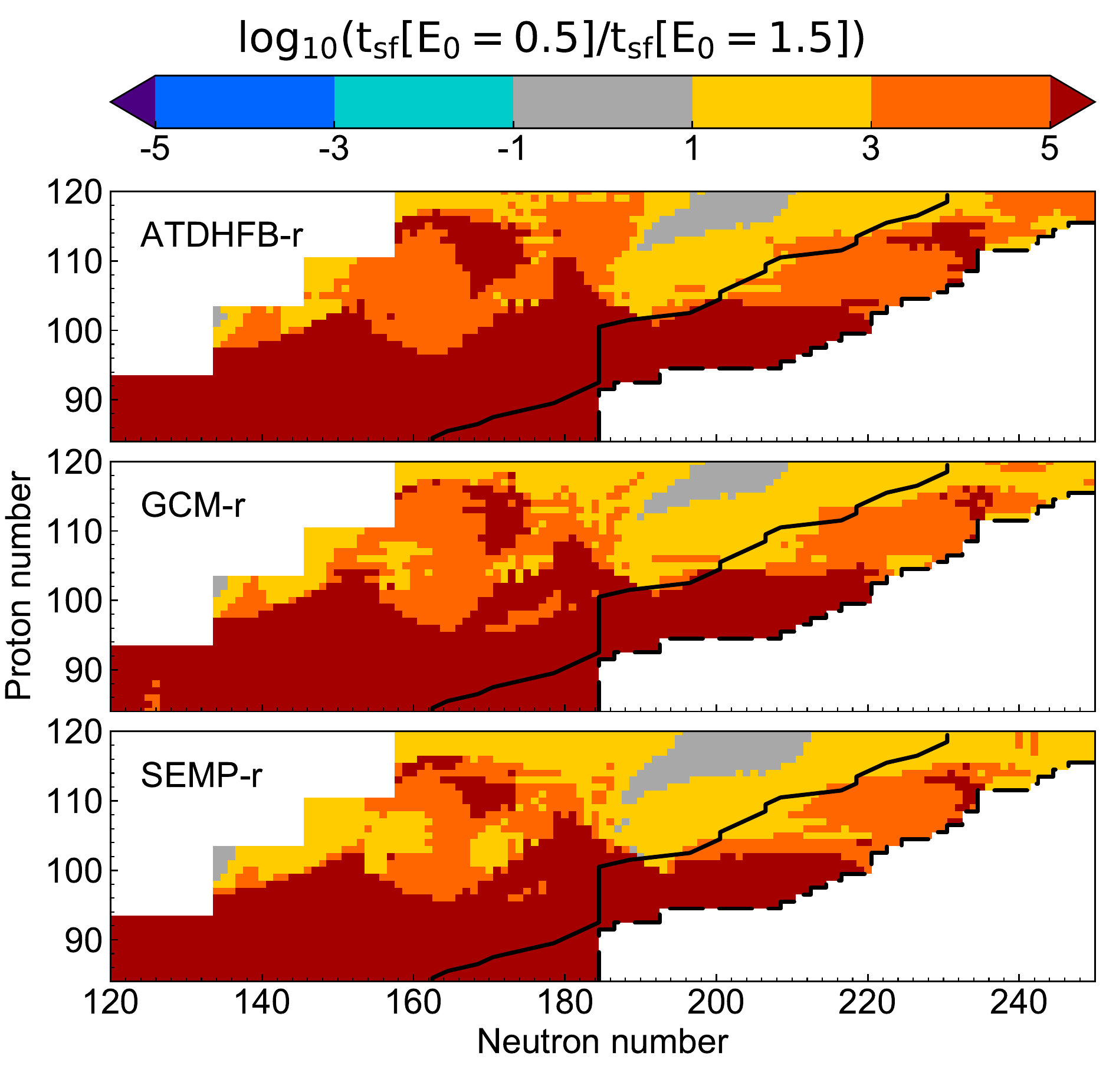}
\caption{Sensitivity of the spontaneous fission lifetimes to different values 
of the ground-state energy 
log$_{10} [\tau_{\rm sf}(E_0 = 0.5\, {\rm MeV})/\tau_{\rm sf}(E_0 = 1.5\, {\rm MeV})]$ 
computed with different collective inertias: ATDHFB (top panels), GCM (middle 
panels), and semi-empirical inertia formula (bottom panels).
Figures reproduced with permission from \cite{giuliani2018fission} 
courtesy of Giuliani; copyright 2018 by The American Physical Society.
}
\label{fig:t_sf_E0}
\end{figure}

A global analysis of spontaneous fission half-lives in heavy nuclei, based on 
quantum tunnelling through one-dimensional potentials including triaxiality 
and reflection-asymmetry, can be found in \cite{baran2015fission}. Earlier 
results were largely focused on superheavy elements with either 
macroscopic-microscopic methods \cite{randrup1976spontaneousfission,baran1981dynamic,
smolanczuk1995spontaneousfission}, Gogny HFB \cite{warda2012fission,
warda2006spontaneous}, Skyrme Hartree-Fock+BCS  \cite{schindzielorz2009fission} 
or Skyrme HFB \cite{staszczak2013spontaneous}. Figure \ref{fig:t_sf} shows an 
example of self-consistent calculation of fission barriers, spontaneous fission 
and $\alpha$-decay half-lives in superheavy elements. Results suggest that in 
superheavy neutron-rich nuclei, spontaneous fission is the dominant channel of 
decay. This is especially important to properly model the r process 
\cite{panov2013influence}. Recent large-scale surveys of fission modes in heavy 
nuclei have attempted to probe the influence of various model parameters such 
as the method to compute the collective inertia or various corrections to the 
ground-state energy \cite{giuliani2018fission}. For example, 
Fig.~\ref{fig:t_sf_E0} shows how the spontaneous fission half-life varies depending 
on the total value $E_0$ of the ground-state energy. As expected because of the 
exponential factor in \eqref{eq:t_sf}, a variation of the ground-state energy 
of only 1 MeV can change half-lives by up to 5 order of magnitude.

Over the years, there have been several attempts to go beyond the 
semi-classical WKB approximation for spontaneous fission rates. 
Functional-integral methods are, formally, the most developed alternatives 
\cite{levit1980timedependenta,levit1980barrier,puddu1987solution,
levit2021variational}. They lead to an imaginary-time evolution equation for 
the nucleus under the barrier, the solutions of which are called instantons. 
This theory has been tested in a few semi-realistic cases 
\cite{skalski2008nuclear,brodzinski2020instantonmotivated}. Other more recent 
approaches include solving directly the time-dependent Schr\"odinger with a 
complex absorbing potential \cite{scamps2015multidimensional}, and using a 
configuration-interaction approach (similar to the nuclear shell model) to test 
the validity of the adiabatic hypothesis, namely the fact that the potential 
energy surface is precomputed \cite{hagino2020microscopic,hagino2020least}. 
Although most of these attempts have been tested with toy model Hamiltonians 
and sometimes involved important numerical simplifications, they clearly 
suggest that the adiabatic approximation may overestimate the spontaneous 
fission half-life significantly.

To finish this section, we should mention a particularly exotic spontaneous 
fission mode called ``cluster'' radioactivity, where one of the fragments is a 
small nucleus of up to mass $A\approx 20$ that is much smaller than its partner 
\cite{rose1984new}. While initial descriptions of this decay mode were based 
on extrapolating the $\alpha$-decay model of Gamow 
\cite{poenaru2002systematics}, it was shown early on that standard fission model 
techniques based on computing tunneling probabilities across multi-dimensional 
potential energy surfaces could be very successful \cite{poenaru2006potential,
mirea2008timedependent}. The main difference is that these PES must probe 
extremely asymmetric shapes with, e.g., a much larger ratio 
$\braket{\hat{Q}_{30}}/\braket{\hat{Q}_{20}}$ than for traditional fission modes. 
This phenomenon was studied extensively in the framework of the SR-EDF approach
\cite{warda2011microscopic,warda2018cluster} and was suggested as the primary 
decay mode of Oganesson ($Z=118$) isotopes \cite{giuliani2019colloquium,
matheson2019cluster}.


\subsection{Neutron-Induced Fission}
\label{subsec:induced}

Neutron-induced fission plays a special role both in basic science and in 
technological applications of fission. As the name indicates, it is a special 
case of fission that is triggered when a neutron interacts with a (heavy) 
target $(Z,N)$ of binding energy $B(Z,N)$. Assuming for now that the neutron is 
absorbed by the target, the energy of the resulting nucleus is then 
$B(Z,N) + E_n$: the resulting nucleus is in an highly excited state of energy 
$E = B(Z,N) + E_n - B(Z,N+1)$ (binding energies are negative). In the case of 
the $^{235}$U(n,f) reaction, for example, the binding energies of $^{235}$U and 
$^{236}$U are $B(^{235}\mathrm{U}) = -1783.102$ MeV and $B(^{236}\mathrm{U}) = 
-1789.648$ MeV leading to an excitation energy of $E = E_n + 6.546$ MeV. In the 
standard view of fission as a deformation process, neutron-induced fission 
occurs or, more precisely, has a large probability to occur whenever $E$ (in 
the compound nucleus) exceeds the height of the highest fission barrier. For 
some naturally-occurring elements such as $^{235}$U, this happens 
regardless of the energy $E_n$ of the incident neutron: these elements are 
called fissile. Odd-mass nuclei are a lot more likely to be fissile than 
even-even ones: pairing correlations induce an additional gain in binding 
energy after the absorption of the neutron, which increases the chance for the 
compound nucleus to be at an energy above the fission barrier. Elements where 
fission can be triggered by higher-energy neutrons, with $E_n > 1$ MeV, are 
called fissionable and a typical case of interest for nuclear technology is 
$^{238}$U. 

\begin{figure}[!ht]
\centering
\includegraphics[width=0.7\textwidth]{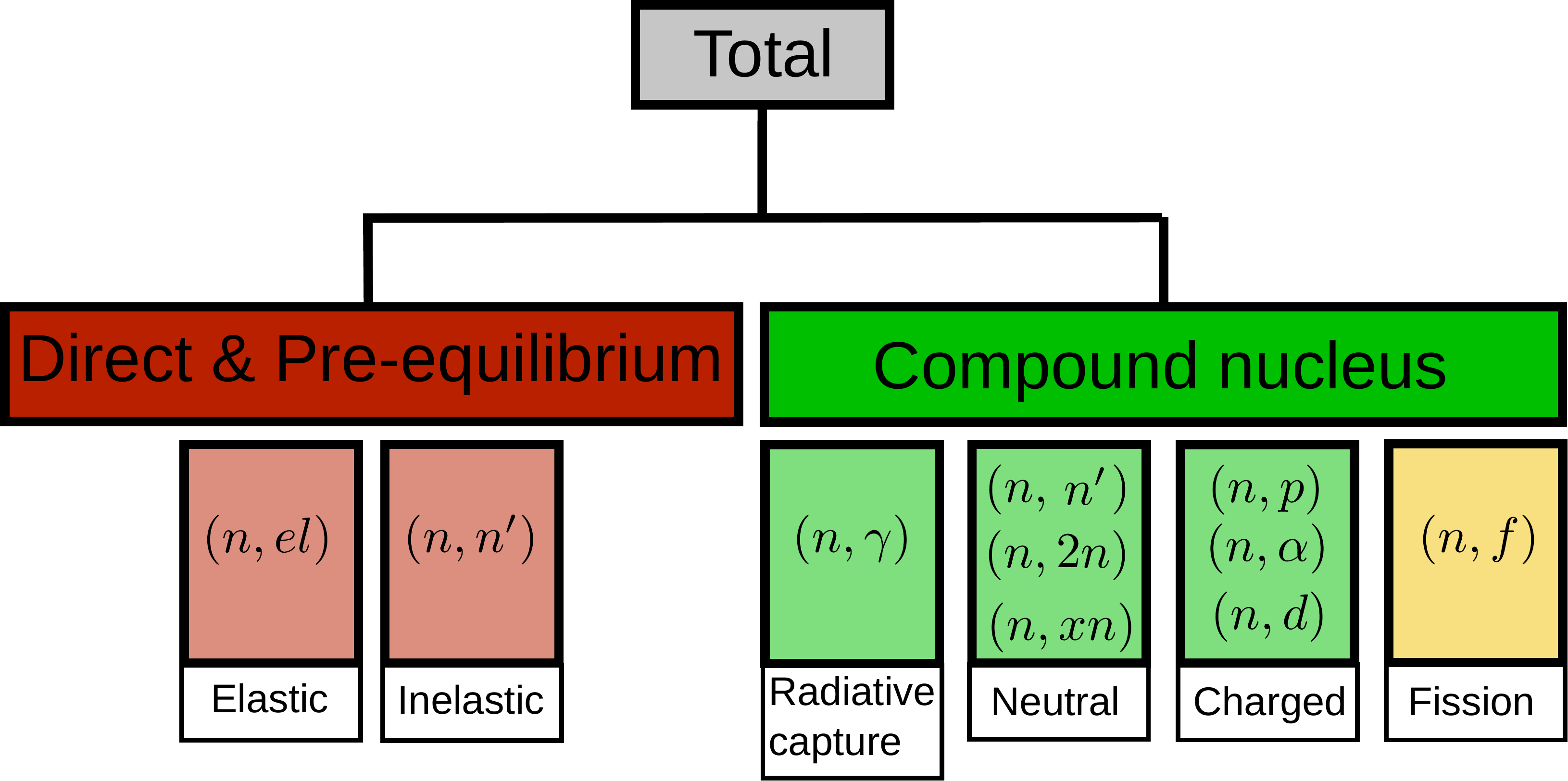}
\caption{Overview of some of the most relevant reaction channels for the 
compound nucleus. Depending on the energy of the incident neutron, other 
reaction mechanisms may be competing. These include direct reactions, where the 
incoming particle interacts with specific rotational/vibrational modes of the target, and 
pre-equilibrium reactions, where several single particle excitations/deexcitations 
take place without a complete thermalization of the compound system.
The total cross section is the 
sum of the cross sections of each process.
}
\label{fig:reactions}
\end{figure}

Qualitatively, the fission cross section $\sigma_{nf}$ can be thought of as 
the probability (expressed as a surface) that an incoming beam's neutron is 
captured by the target 
and that the resulting nucleus decays through fission. 
It is obviously a function of the energy of the incident neutron. 
A very pedagogical introduction to this topic can be found in 
\cite{younes2021introduction}.
In practice, the 
theoretical treatment of neutron-induced fission is thus markedly more complex 
than that of spontaneous fission. Because of the substantial amount of 
excitation energy in the fissioning nucleus many decay channels such as neutron 
emission $(n,xn)$, radiative capture $(n,\gamma)$, etc. can be open. Crucially, 
the fission cross section also depends on the entrance channel, i.e., the 
probability that the neutron is absorbed in the first place -- a process that 
competes with scattering, both elastic $(n,el)$ and inelastic  $(n,n')$. 
Therefore, the variations of the fission cross section with the nucleus $(Z,N)$ 
or the energy $E_n$ of the incident neutron depend not only on how the 
characteristics of the fission process itself vary with these quantities, but 
also on how every other channel is affected. 
For example, neutron emission from 
an excited nucleus cannot occur if the energy is lower than the neutron 
separation energy $S_n$. 
This channel is thus entirely closed for all $E < S_n$ 
(the cross section is zero), a situation that 
can happen for instance in low energy gamma-induced fission.
This will in turn 
impact in a non-trivial way how the fission cross section will change as a 
function of $E$. 

Figure \ref{fig:reactions} lists some of the various channels 
that should typically be considered in a heavy nucleus such as actinides and 
transactinides.
It is important to mention that the various output channels may be
populated by processes with very different time scales. 
In short-time processes, the neutron may simply scatter without 
interacting with the internal region of the target nucleus. 
This corresponds to 
what is called direct reactions,
which constributes to the elastic scattering channel and possibly the inelastic
one when low energy vibrational/rotational modes of the target are excited.
In intermediate-time processes at higher energies, the neutron can lead to successive excitations/deexcitations of 
single particles degrees of freedom (DoFs) of the target with not enough time for the whole system to thermalize. 
This is the pre-equilibrium part of the cross
section~\cite{bonetti1991feshbachkermankoonin}, which populates mostly the inelastic
scattering channel. Finally, the longer-time processes
consist in the absorption of the neutron leading to a thermalized compound nucleus.
In this case, one assumes 
that the compound nucleus is in a strongly-damped regime 
\cite{bertsch1983damping,lauritzen1986damping,matsuo1993poisson,
dossing1996fluctuation,matsuo1999microscopic}: its wave function 
is a very complex, incoherent linear superposition of many 
different types of intrinsic, collective and coupled wave functions.
The compound nucleus itself emits particles and populate a variety of output
channels. Due to its typically large time scale, neutron-induced fission falls 
into this category of compound nucleus processes.

\begin{figure}[!ht]
\centering
\includegraphics[width=0.45\textwidth]{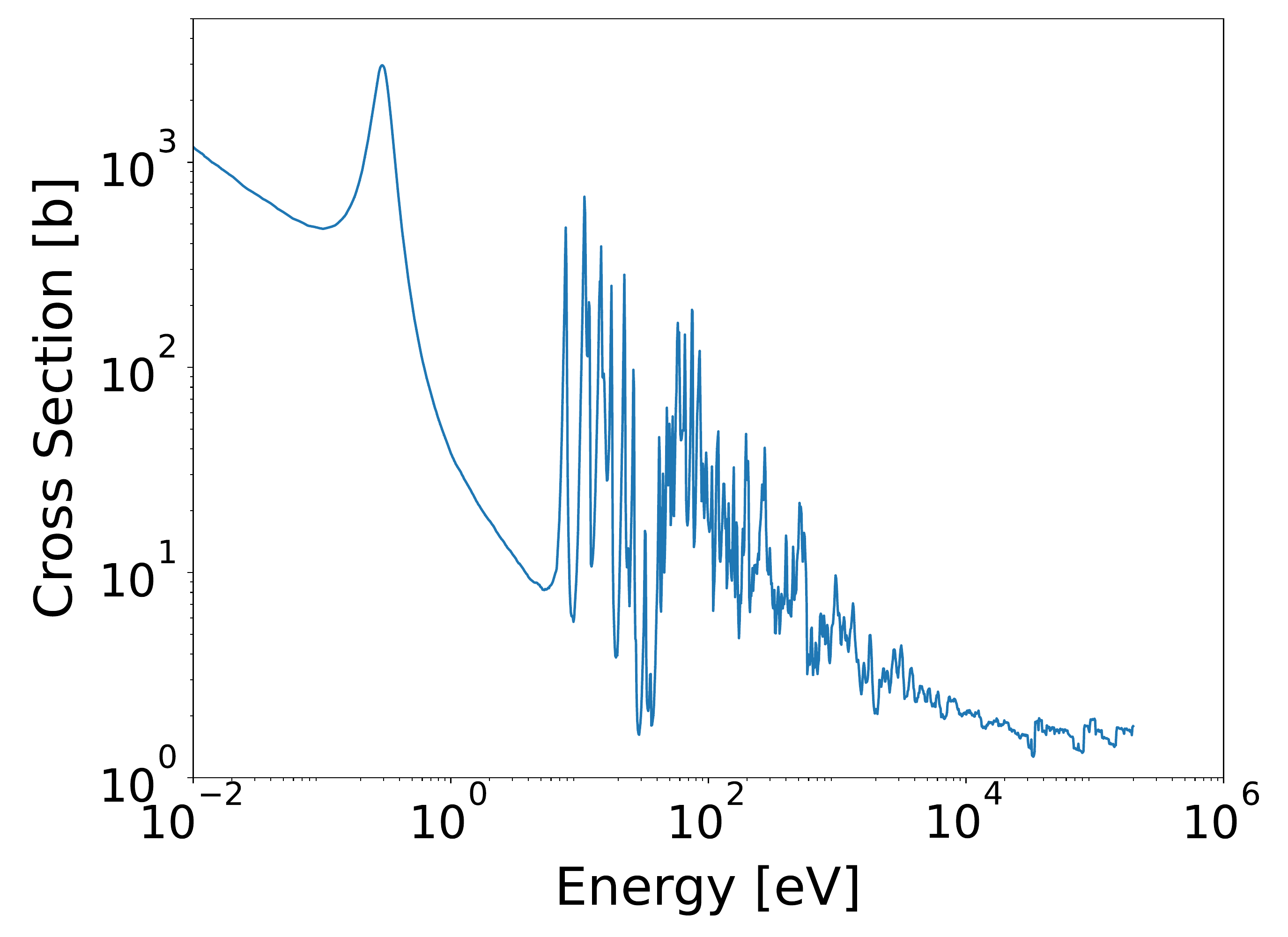}
\includegraphics[width=0.45\textwidth]{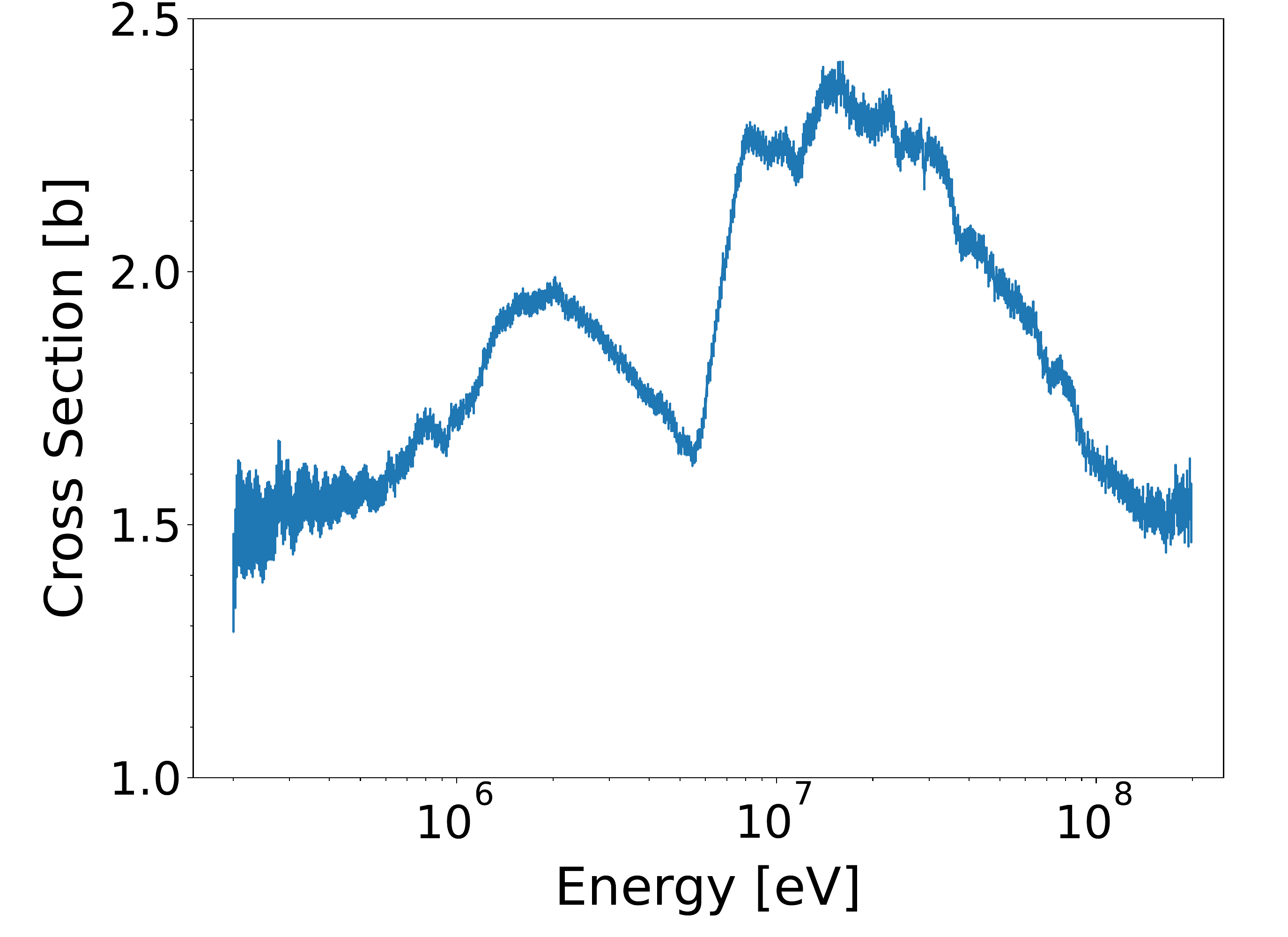}
\caption{Fission cross sections in the reaction $^{239}$Pu(n,f) as a function 
of the energy of the incident neutron. Data taken from the EXFOR database 
\cite{otuka2014more,zerkin2018experimental}; measurements from 
\cite{tovesson2010cross}. Left: thermal and resonance region $E_{n} < 0.1$ MeV; 
both axes are in log scale. Right: fast region $E_n > 0.1$ MeV; the x-axis is 
in log scale.}
\label{fig:cross_sections}
\end{figure}

As an example, Fig.~\ref{fig:cross_sections} shows the variations of the fission 
cross section as a function of the incident neutron energy for the case of
$^{239}$Pu(n,f). At very low energies $E_n < 0.1$ eV (the {\it thermal} regime), 
the cross section varies smoothly with the energy and is roughly inversely
proportional to the neutron velocity. At such low-energies, the $(n,n')$ and 
$(n,xn)$ channels are closed and the fission cross section results from the 
competition between elastic scattering, photodeexcitation and actual fission. 
In fact, the smoothly decreasing trend mostly reflects the fact that the 
probability of absorption decreases with neutron energy (or equivalently: the 
scattering probability increases). The epithermal or resolved resonance region 
ranges from 1 to about $10^4$ eV. In this range of neutron energies, the 
resonances associated with the capture of the neutron by the target can be 
resolved experimentally. In this maze of resonances, only a few actually
corresponds to the physics of fission while most of them are related with the 
probability of capturing the neutron. Finally, the region of $E_n > 100$ keV corresponds to what 
is defined as fast neutrons. At such energies, most decay channels are open: 
$(n,\gamma)$, $(n,n')$ and $(n,xn)$. The rapid increase of $\sigma_{nf}$ at 
around 5-6 MeV corresponds to second-chance fission, i.e., a neutron is emitted 
from $^{240}$Pu but leaves enough energy in the $^{239}$Pu daughter that this 
nucleus fissions. Note that several features of the resonance and fast-neutron 
region are markedly different in fissile and fissionable elements, as 
illustrated in Figure \ref{fig:cross_sections_1} page 
\pageref{fig:cross_sections_1}.

\subsubsection{Fission Rates at High Energy}
\label{subsubsec:rates}

One of the earliest theories attempting to compute the probability of induced 
fission was proposed by Kramers \cite{kramers1940brownian}. The main hypothesis is to assume that the fission 
{\it rate} can be obtained by computing the flux of a current of probability 
$\gras{j}(\gras{q},\gras{p},t)$ through the saddle point. 
This semi-classical and statistical model depicts the nucleus as a few shape collective
coordinates $\gras{q}$ and their associated momenta $\gras{p}$ interacting
with a thermal bath of residual DoFs. 
The current $\gras{j}(\gras{q},\gras{p},t)$ is computed from the probability 
distribution function $f(\gras{q},\gras{p},t)$ representing the
probability for the nucleus to be at point $(\gras{q},\gras{p})$ at time $t$.
The equation of motion for this quantity is the Kramers equation 
\cite{abe1996stochastic},
\begin{align}
\label{eq:kramers}
\frac{\partial f}{\partial t}(\gras{q},\gras{p},t)
= \sum_{\alpha\beta}
\left[
- B_{\alpha\beta}p_{\beta}\frac{\partial}{\partial q_{\alpha}}
+ \left(  \frac{\partial V}{\partial q_{\alpha}} + \sum_{\delta}\frac{1}{2}\frac{\partial B_{\beta\gamma}}{\partial q_{\alpha}} p_{\beta}p_{\delta}
\right)\frac{\partial}{\partial p_{\alpha}} \right.\nonumber\\
\left.
+ \sum_{\gamma}\Gamma_{\alpha\beta}B_{\beta\gamma}\frac{\partial}{\partial p_{\alpha}}p_{\delta}
+  D_{\alpha\beta}\frac{\partial^{2}}{\partial p_{\alpha}\partial p_{\beta}}
\right] f(\gras{q},\gras{p},t),
\end{align}
where $D_{\alpha\beta}(\gras{q}) = \Gamma_{\alpha\beta}(\gras{q})T$ is the dissipation tensor,
$T$ the temperature of the bath, $\Gamma_{\alpha\beta}$ the friction tensor and 
$B_{\alpha\beta}(\gras{q})$ the collective inertia tensor. We define the 
reduced friction tensor $\gras{\beta} = \newtensor{\Gamma} \newtensor{B}$, that is, 
$\gras{\beta} \equiv \beta_{\alpha\gamma} = \sum_{\gamma}\Gamma_{\alpha\beta}B_{\beta\gamma}$.
The Kramers equation 
is the extension of the Fokker-Planck equation in presence of a potential 
$V(\gras{q})$ \cite{risken1989fokker,hanggi1990reactionrate}. In addition to 
not assuming a stationary process, the advantage of such approaches based on 
diffusion is that they include a mechanism to couple intrinsic excitations to 
collective motion through the dissipation tensor $D_{\alpha\beta}$. 

In its original formulation, Kramers used a one-dimensional version of 
\eqref{eq:kramers} where he further assumed that the potential well near the 
ground state was approximately parabolic with a frequency $\omega'$ and that 
the top of the barrier was also well described by an inverted parabola of 
frequency $\omega_0$. In such a case, he obtained the following fission rate $\lambda_f$ in 
the limit of large dissipation,
\begin{align}
\lambda_f =  \frac{\omega'}{2\pi\omega_0}
\left( \sqrt{\frac{\beta^2}{4} + \omega'^2} - \frac{\beta}{2} \right) e^{-E_f/kT} ,
\end{align}
where $T$ is the nuclear temperature, which is given (approximately) by 
$T \approx \sqrt{E^{*}/a}$ with $a$ the level-density parameter,  $E^{*}$ the 
excitation energy, and $\beta$ is the one-dimensional version of the reduced friction tensor 
introduced earlier. This approach was later generalized in multi-dimensional 
cases \cite{grange1980fission,grange1983induced,jing-shang1983generalization,
grange1984effects,weidenmuller1984stationary}. In this case, assuming that the 
local minimum of the potential $V(\gras{q})$ characterizing the ground state is 
at $\gras{q}_{\rm gs}$ and the saddle point at $\gras{q}_{\rm B}$, the formula for 
the fission rate becomes
\begin{equation}
\lambda_f = \frac{1}{2\pi} e^{-V_{B}/kT} \left( \frac{\det W^{\rm (gs)}}{|\det W^{\rm (B)}|} \right)^{1/2}\Lambda ,
\end{equation}
with 
\begin{equation}
W^{\rm (gs)}_{\alpha\beta} = \left. \frac{\partial^2 V}{\partial q_{\alpha}q_{\beta}}\right|_{\gras{q} = \gras{q}_{\rm gs}} ,
\qquad
W^{\rm (B)}_{\alpha\beta} = \left. \frac{\partial^2 V}{\partial q_{\alpha}q_{\beta}}\right|_{\gras{q} = \gras{q}_{\rm B}} ,
\end{equation}
and $\Lambda$ given by the only positive eigenvalue of the matrix equation
$\det |\newtensor{M}\Lambda^2 + \gras{\beta}\Lambda + \newtensor{W}^{\rm (B)}| = 0$, where $\newtensor{M}$
is the collective mass tensor, $\newtensor{M} = \newtensor{B}^{-1}$. Projecting 
such multi-dimensional calculations on one-dimensional trajectories is 
equivalent to introducing a mass correction and an additional friction term 
\cite{brink1986decay}. 

\begin{figure}[!ht]
\centering
\includegraphics[width=0.6\textwidth]{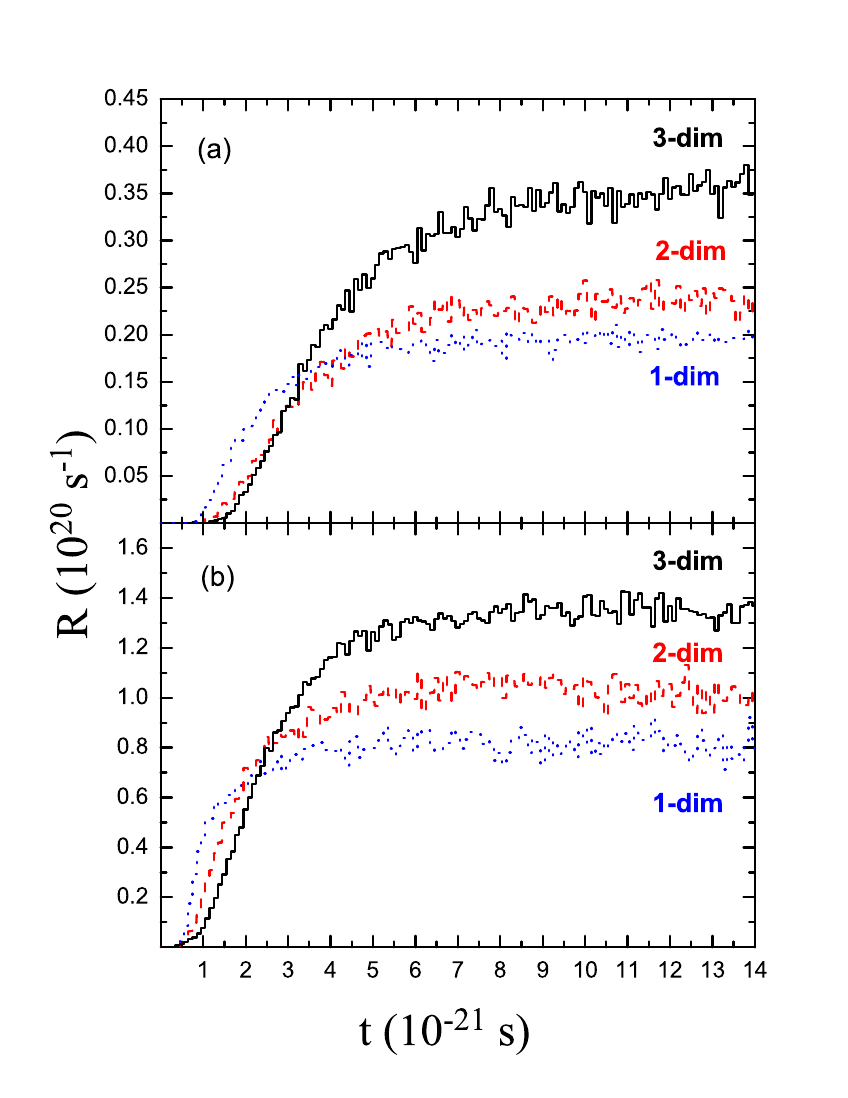}
\caption{Fission rate calculated for the nucleus $^{248}$Cf in the case of 
two-body dissipation for excitation energy $E^{*} = 30$ MeV (a) and 
$E^{*} = 1500$  MeV (b). The solid, dashed, and dotted curves correspond to the 
Langevin calculations in three-, two-, and one-dimensional collective potential 
energy surfaces, respectively.
Figures reproduced with permission from \cite{nadtochy2007fission} courtesy of 
Nadtochy; copyright 2007 by The American Physical Society.
}
\label{fig:rate}
\end{figure}

Early studies of fission rates were limited since numerically solving the 
Kramers equation \eqref{eq:kramers} is notoriously difficult \cite{gontchar2011integral}. However, 
the solutions to this equation, which are probability distribution functions, 
can also be sampled by solving the stochastic Langevin equation 
\eqref{eq:langevin}. Given a large enough number of full Langevin trajectories, 
the time-dependent fission rate is given by $\lambda_f(t) = - 1/N(t) dN/dt$ 
where $N(t)$ is the number of trajectories that did not escape the saddle point 
at time $t$ \cite{boilley1993nuclear,nadtochy2007fission}. At small $t$, this 
fission rate is nearly 0, while at larger $t$ calculations show that the rate 
converges to a nearly constant, quasistationary value 
\cite{wada1992multidimensional,gontchar1993nuclear,nadtochy2007fission} that 
matches what one would obtain from the solution of the Kramers equation 
\cite{frobrich1992pathintegral}, as illustrated in Fig.~\ref{fig:rate}. The 
transient time needed for the system to 
reach quasistationary flow plays an important role when other decay channels 
such as neutron or $\gamma$ emission are taken into account: during that time, 
fission plays a relatively smaller role compared with these other processes, 
and this can affect, e.g., the number of pre-scission neutrons 
\cite{frobrich1993langevin}. As shown in Fig.~\ref{fig:rate}, the fission rate 
is very sensitive to the size of the collective space and the excitation 
energy, but it also depends strongly on the value of the dissipation 
coefficient \cite{sargsyan2007fission}. 
Note that because of its classical character, this approach cannot describe lower-energy effects
related to resonances or even tunneling. 
Until now, it has therefore mostly 
been applied to calculations of fission rates at very high energies such as 
encountered in heavy-ion collisions.



\subsubsection{Fission Cross Sections for Fast Neutrons}
\label{subsubsec:fast_neutrons}

The approach presented above allows computing fission rates (or probabilities) 
but has not been used to the calculation of fission cross sections yet. 
Predicting a fission cross section is actually more demanding as it requires
the proper description of both the input channel and of the competition 
mechanism between all the possible output channels (\textit{e.g.} $\gamma$
and neutron emission).
In the fast-neutron range ($E_n \gtrsim 100$ keV),
such calculations are usually performed within the Hauser-Feshbach theory~\cite{hauser1952inelastic} that relies
on a statistical picture of nuclear structure.

\paragraph{General Theoretical Framework}

In the high-energy range the density of resonances is so high that it is impossible
to resolve them experimentally. 
As a consequence, theoretical approaches focus on estimating the average 
cross sections over a small sliding window in energy.
The standard theory to describe the average cross section for compound-nucleus
processes in this energy range is the Hauser-Feshbach theory. 
A pedagogical review of this approach can be found in~\cite{hodgson1987compound}.
The Hauser-Feschbach framework relies heavily on the Bohr hypothesis stating that the probability
for an output channel depends only on the energy, spin and parity of the compound 
nucleus and is independent of the details of its formation process~\cite{bohr1936neutron,weisskopf1937statistics,bethe1937nuclear,
weisskopf1940yield,hodgson1987compound,brink1990compound}.
Its goal is to compute 
the average cross section $\sigma_{\alpha\beta}(E,J,\pi)$ that describes the formation 
through an entrance channel generically noted $\alpha$ of a compound nucleus of 
energy $E$, spin $J$ and parity $\pi$ that subsequently decays through a 
channel $\beta$. Channels refer both to a certain combination of projectile 
and targets, e.g., $\alpha \equiv n + ^{239}\mathrm{Pu}$, but also to specific 
combinations of quantum numbers for each of the constituents. 
In this context, fission is depicted by one macroscopic channel without details about
the exact fragmentations. For example both the $^{138}$Fe+$^{102}$Ca
and $^{140}$Cs+$^{100}$K mass splits will be part of what is considered the fission channel of $^{240}$Pu.
Given an entrance 
channel $\alpha = a+A$ corresponding to an incident particle $a$ and a target 
$A$, and an exit channel $\beta$ (e.g. neutron or gamma emission), the cross section for the process,
assuming no width fluctuations, reads~\cite{hodgson1987compound}
\begin{equation}
\sigma_{\alpha\beta}(E,J,\pi) 
= \pi\lambdabar^2
\sum_{J\pi} \frac{2J+1}{(2j_{a} + 1)(2J_{A}+1)}
\sum_{l_{\alpha}j_{\alpha} l_{\beta}j_{\beta}} 
\frac{T_{\alpha l_\alpha j_\alpha}^{J\pi}(E) 
      T_{\beta l_\beta j_\beta}^{J\pi}(E)}{\displaystyle\sum_{\delta} T_{\delta}(E)} ,
\label{eq:hauser-feshbach-1}
\end{equation}
where $j_a$ and $J_A$ are the total angular momentum of the incident and target nuclei,
$J$ is the total compound-nucleus angular momentum, $l_{\alpha}$, $j_{\alpha}$ 
and $l_{\beta}$, $j_{\beta}$ are the orbital and angular momentum coupling in the channels $\alpha$
and $\beta$,
and $\lambdabar$ is the reduced wavelength of the incident particle $a$.

The coefficients $T_{\alpha}(E) \equiv T_{\alpha l_\alpha j_\alpha}^{J\pi}(E)$ are the transmission coefficients for the 
channel $\alpha$ at energy $E$. In neutron-induced fission, the typical decay
channels involve neutron emission, $\gamma$ emission, and actual fission, while 
the entrance channel is neutron absorption. In the case of resolved resonances, $T_{\alpha}(E) = 2\pi 
\Gamma_{\alpha}/D$ where $\Gamma_{\alpha}$ is called the width of the channel 
$\alpha$, which is inversely proportional to the lifetime $\tau$ of the state 
through $\Gamma_{\alpha} = \hbar / \tau_{\alpha}$, and $D$ is the level spacing 
at energy $E$ \cite{blatt1979theoretical}. 
In practice, the partial widths fluctuate rapidly with the energy.
The average cross section is obtained by averaging \eqref{eq:hauser-feshbach-1}
over a small energy window centered on $E$ by using the relation
\begin{equation}
\left\langle
\frac{\Gamma_{\alpha}(E) \Gamma_{\beta}(E)}{\displaystyle\sum_{\delta} \Gamma_{\delta}(E)} 
\right\rangle
= W_{\alpha\beta}
\frac{\braket{\Gamma_{\alpha}(E)}\braket{\Gamma_{\beta}(E)}}{\displaystyle\sum_{\delta} \braket{\Gamma_{\delta}(E)}} ,
\end{equation}
where $\braket{\dots}$ refers to the energy averaging. The actual transmission 
coefficients used in \eqref{eq:hauser-feshbach-1} are thus 
defined as the {\it average} quantities $T_{\alpha}(E) = 2\pi 
\braket{\Gamma_{\alpha}}/D$ and $W_{\alpha\beta}$ is called 
the width-fluctuation correction. If $\braket{\Gamma} / D \ll 1$, then this 
correction can be calculated semi-analytically \cite{moldauer1961theory}. Let 
us emphasize that we have discussed here only the simplest version of the 
Hauser-Feshbach theory and have restricted ourselves to the formula most 
relevant for fission. In-depth discussions and additional references can be 
found in several review articles \cite{mahaux1979recent,hodgson1987compound}. 
One of the main limitations of Hauser-Feshbach theory is its assumption that 
channels do not interfere, i.e., direct interactions cannot take place. 
Several extensions of the theory were developed in the 1970ies to address these 
issues but they are not applied very frequently \cite{engelbrecht1973hauserfeshbach,kawai1973modification,
agassi1975statistical}. 

From a practical perspective, the application of Hauser-Feshbach theory to 
fission implies that we ``only'' have to determine the transmission 
coefficients of all the relevant processes. Their expression depends on the 
type of channel involved, e.g., the neutron optical model gives the neutron transmission $T_n(E)$. 
In this review, we are only going to discuss the 
calculation of $T_f(E)$; see \cite{talou2021fission} and references therein for 
additional details about the neutron-capture, neutron-emission and $\gamma$ 
transmission coefficients. 

\paragraph{Calculation of Transmission Coefficients}

The standard method to determine fission transmission coefficients $T_f(E)$ 
consists in computing the penetrability through a one-dimensional fission barrier
~\cite{bjornholm1980doublehumped}.
This involves first the construction of a collective potential and inertia
as a function of a single collective coordinate associated with the elongation of the
fissioning system.
The probability that the compound nucleus decay through fission 
derives from the transmission of a collective plane wave from the 
low-deformations to the high-deformation domain.
In other words, the transmission coefficient is extracted from the solution of 
a stationary version of an equation of the type \eqref{eq:evolution0} 
(see Section~\ref{subsubsec:TDGCM_GOA}).
It is important to realize that such a wave equation is not the Schr\"odinger 
equation for a particle, but the equation of motion for a collective wave 
packet traveling through the collective space (=deformation space) 
characterizing the fission process. 
Although this approach could in principle be rooted in first-principle theories
~\cite{bjornholm1980doublehumped} it is still mostly guided by 
phenomenological considerations.
When the energy of the system lies above the fission barrier, this approach 
is often combined with the transition-state model 
\cite{wigner1938transition}: at the top of each barrier are a sequence of 
discrete excited states of energies $\epsilon_{k}$. For each such state, there 
is an associated transmission coefficient \eqref{eq:Tf} determined from
a collective potential up-shifted by an energy $\epsilon_k$.
These states are typically modeled as members 
of collective rotational bands built on top of vibrational bandheads. The total 
fission transmission coefficient thus becomes
\begin{equation}
T_f(E) = \sum_{k} f_k T_f(E-\epsilon_k) ,
\label{eq:Tf_total}
\end{equation}
where $f_k = 1$ if the spin and parity of the transition state equals that of 
the compound nucleus, and is zero otherwise. Note that \eqref{eq:Tf_total} can  
be extended by additionally considering a continuum of excited states 
\cite{koning2019talys}.

The main ingredients of this approach are the collective potential and the associated
inertia. For fissile nuclei such 
as $^{239}$Pu and $^{235}$U, the energy $E$ of the compound nucleus always 
exceeds the highest fission barrier irrespective of the energy of the 
incident neutron. 
In this regime, transmission coefficients are most of the time determined from the formula 
given by Hill and Wheeler \cite{hill1953nuclear}. It is 
obtained by assuming that the collective potential is an inverted parabola 
of frequency $\omega_0$ and height $E_f$. If we work within a 
one-dimensional collective space and further assume that the collective inertia 
does not depend on the collective variable, then the wave equation can be 
solved analytically and yields the following Hill-Wheeler formula for the 
transmission coefficient,
\begin{equation}
T_f(E) = \frac{1}{1 + \exp \big[ 2\pi(E_{f} - E)/\hbar \omega_0 \big] } .
\label{eq:Tf}
\end{equation}
Note that the resulting transmission coefficient is independent of the 
collective inertia.
In practice, \eqref{eq:Tf} is often applied by using semi-realistic estimates 
of the fission barrier height $E_f$ from either 
macroscopic-microscopic~\cite{sierk1986macroscopic,mamdouh2001fission} or fully 
microscopic calculations~\cite{goriely2009prediction} and by choosing $\omega_0$ as an adjustable parameter. 
In the case of the double-humped fission barrier typical of actinide nuclei, the 
full transmission coefficient is computed by coupling the coefficients of 
the first ($T_{A}$) and second ($T_B$) barriers according to
\begin{equation}
\label{eq:tatb}
T_f = \frac{T_{A}T_{B}}{T_{A} + T_{B}} ,
\end{equation}
if we assume a statistical equilibrium in the class II states~\cite{tamagno2015challenging}.

\begin{figure}[!ht]
\centering
\includegraphics[width=0.57\textwidth,align=c]{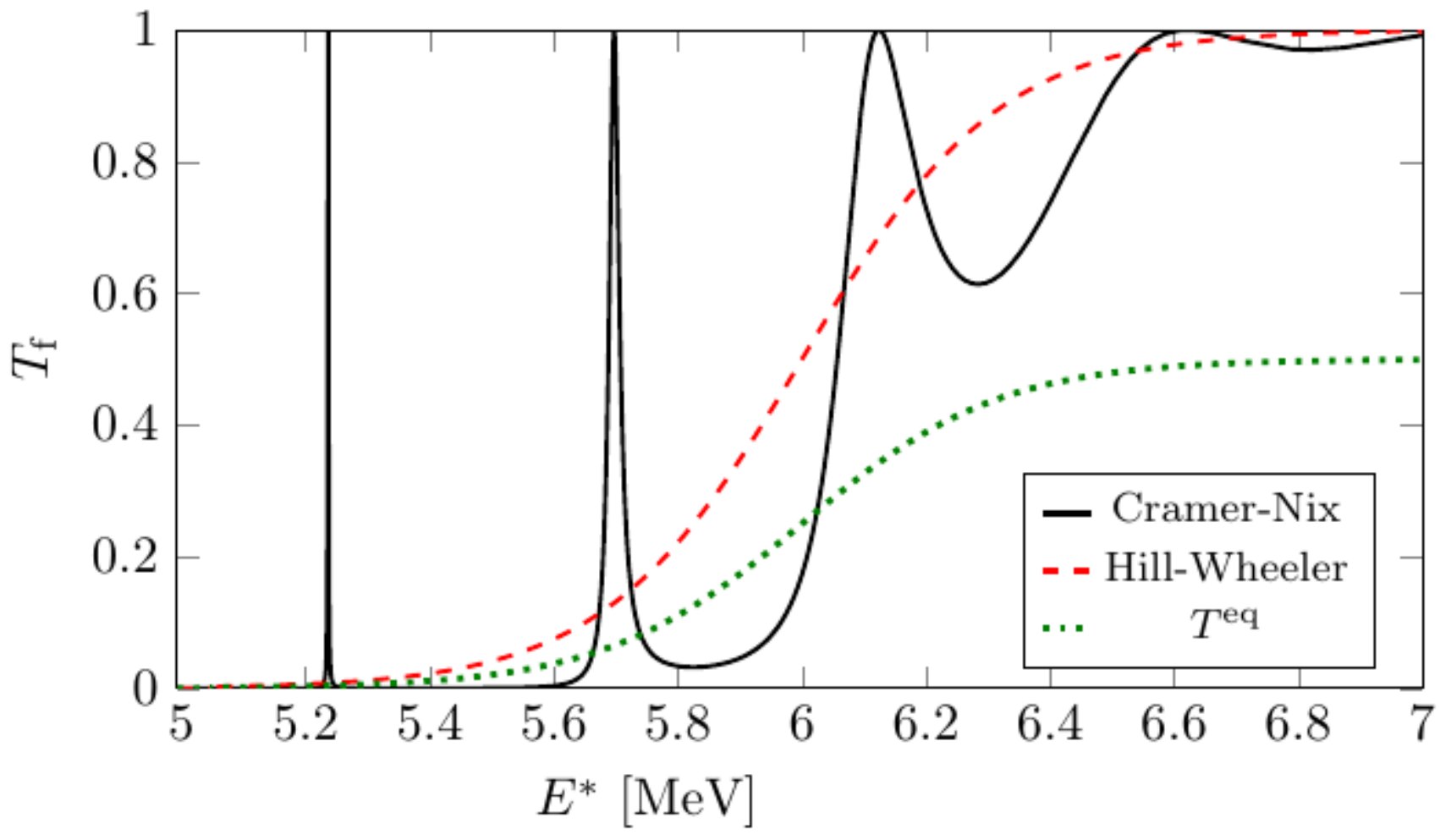}
\includegraphics[width=0.42\textwidth,align=c]{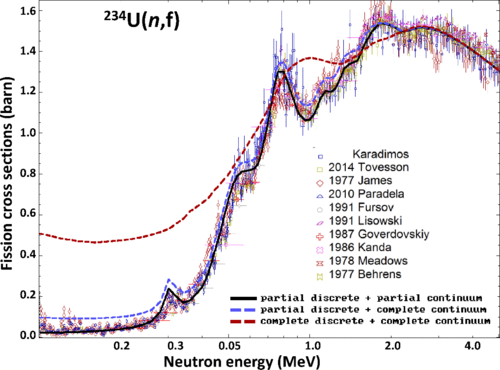}
\caption{Left: Fission transmission coefficient as a function of the excitation energy
of the fissioning system. The one humped Hill-Wheeler estimation (red curve) is compared
to its counterpart when coupling two barriers as in Eq.~\eqref{eq:tatb} 
(green dotted line) as well as the one obtained from a Cramer-Nix potential (full
black line). While the Hill-Wheeler formula gives a smooth evolution with energy,
the Cramer-Nix approach predicts resonances in qualitative agreement with experiment.
Courtesy of P.~Tamagno~\cite{tamagno2015challenging}.
Right: $^{234}$U cross section predicted by the extended optical model for fission 
(full black line) compared to various experimental data. Figure reproduced with permission
from \cite{sin2016extended} courtesy of M. Sin; copyright 2007 by The American Physical
Society.
}
\label{fig:fission_transmission}
\end{figure}

Equation \eqref{eq:Tf} could in principle be generalized to a more realistic collective space. 
Along this line, we can mention the early work of Cramer and Nix
solving the case of two inverted parabola on a finite support connected by 
a parabolic shape~\cite{cramer1970exact}. The subsequent description of
quasi-bound states in the outer potential qualitatively explains the 
presence of resonances in the fission cross section at energies below the 
barrier height; see Fig.\ref{fig:fission_transmission}.
Similar studies are performed in Ref.~\cite{leboeuf1973fmatrix}
Going even further, one can add an imaginary part to the collective potential,
typically in the second well, to account for the damping of collective vibrations.
This idea leads to the fission optical model and is for instance discussed in
Ref.~\cite{sin2016extended,vladuca1994probfis}.
As shown in the right panel of Fig.\ref{fig:fission_transmission}, such an approach
reproduces nicely the large (few 100 keV) resonances just below the fission threshold.

It could also be possible to solve Eq.\eqref{eq:Tf} numerically for 
an arbitrary potential and inertia functions.
Doing so still poses a number of practical challenges. Encoding the physics of 
fission requires collective spaces with more than one deformation degree of 
freedom. 
Most importantly, the Hill-wheeler equation was obtained by neglecting 
the deformation dependence of the collective inertia: in this case, the final 
formula is independent of it. In a more realistic calculation, the 
transmission coefficient would most certainly depend on it -- like the 
spontaneous fission rate of \eqref{eq:t_sf}. The problem of which boundary 
conditions to adopt is also not trivial. 
Finally, the extreme sensitivity of the fission cross section to the barrier height
as well as to the fine structure of the resonant states in the secondary potential
well makes such approaches highly challenging.


\subsubsection{Resolved Resonance Region: the R-matrix Formalism}
\label{subsubsec:resonances}

When the energy of the compound nucleus becomes close to the top of the fission 
barrier, or is even lower than it, the situation becomes more complex as both 
the level structure of the two potential wells and their coupling begin playing 
a role \cite{weigmann1968interpretation,bjornholm1969intermediate}. In 
particular, the excited states associated with the second potential well, which 
correspond to more elongated shapes and have a higher probability to fission, 
play the role of {\it doorway states} for the fission exit channel 
\cite{feshbach1967intermediate,lynn1974subbarrier,goldstone1978doorwaystate}. 
The presence of such intermediate structures -- states that are neither 
continuum states with a very broad width nor are pure single-particle or 
collective states with a very narrow width -- in the cross sections partly 
explains the structure of the resolved resonance region at $10\, \mathrm{eV} 
< E_n < 10\, \mathrm{keV}$ in Fig. \ref{fig:cross_sections}, together with 
the resonances that come from the entrance channel and reflect an enhanced 
probability that the neutron is absorbed.

While the Hauser-Feschbach theory captures well the smooth-energy trends of the 
cross sections, it is, by design, not able to describe the narrow resonances observed
in the intermediate-energy range.
The description of these fine structures relies on another formalism: the R-matrix theory.
The R-matrix theory is a general, multi-channel approach especially designed 
to handle resonances \cite{lane1958rmatrix,descouvemont2010rmatrix}.
More specifically, it is the reference method used for evaluating the neutron cross sections
in the resolved resonance region (RRR) and is extensively used to fit 
experimental cross section data~\cite{jean2021conrad,larson2008updated}.

We recall that in the standard R-matrix theory of a reaction between a projectile and a target, 
space is divided into two regions, the interaction (or internal) and asymptotic 
regions, and the system target+projectile is also divided into two subsystems 
with different numbers of particles and quantum numbers (=the channels $\alpha$) 
\cite{lane1958rmatrix,descouvemont2010rmatrix}. The wave function in the 
external region is the product of the intrinsic wave functions of each 
subsystem and an asymptotic wave function for the relative motion of the 
center of mass; in the internal region, it is a many-body wave function. The 
actual R matrix is a mathematical object that connects the wave functions of
all possible output channels at the boundary between the internal and 
external regions. 

Fission requires a special treatment in the context of the R-matrix due to the 
fact that it is not by itself a well defined output channel but can actually lead
to a large
number of different fragmentations. 
The extension of the R-matrix formalism to the case of fission was first 
formalized by Lynn \cite{lynn1968structure,lynn1969structure,lynn1973fission};
a comprehensive and pedagogical presentation of his model can also be found in 
Ref.~\cite{vitturi1975doorwaystate}. The basic idea is to identify a set of 
collective variables $\gras{\eta}$ that defines the fission channel and separate 
it from all the other degrees of freedom $\gras{\xi}$ of the nucleus. 
In all applications of this formalism, the vector 
$\gras{\eta}$ has only one component $\eta$ that typically coincides with the axial 
quadrupole deformation $\beta$ of the nucleus -- which is also one of the 
collective variables used to define PES; see Section \ref{subsec:deformation}. 
The associated collective motion will only 
include the corresponding $\beta$-vibrational modes, i.e., the shapes vibrations
of the axial quadrupole moment of the nucleus.
In this approach the many-body Hamiltonian is expanded as \cite{vitturi1975doorwaystate}
\begin{equation}
\hat{H} = 
\hat{H}_{\rm vib}(\gras{\eta}) 
+ 
\hat{H}_{\rm int}(\gras{\xi}) 
+ 
\hat{H}_{c}(\gras{\eta},\gras{\xi}) ,
\label{eq:H_rmatrix}
\end{equation}
where the collective Hamiltonian $\hat{H}_{\rm vib}(\gras{\eta})$ acts only on 
the fission mode variables $\gras{\eta}$. The intrinsic Hamiltonian operator
$\hat{H}_{\rm int}(\gras{\xi})$ acts only on the intrinsic variables 
$\gras{\xi}$, which includes both single- or quasi-particle degrees of freedom, 
rotational modes and any non-$\beta$-vibrational mode. 
$\hat{H}_{c}(\gras{\eta},\gras{\xi})$ encodes the coupling between the fission 
modes and all intrinsic excitations. In analogy to standard R-matrix theory, 
the total space, here the collective space of deformations $\gras{\eta}$, is 
divided into two regions, internal and external. Traditionally, the fission 
channel ``radius'' that defines the boundary is put near, or at the outer 
barrier denoted $\gras{\eta}_0$ \cite{lynn1973fission}.

\begin{figure}[!ht]
\centering
\includegraphics[width=0.48\textwidth]{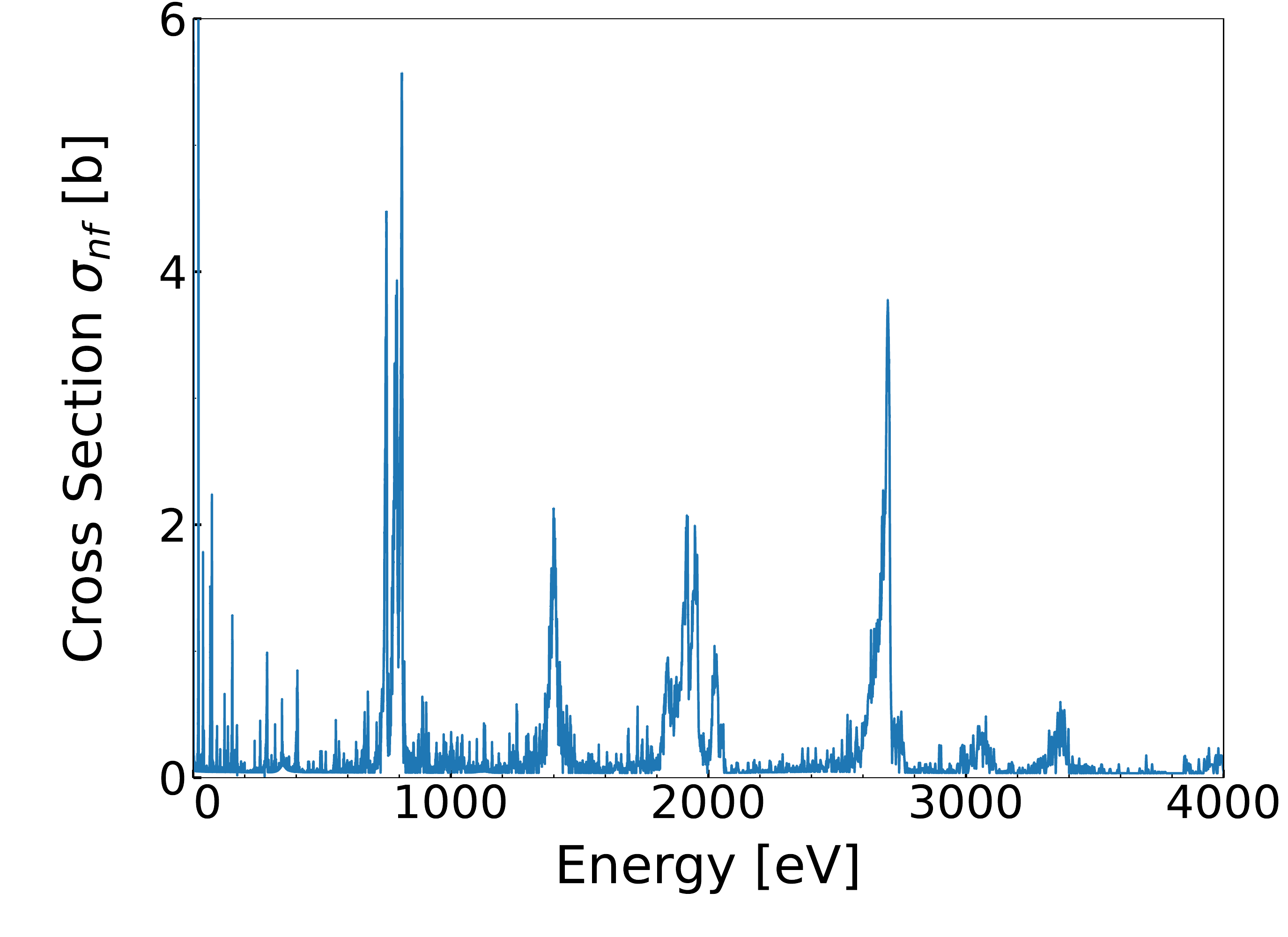}
\includegraphics[width=0.48\textwidth]{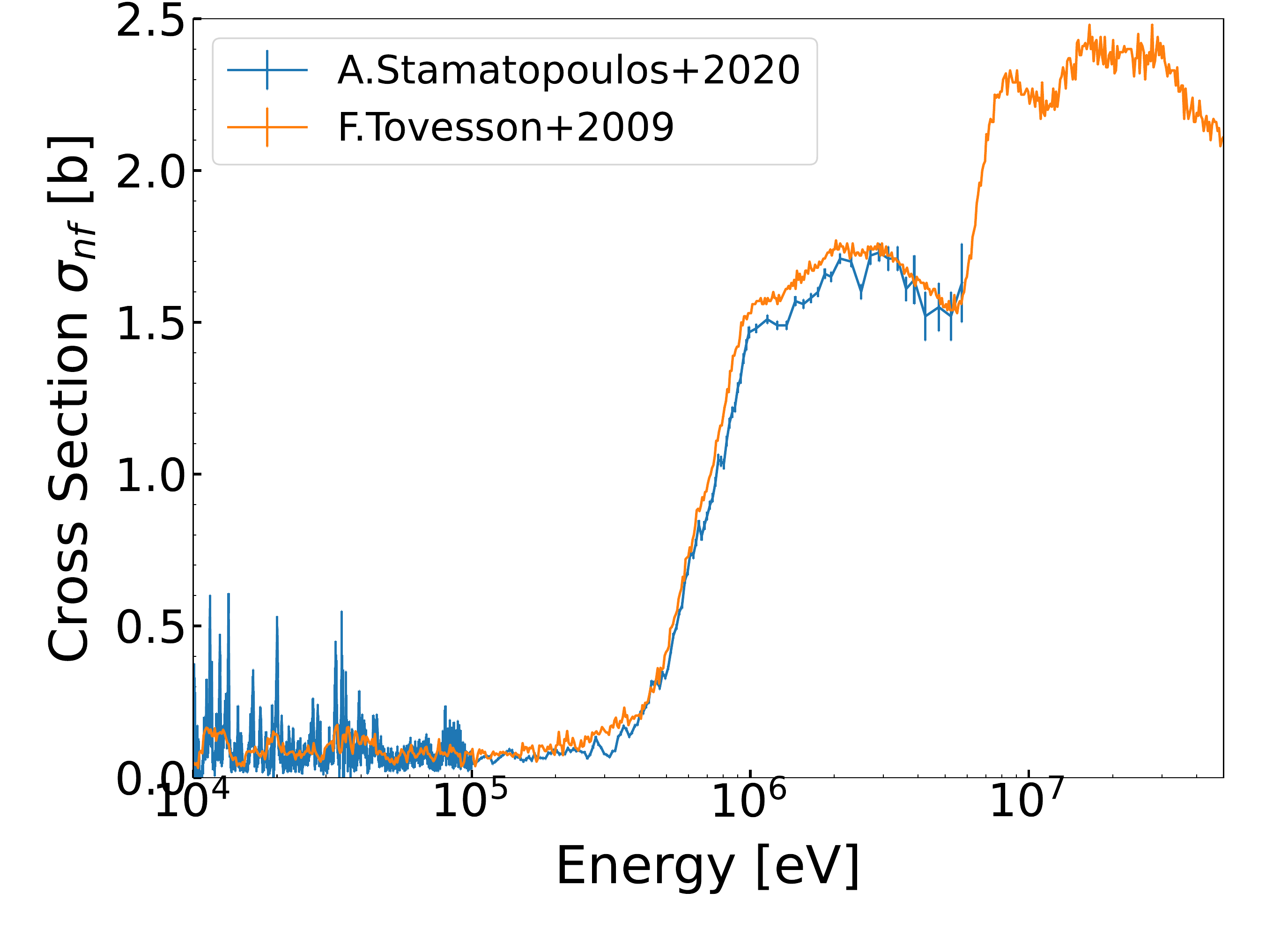}
\caption{Left: Fission cross section for the $^{240}$Pu(n,f) reaction in the 
resonance region for $0\leq E_n \leq 4$ keV. The data is taken from 
\cite{stamatopoulos2020investigation}. 
Right: Same quantity for neutrons energies $10\, \text{keV}\leq E_n \leq 
50\,\text{MeV}$. Data taken from \cite{stamatopoulos2020investigation,
tovesson2009neutron}.
}
\label{fig:cross_sections_1}
\end{figure}

In the internal region, the eigenstates $X_{\lambda}$ of the complete Hamiltonian 
\eqref{eq:H_rmatrix} are expanded as
\begin{equation}
X_{\lambda}
=
\sum_{n}\sum_{\nu} C^{\lambda}_{n\nu}(\gras{\eta}_0) 
\Phi^{\gras{\eta}_0}_{n}(\gras{\eta})\chi_{\nu}(\gras{\xi}) ,
\label{eq:expand_internal}
\end{equation}
where the basis wave functions of the vibrational modes are denoted by 
$\Phi^{\gras{\eta}_0}_{n}(\gras{\eta})$. The notation indicates that these are 
eigenfunctions of the collective, vibrational Hamiltonian that only depend on 
$\gras{\eta}$ but with a specific boundary conditions set at $\gras{\eta}=\gras{\eta}_0$. 
The basis wave functions for the intrinsic excitations are $\chi_{\nu}(\gras{\xi})$ 
and correspond to eigenstates of the intrinsic Hamiltonian.
They define the Bohr fission channels. 
For the double-humped barrier typical of actinides, collective vibrational 
states can only be built inside the two potential wells corresponding to the 
ground state and the fission isomer. Therefore, one can adapt the expansion 
\eqref{eq:expand_internal} as follows,
\begin{equation}
X_{\lambda}
=
\sum_{n}\sum_{\nu} C^{\lambda;I}_{n\nu}(\gras{\eta}_0) 
\Phi^{\gras{\eta}_0;I}_{n}(\gras{\eta})\chi_{\nu}(\gras{\xi}) 
+
\sum_{n}\sum_{\nu} C^{\lambda;II}_{n\nu}(\gras{\eta}_0) 
\Phi^{\gras{\eta}_0;II}_{n}(\gras{\eta})\chi_{\nu}(\gras{\xi}) ,
\label{eq:expand_internal_1}
\end{equation}
where the additional labels $I$ and $II$ refer to the first and second well 
respectively. The computation of the matrix elements of the total Hamiltonian 
with the ansatz \eqref{eq:expand_internal_1} makes manifest the potential couplings 
between the two wells. Different regimes of coupling (weak coupling, uniform mixing)
lead ultimately to different resonant structures in the fission cross section which
are presented in great details in ~\cite{bjornholm1980doublehumped}.
It includes phenomena going from the large (100 keV) sub-barrier vibrational resonances
observed for instance in $^{230}$Th(n,f) to the clusters of low-energy (a few keV)
fission resonances illustrated in Fig.~\ref{fig:cross_sections_1}.
In this formalism it is possible to interpret such structures as the interplay
between class I and class II states.
Upon formation, the nucleus has some probability (and associated width) 
to pass the first fission barrier and to exit in the second potential well. 
In that second well, the system has a (much higher) probability of 
fissioning. The amount of couplings depends on the energy of the nucleus and 
on all other possible decay modes inside the two wells. 

The final product of the formal R-matrix theory is a link between the scattering matrix, 
hence the cross section, and a set of parameters related to the internal
wave functions $X_\lambda$.
The theory was initially successful at explaining qualitatively the large number 
of narrow resonances (intermediate structures) in the fission sub-barrier cross sections \cite{lynn1968structure,
lynn1969structure,lynn1973fission,lynn1974subbarrier,back1974strength}. It is now 
used routinely as an advanced theoretical framework in the context of
cross section evaluation to fit existing measurements~\cite{bouland1997rmatrix,leal1999rmatrix}. 
To our knowledge, the R-matrix formalism has not yet been used to predict
(instead of fitting or analyzing)
features of resonances starting from a first-principle Hamiltonian. 
Although possible in principle, such predictions would require the knowledge of the
internal region at a level of precision that is most likely beyond what is possible 
with state-of-the-art many-body theories.

\begin{figure}[!ht]
\centering
\includegraphics[width=0.7\textwidth]{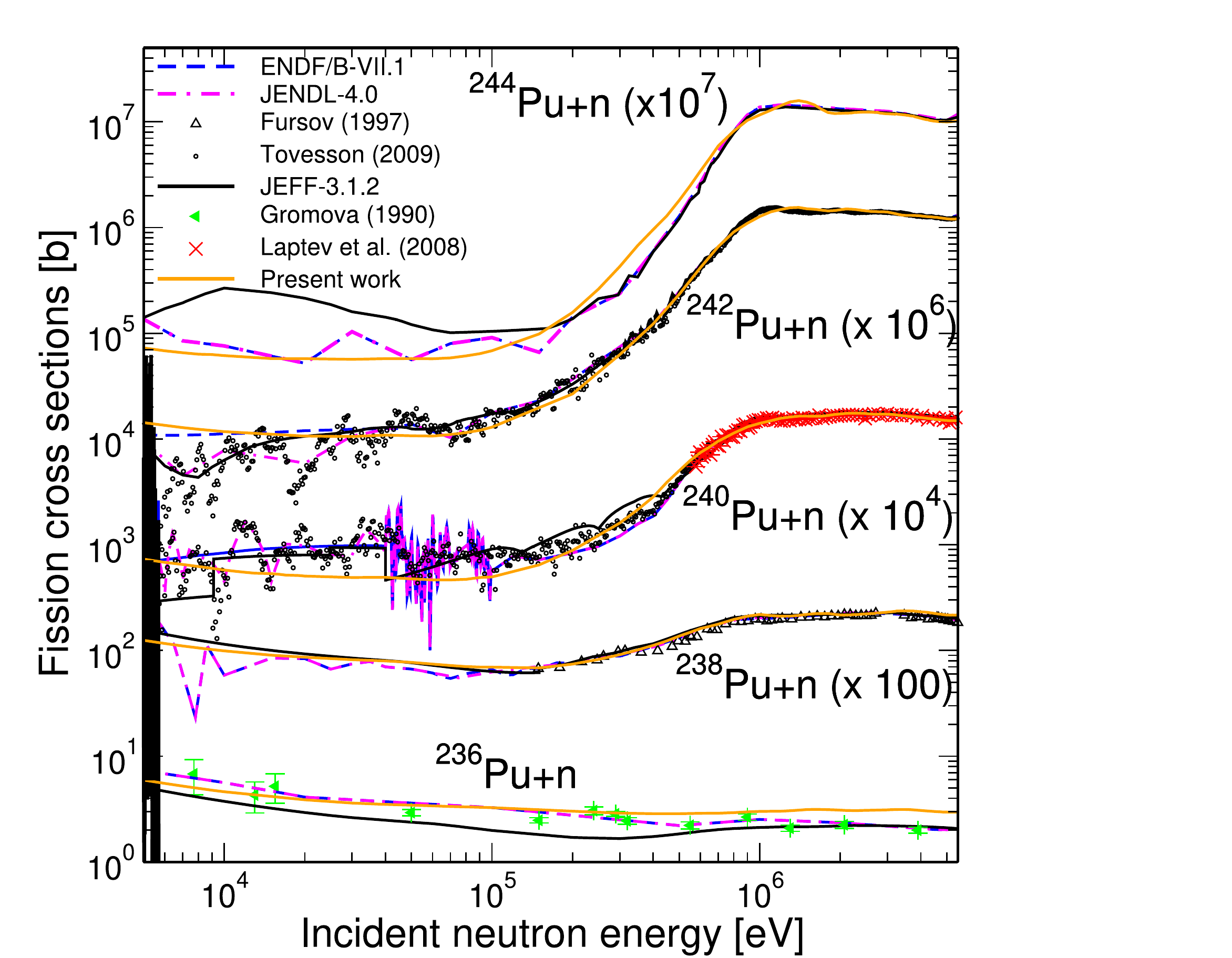}
\caption{Fission cross-sections as a function of the energy of the incident 
neutron for a chain of Plutonium isotopes. Calculations were performed using 
R-matrix theory for sub-barrier fission and are compared to experimental 
measurements and reference evaluations.
Figures reproduced with permission from \cite{bouland2013rmatrix} courtesy of 
Bouland; copyright 2013 by The American Physical Society.
}
\label{fig:pu_cs}
\end{figure}

In recent years, the focus of such 
approaches has also shifted to the prediction of the average fission cross section
in the continuum energy range~\cite{bouland2013rmatrix,bouland2014impact,
bouland2019reexamining}.
Figure \ref{fig:pu_cs} shows an 
example of the fission cross sections in several Plutonium isotopes obtained 
within this formalism.
The idea is to consider a stochastic ensemble of 
realistic R-matrices that account for a statistical ensemble of resonances.
One then determines the cross section as an average over all the cross sections 
obtained with the R-matrix samples. This kind of approach somehow bridges 
the R-matrix formalism with the Hauser-Feshbach approach.


\section{Formation of Primary Fission Fragments}
\label{sec:lacm}

In the previous two sections, we have described how the concept of nuclear 
fission emerges naturally from a description of the atomic nucleus in a 
symmetry-breaking mean field, where fission probabilities are related to 
quantum tunnelling through a multi-dimensional potential energy surface and 
fission cross sections to the presence and structure of deformed excited 
states. In 
this section, we assume that fission is happening and we review the methods 
capable of predicting the primary fragments properties.
One may distinguish two classes of theoretical approaches. The first one
relies on defining an ensemble of scission configurations and populating them
according to a statistical distribution. The so-called scission point
models that we discuss in Section~\ref{subsec:statistical} are prime examples of this class of methods.
A second option is to describe explicitly the dynamics of the process as the nucleus evolves 
from its weakly-deformed initial state toward configurations where two primary fragments
can be identified.
Such models of fission dynamics can be further subdivided into two sub-categories:
\begin{itemize}
\item Real-time, time-dependent calculations simulate a {\it single fission event}, 
that is, the evolution in time of a single configuration of the fissioning 
nucleus. A single solution of the Langevin equation or a single random walk 
(Section \ref{subsubsec:t_classical})
or a fully self-consistent time-dependent Hartree-Fock (TDHF) or time-dependent 
Hartree-Fock-Bogoliubov (TDHFB) trajectory (Section 
\ref{subsubsec:t_diabatic}) are examples of such techniques;
\item Time-dependent collective models simulate the time-dependent evolution of 
the {\it probability distribution} of populating a specific configuration. The 
most common example of such an approach is the quantum-mechanical time-dependent generator 
coordinate method (TDGCM) presented in Section \ref{subsubsec:TDGCM_GOA}
\cite{verriere2020timedependent}. Although it has 
been applied very rarely in practical calculations, its classical equivalent would 
be the Kramers equation \cite{abe1996stochastic}.
\end{itemize}
The two classes of methods are obviously related. For example, sampling many 
different single-configuration trajectories allows reconstructing probability 
distributions: this is how fission fragment distributions can be extracted from 
Langevin or random walks simulations and could, in principle, be compared with 
the equivalent time-dependent collective model. As we will discuss in more 
details in Section~\ref{sec:initial_fragments}, the main goal of time-dependent 
models of fission dynamics is to predict actual observables related to the primary fission
fragments such as their distribution in charge and 
mass, the \definition{total kinetic energy} (TKE) distribution, or the sharing of 
excitation energy between the fragments. These initial conditions are 
important inputs to be used 
in fission event generators to simulate the deexcitation of the fragments and 
compare with experimental measurements.


\subsection{Statistical Models for Fission Fragment Distributions}
\label{subsec:statistical}

Historically, solving explicitly time-dependent differential equations to 
directly simulate fission dynamics often exceeded the available computational 
resources. At the same time, there was a consensus that the characteristics of 
the potential energy surface of the fissioning nucleus at scission played a key 
role in determining the fission fragment mass and charge distributions. 
In particular, it had 
become clear that the nuclear shell structure and, more generally, the 
deformation properties of the fission fragments were important drivers in 
setting their relative abundance \cite{fong1964fission,wilkins1976scissionpoint}. What is known 
as the scission-point model is an attempt to take advantage of these 
observations to obtain realistic estimates of the primary fission fragment 
distributions without simulating the dynamics explicitly. 
Scission-point models rely on the hypothesis of quasi-statistical equilibrium at scission:
it is assumed that 
the dynamics of fission populates scission configurations in a statistical manner.
Scission-point models have 
been relatively successful and have the clear advantage of being 
computationally very cheap since all that is needed is a ``good'' potential 
energy surface: it entirely avoids simulating fission dynamics.

In most implementations of scission-point models, the PES is not computed from 
the potential energy of the fissioning nucleus, but instead defined as the sum 
of the potential energy of each fragment $V_{i}(\gras{q}_i)$, $i=1, 2$ and of the 
interaction between them that depends mostly on their relative distance $d_{12}$. The latter term contains both a Coulomb and nuclear 
component,
\begin{align}
V(\gras{q}) = & V_{1}(Z_1, N_1, \gras{q}_{1}) + V_{2}(Z_2, N_2, \gras{q}_{2}) \nonumber \\
& + V_{\rm Coul}(Z_1, N_1, \gras{q}_{1}, Z_2, N_2, \gras{q}_{2}, d_{12}) 
+ V_{\rm nucl}(Z_1, N_1, \gras{q}_{1}, Z_2, N_2, \gras{q}_{2}, d_{12})
\end{align}
After the total potential energy $V(\gras{q})$ of the system at scission has 
been tabulated, the probability of populating a given fragment at a given 
excitation energy $E^{*}$ can be computed in several ways. One option is to 
relate it to a statistical Boltzman factor, i.e. $\propto e^{-V(\gras{q})/T}$ 
where $T$ is a ``collective temperature'', which can be either a parameter of 
the model or related to excitation energy via the usual $E^{*} \approx aT^2$ 
expression \cite{wilkins1976scissionpoint,pasca2016energy}. Another option 
consists in relating it to the level density in each fragment at the current 
excitation energy $E_{i}^{*}$ of the fragment $i$ \cite{lemaitre2015new,
lemaitre2019fully}. This latter prescription requires introducing a mechanism 
to share the \definition{total excitation energy} (TXE) available among the fragments; see 
discussion in Section \ref{subsec:initial_e}. Obviously, different 
assumptions can be made about the ingredients of the calculation, such as the 
size of the deformation space for each fragment, the specific model of the 
interaction between the fragments, and the method used to populate each 
fragmentation. Overall, scission-point models reproduces fairly well the mass 
yields, for example in the neutron-induced fission of major actinides 
\cite{pasca2016energy}, or the transition between symmetric and asymmetric 
fission in heavy actinides such as Fm isotopes \cite{pasca2018transitions,
pasca2018induced}. The models can be extended to predict other important 
observables such as the charge, mass, the TKE of the reaction and neutron emission in actinides 
\cite{pasca2021simultaneous}.

\begin{figure}[!ht]
\includegraphics[height=6cm]{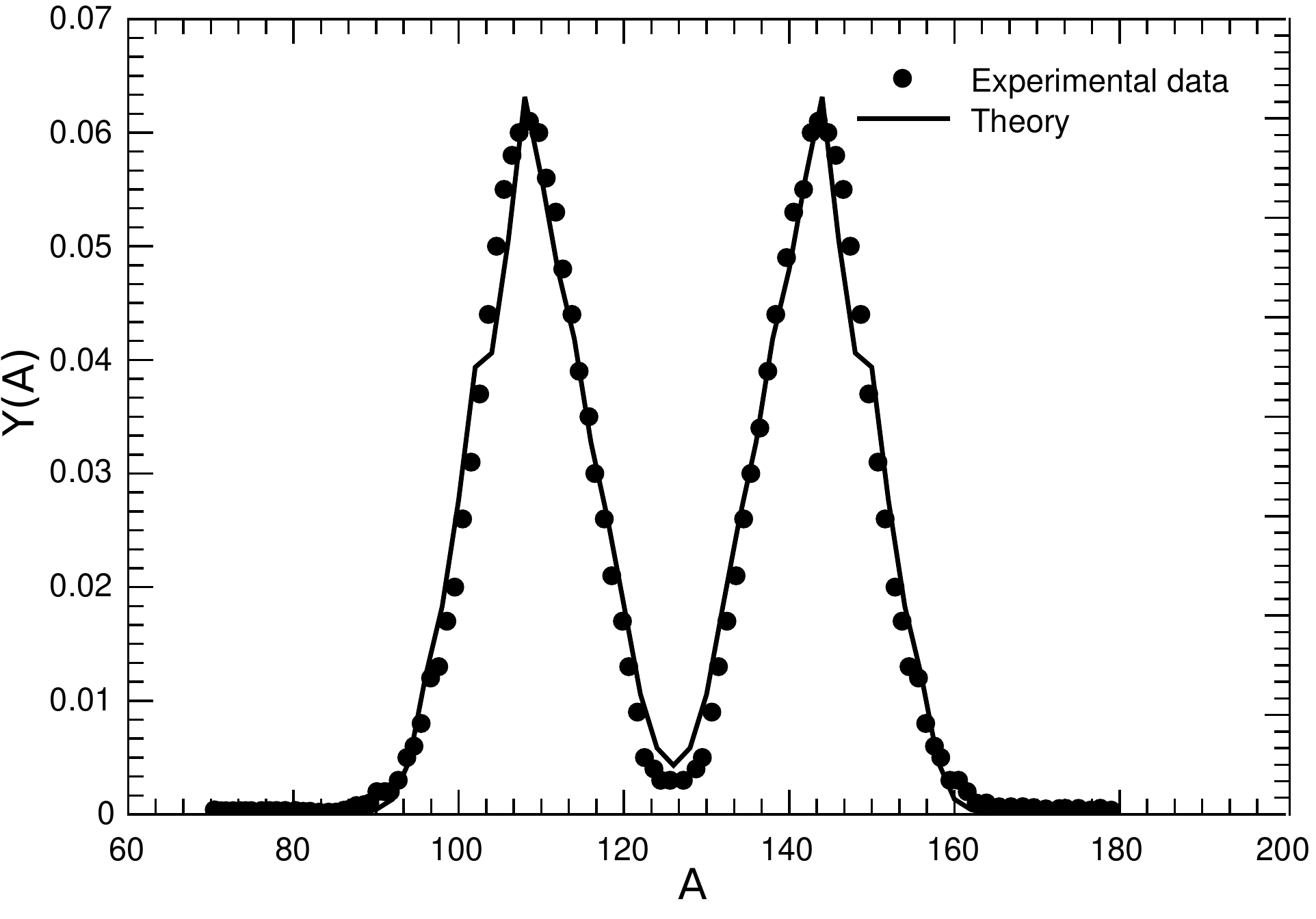}
\includegraphics[height=5.9cm]{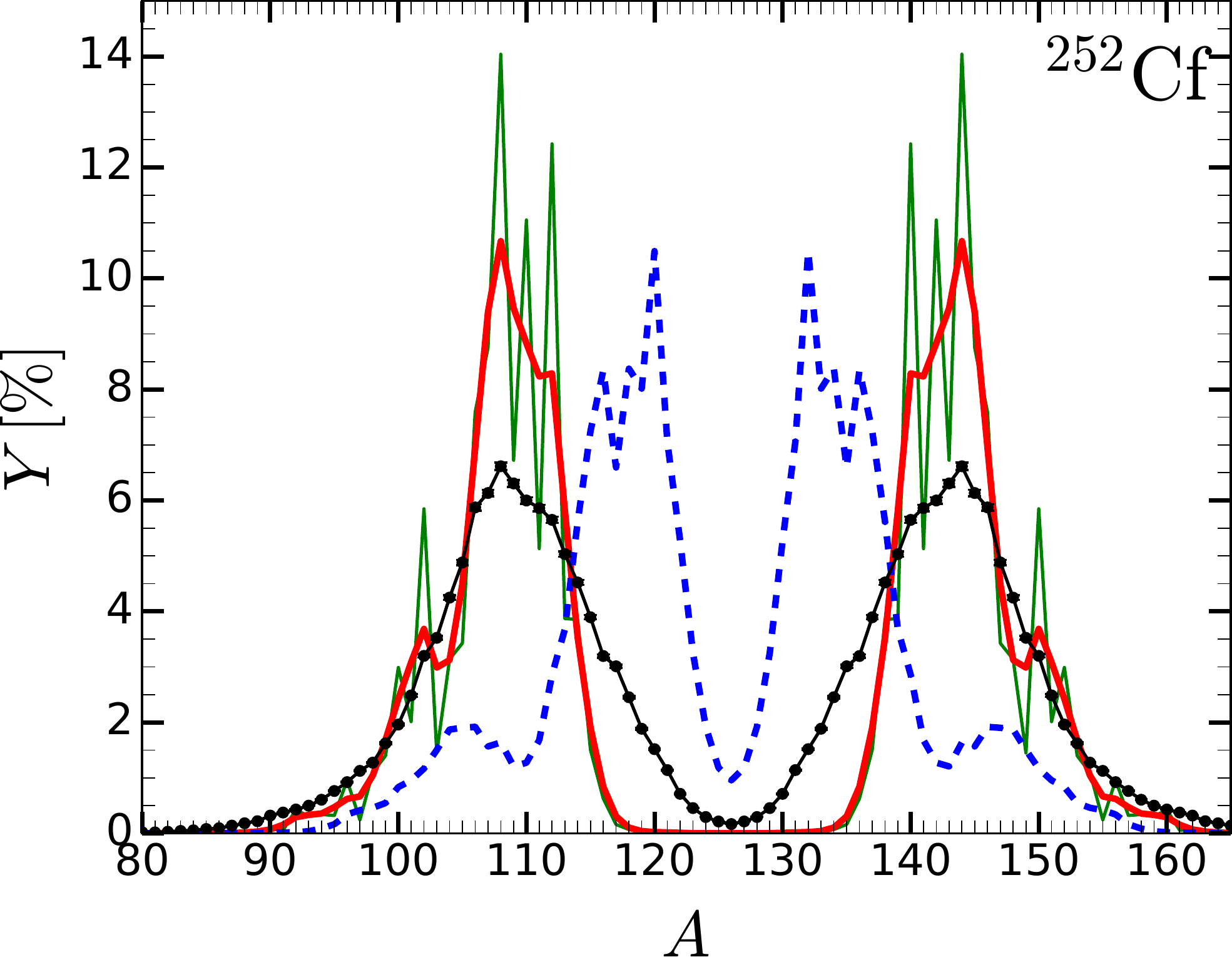}
\caption{
Primary fragments mass distribution for the spontaneous fission of $^{252}$Cf.
Left: Comparison of the scission point model of Pasca \textit{et al.} (full line)
to experimental data (black dots).
Figure reproduced with permission from \cite{pasca2019change} courtesy of 
H.~Pasca; copyright 2013 by The American Physical Society.
Right: Comparison of the scission point model SPY-2 with a smoothing (full red line)
to experimental data (black dotted line). 
The raw predictions of SPY-2 (green full line) along with the predictions
of SPY-1 (blue dotted line) are also plotted for the record.
Figures reproduced with permission from \cite{lemaitre2019fully} courtesy of 
J.-F.~Lemaitre; copyright 2013 by The American Physical Society.
}
\label{fig:pasca_prc_2019_fig1}
\end{figure}

The low computational cost of scission-point models enables performing systematic studies over 
a large range of fissioning systems (e.g. in Ref. \cite{lemaitre2019fully}).
A key ingredient of these approaches is the ensemble of configurations considered
in the neighborhood of scission. This ensemble is inevitably chosen
in a somewhat arbitrary way since the very concept of scission remains ill-defined, 
as mentioned in Section \ref{subsec:scission}.
As a consequence, different scission-point models may predict significantly
different primary fragments distributions as illustrated in the right panel of Fig.~\ref{fig:pasca_prc_2019_fig1}.

\subsection{Classical Dynamics}
\label{subsubsec:t_classical}

Fission is by nature a time-dependent process and one would be forgiven to 
think that static approaches such as the scission-point model may not capture 
all facets of the phenomena. 
Going beyond such models therefore requires describing the 
time evolution of the nucleus starting from its nearly spherical shape
up to well-separated primary fragments.
A fully microscopic theory of this process involves a quantum and real-time 
evolution of more than 200 nucleons characterized by their position
and spin/isospin quantum numbers. The complexity of this task scales 
exponentially 
with the number of DoFs and prevents solving exactly the associated
Schr\"odinger equation on classical computers. A long-term goal of fission 
theory is thus to find reliable approximations, or reductions,
of this quantum many-body dynamics that capture the physics of fission 
while staying computationally tractable.

The first possible level of approximation is based on a classical 
and statistical picture of the dynamics, which leads to the well-known
Langevin equations. These equations have a long history in physics and we refer to the 
review articles of Abe {\it et al.} \cite{abe1996stochastic} and Fr\"obrich \& 
Gontchar \cite{frobrich1998langevin} for a detailed discussion of their various 
forms in the context of nuclear physics. In this section, we simply want to 
recall the basic form of the Langevin equations and highlight some of the 
results that have been obtained from their application to fission.

The Langevin equations originate from a transformation of the classical nucleonic
degrees of freedom into (i) a few collective coordinates 
$\gras{q}(t) = (q_1(t),\dots, q_{N}(t))$ and their associated momenta 
$\gras{p}(t) = (p_1(t),\dots, p_{N}(t))$
(ii) the set of all remaining degrees of freedom usually dubbed intrinsic DoFs.
The idea is to choose properly the collective variables such that the formation of the 
primary fragments is mostly independent of the detailed characteristics of the intrinsic DoFs.
With this in mind, the intrinsic DoFs are often interpreted as a thermal bath 
interacting with the collective DoFs.
The resulting Langevin equations of the system describe the dynamics of the 
collective DoFs and read
\begin{subequations}
\begin{align}
\dot{q}_{\alpha} = & \sum_{\beta} B_{\alpha\beta}p_{\beta} ,\\
\dot{p}_{\alpha} = &
- \frac{1}{2}\sum_{\beta\gamma} \frac{\partial B_{\beta\gamma}}{\partial q_{\alpha}}p_{\beta}p_{\gamma}
- \frac{\partial V}{\partial q_{\alpha}}
- \sum_{\beta\gamma} \Gamma_{\alpha\beta} B_{\beta\gamma}p_{\gamma}
+ \sum_{\beta} \Theta_{\alpha\beta}\xi_{\beta}(t).
\end{align}
\label{eq:langevin}
\end{subequations}
The scalar field $V(\gras{q})$ is the potential energy, 
$\newtensor{B}(\gras{q}) \equiv B_{\alpha\beta}$ is the collective inertia tensor 
and $\newtensor{\Gamma}(\gras{q}) \equiv \Gamma_{\alpha\beta}(\gras{q})$ the 
coordinate-dependent friction tensor. A dissipation term is encoded in the random 
force $\gras{\xi}(t)$. Its strength $\newtensor{\Theta}\equiv\Theta_{\alpha\beta}$ 
is related to the friction tensor through the fluctuation-dissipation theorem, 
$\sum_{\beta}\Theta_{\alpha\beta}\Theta_{\beta\gamma} = \Gamma_{\alpha\gamma}T$, with $T\equiv T(\gras{q})$ the 
local temperature of the bath of intrinsic DoFs at point $\gras{q}$. 

There are many different ways to compute 
the ingredients entering \eqref{eq:langevin}, that is, the fields 
$V(\gras{q})$, $\newtensor{B}(\gras{q})$, $\Gamma(\gras{q})$. These quantities 
must be pre-computed by a nuclear model capable of giving realistic estimates 
of nuclear deformation properties. Traditional applications of the Langevin 
equation often rely on the macroscopic-microscopic model described in Section 
\ref{subsec:micmac} to compute them. However, there have been recently a few 
attempts to couple microscopic inputs computed from the EDF approach with the 
semi-classical dynamics given by the Langevin equation 
\cite{sadhukhan2016microscopic,sadhukhan2017formation,
matheson2019cluster,sadhukhan2020efficient}. In this case, $V(\gras{q})$ is the 
HFB energy and $\newtensor{B}(\gras{q})$ the collective inertia tensor briefly 
discussed in Section \ref{subsec:sf}. Along the same lines, it is very 
important to keep in mind that the formal and technical difficulties in 
determining the scission configurations discussed in Section 
\ref{subsec:scission} will be relevant when solving the Langevin equation since 
the end-point of Langevin trajectories typically correspond to scissionned 
configurations.

\begin{figure}[!ht]
\centering
\includegraphics[width=0.5\textwidth]{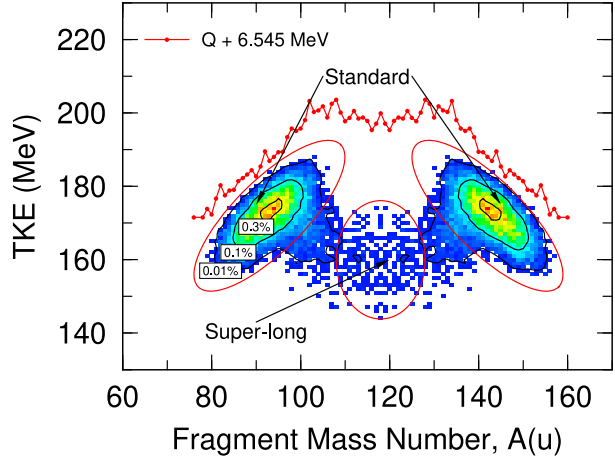}
\caption{Distribution of TKE as a function of the fragment mass number for the 
fission of $^{236}$U at excitation energy $E^{*} = 6.545$ MeV. The red curve 
represents the kinematically allowed maximal value of TKE given by $Q + E^{*}$, 
where $Q$ is the $Q$-value of the reaction $^{235}$U(n,f).
Figures reproduced with permission from \cite{usang2017analysis} courtesy of 
Usang; copyright 2017 by The American Physical Society.
}
\label{fig:TKE_A}
\end{figure}

To obtain meaningful estimates of fission observables, the equations 
\eqref{eq:langevin} must be solved many times, possibly with 
different initial conditions. These statistical fluctuations of the 
initial state mimic here the quantum fluctuations of the system.
Because of the stochastic nature of the equation 
of motion, each trajectory is unique. If sufficiently many trajectories are 
sampled, probability distributions can be reconstructed. The full Langevin 
equation has thus been applied to studies of fission rates  
\cite{nadtochy2007fission}, to the determination of the charge and mass 
distribution of fission fragments \cite{aritomo2015independent,
sierk2017langevin,pomorski2020mass}, and to the evaluation of the TKE
\cite{usang2016effects,sierk2017langevin,usang2017analysis,liu2021analysis,
albertsson2020calculated}. 
Figure \ref{fig:TKE_A} shows an example of the latter calculations.
Finally, it is possible to include into the Langevin framework the competition
between fission and the emission of light particles such as neutrons or $\gamma$
rays. This is of particular interest to study multi-chance fission at high 
energies~\cite{tanaka2019effects}.

\begin{figure}[!ht]
\centering
\includegraphics[width=0.6\textwidth]{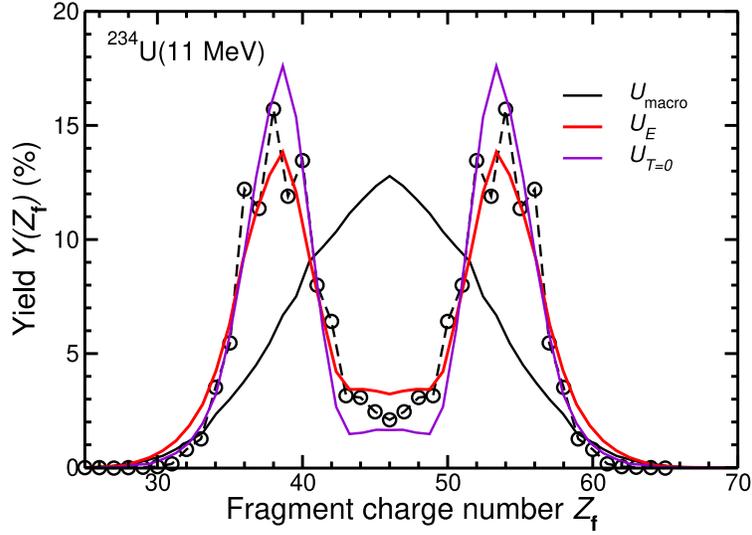}
\caption{Relative fragment charge distribution fo the fission of $^{234}$U at 
$E^{*} = 11$ MeV excitation energy. Each curve was obtained by solving 
\eqref{eq:smoluchowski} with different prescriptions for the PES: a pure 
liquid drop formula ($U_{\rm macro}$), the standard macroscopic-microscopic 
potential defined by \eqref{eq:micmac} ($U_{T=0}$) and a modified version of 
the latter where the shell correction is multiplied by a damping factor 
$S(E^{*})$ ($U_E$).
Figures reproduced with permission from \cite{randrup2013energy} courtesy of 
Randrup; copyright 2013 by The American Physical Society.
}
\label{fig:damping}
\end{figure}

If the motion across the potential is strongly dissipative, it makes sense to 
assume that collective velocities are very small. 
In this case, we can neglect terms proportional to the square of the velocities as 
well as the acceleration terms \cite{randrup2011brownian,
randrup2011fissionfragment}. The Langevin equation takes the 
much simpler form of the Smoluchowski equation
\begin{equation}
\sum_{\beta\gamma} \Gamma_{\alpha\beta} \dot{q}_{\beta}
=
- \frac{\partial V}{\partial q_{\alpha}}
+ \sum_{\beta} \Theta_{\alpha\beta}\xi_{\beta}(t) .
\label{eq:smoluchowski}
\end{equation}
The Smoluchowski equation \eqref{eq:smoluchowski} was used to compute estimates 
of primary fission fragment distributions \cite{randrup2011brownian,
randrup2011fissionfragment,moller2012calculated,randrup2013energy,
moller2015calculated,mumpower2020primary}. The formalism was extended in 
\cite{randrup2013energy} to describe fission dynamics at higher excitation 
energies. In the case of the macroscopic-microscopic model, the main effect of 
increasing excitation energy is to destroy shell corrections 
\cite{jensen1973shell,ignatyuk1980shape,dudek1988pairing}, although the 
macroscopic energy should in principle also be modified 
\cite{diebel1981microscopic}. One can incorporate this effect by computing, at 
each deformation point $\gras{q}$ of the standard macroscopic-microscopic PES, 
a local temperature $T(\gras{q})$ based on the value of the relative excitation 
energy at this point. This local temperature can then be used to modify the 
Strutinsky shell correction and generate PES at finite excitation energy; see 
examples in \cite{dudek1988pairing,schunck2007nuclear}. This approach was used 
in \cite{usang2016effects,usang2017analysis} to analyze the impact of 
excitation energy on dissipation. A simplified version of temperature-dependent shell corrections 
consists in multiplying the shell 
correction by a phenomenological damping factor. Figure \ref{fig:damping} shows the 
impact of the PES underpinning the random walk calculations at finite 
excitation energies. At temperatures $T > 1$ MeV, 
such an approximation works rather well and has been used in large-scale 
calculations in \cite{randrup2013energy}, but it fails to account for the more 
complex dependency of the yields at low excitation energies. In this regime, 
an alternative to temperature-dependent shell correction is to use level 
densities to guide the evolution of the Metropolis walk on the surface 
\cite{ward2017nuclear}.

\begin{figure}[!ht]
\centering
\includegraphics[width=0.55\textwidth]{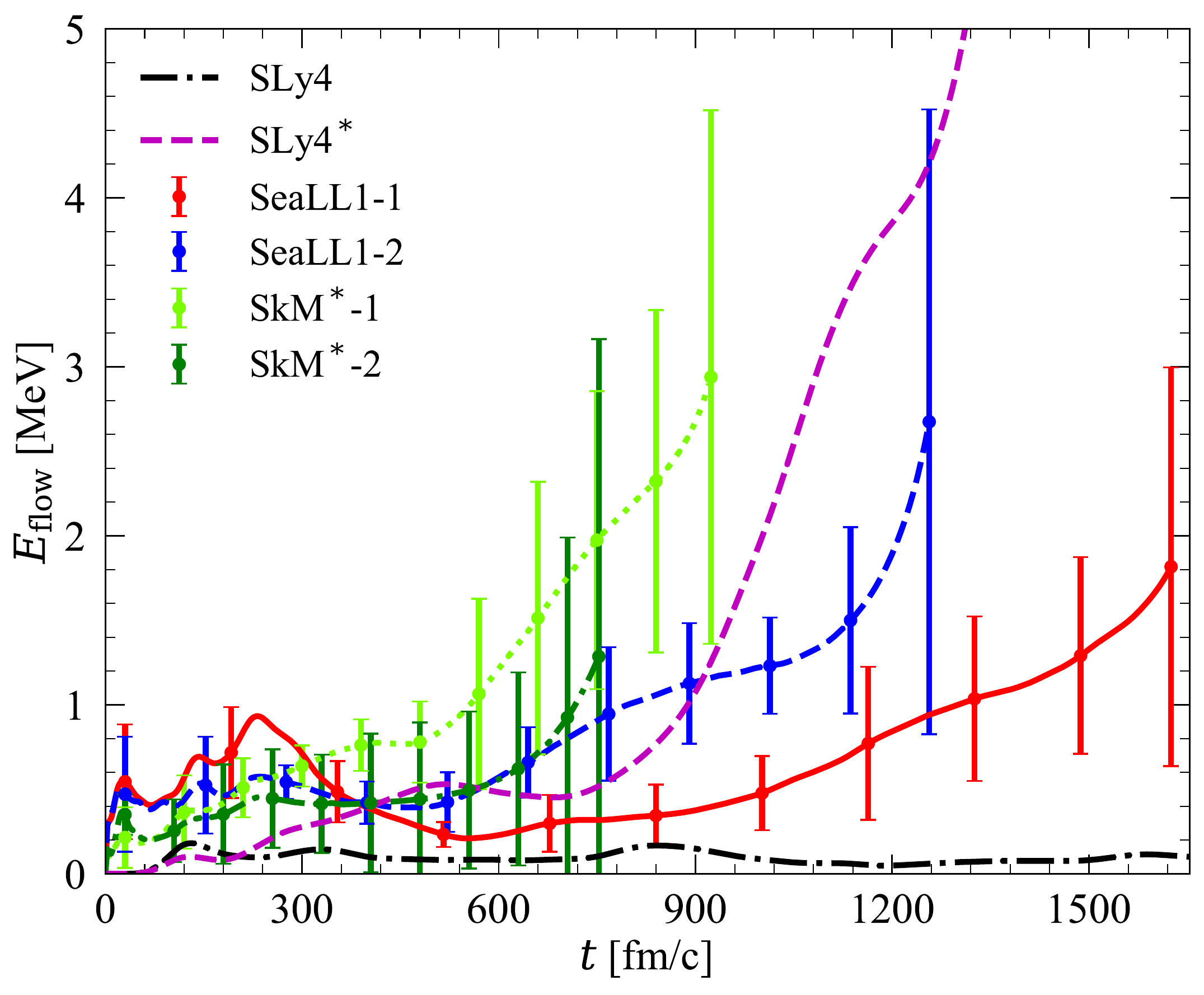}
\caption{Collective flow energy $E_{\rm flow}(t)$ as a function of time in the 
fission of $^{240}$Pu. Each curve correspond to a simulation with different 
energy functional and/or pairing characteristics. Error bars indicate 
variations with respect to different initial conditions.
Figures reproduced with permission from \cite{bulgac2019fission} courtesy of 
Shi; copyright 2019 by The American Physical Society.
}
\label{fig:dissipation}
\end{figure}

The hypothesis of strongly-damped motion was originally introduced with 
semi-phenomenological arguments based on the wall-and-window model for 
dissipation \cite{blocki1978onebody,sierk1980fission}. Early attempts at 
verifying this hypothesis from more microscopic arguments related to the 
nuclear shell structure suggested that dissipation may be even stronger than 
anticipated \cite{schutte1980fission}. Theories of dissipation based on 
transport models were introduced to provide a better description of the 
dissipation tensor in either the Langevin or Smoluchowski equations 
\cite{hofmann1997quantal,ivanyuk1999pairing,hofmann2001nuclear}. In recent 
years, fully microscopic simulations of fission based on TDHF+BCS and TDHFB 
also predict a strongly dissipative dynamics along with a very small collective 
kinetic energy \cite{tanimura2015collective,
bulgac2016induced,bulgac2019fission}. This is illustrated in 
Fig.~\ref{fig:dissipation}, which shows the collective kinetic energy in the descent 
from saddle to scission in the real-time simulation of $^{240}$Pu,
\begin{equation}
E_{\rm flow}(t) = \frac{1}{2}\int d^{3}\gras{r}\, m\gras{v}^2(t)\rho(\gras{r},t) .
\end{equation}
For most of the evolution, this collective energy is of the order of 1-2 MeV, 
which is less than 10\% of the variations in potential energy and less than 
1\% of total energy released at scission.

\subsection{Time-Dependent Density Functional Theory}
\label{subsubsec:t_diabatic}

The Langevin equations and their different variants provide a computationally 
tractable method to study the real-time evolution of a fissioning nucleus in 
presence of dissipation and fluctuations but they present a few limitations. 
For instance, a classical treatment of dynamics cannot account for
tunneling effects, which is crucial for spontaneous fission. In addition, going beyond
a thermal bath description of the intrinsic DoFs is necessary if one wants to
predict detailed properties of the emerging fragments such as, e.g. the energy or spin of primary fragments.
The goal of microscopic approaches is to overcome these limitations by 
describing the quantum dynamics of the many-body wave function associated
with all the nucleonic DoFs. A reduction in complexity is enforced by
a variational principle that limits the possible many-body wave functions to a
subspace of the complete Hilbert space.
Assuming that the many-body wave function remains a coherent state at all times
leads to the time-dependent density functional theory (TDDFT). 
In this article, we include under the label TDDFT an arsenal of different methods 
including TDHF, THDF+BCS, TDHFB. A detailed discussion of how 
these methods are related can be found, e.g. in 
\cite{lacroix2010quantum,simenel2018heavyion}. Several important aspects of the 
general formalism have been presented in review articles and textbooks; see 
\cite{balian1985timedependent,blaizot1985quantum,balian1988static,
balian1992correlations,nakatsukasa2012density,nakatsukasa2016timedependent,
schunck2019energy}. A more succinct presentation of the theory specifically 
geared toward fission theory can be found in \cite{bulgac2020nuclear}. 

\begin{figure}[!ht]
\centering
\includegraphics[width=0.50\textwidth]{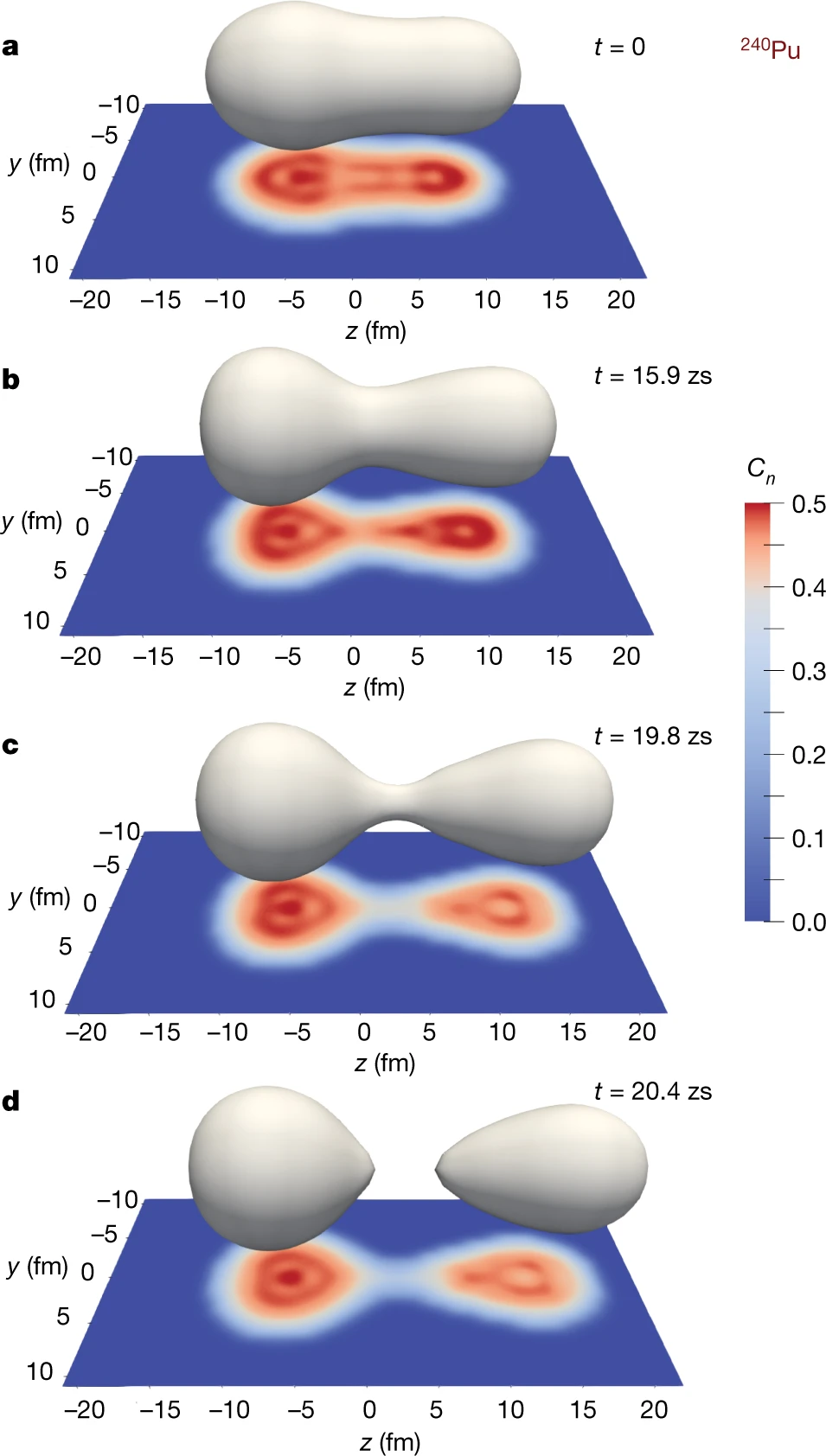}
\caption{ 
Real-time evolution of the one-body density of the $^{240}$Pu asymmetric fission.
The predictions are obtained from a TDHF+BCS calculation. The 3D surface 
highlights the half-saturation density (0.08 fm$^{-3}$) isosurface whereas
the projected color map corresponds to a localization function of the nucleons.
Figures reproduced with permission from \cite{scamps2018impact} courtesy of 
Scamps; 
 }
\label{fig:scamps_nature_2018_fig1}
\end{figure}

Time-dependent density functional theory is especially relevant for fission: given an initial state, which is 
typically a deformed, SR-EDF solution, TDDFT describes the real-time evolution 
of the nucleus as it deforms all the way until the point of scission and even beyond; 
see Fig.~\ref{fig:scamps_nature_2018_fig1}.
For instance, in Ref.~\cite{bulgac2021fission} the dynamics is simulated up 
to fragments separated by a distance of 30 fm. At this stage the nuclear
interaction between the fragments is negligible and the properties of
the primary fragments such as their excitation energy and spin can be reliably
estimated.
The simplest version 
of TDDFT is the time-dependent Hartree-Fock (TDHF) theory which leads to 
the equation of motion
\begin{equation}
i\hbar\frac{\partial\rho}{\partial t} = \big[ h[\rho(t)], \rho(t) \big],
\label{eq:tdhf}
\end{equation}
where $\rho(t)$ is the time-dependent one-body density matrix and $h[\rho(t)]$ 
the Hartree-Fock mean field -- the time-dependent version of the mean field in 
$h$ in \eqref{eq:HFB_matrices}. The time-dependent Hartree-Fock-Bogoliubov 
(TDHFB), which includes the pairing degrees of freedom in the dynamics, is 
obtained from \eqref{eq:tdhf} by doing the substitutions $\rho(t) \rightarrow \mathcal{R}(t)$ 
and $h[\rho(t)] \rightarrow \mathcal{H}[\mathcal{R}(t)]$; see Eq.~\eqref{eq:HFB_matrices}. 
Note that practical implementations of these theories in nuclear physics always
rely on an energy density functional to encode the nucleon-nucleon interaction.

\begin{figure}[!ht]
\centering
\includegraphics[width=0.45\textwidth]{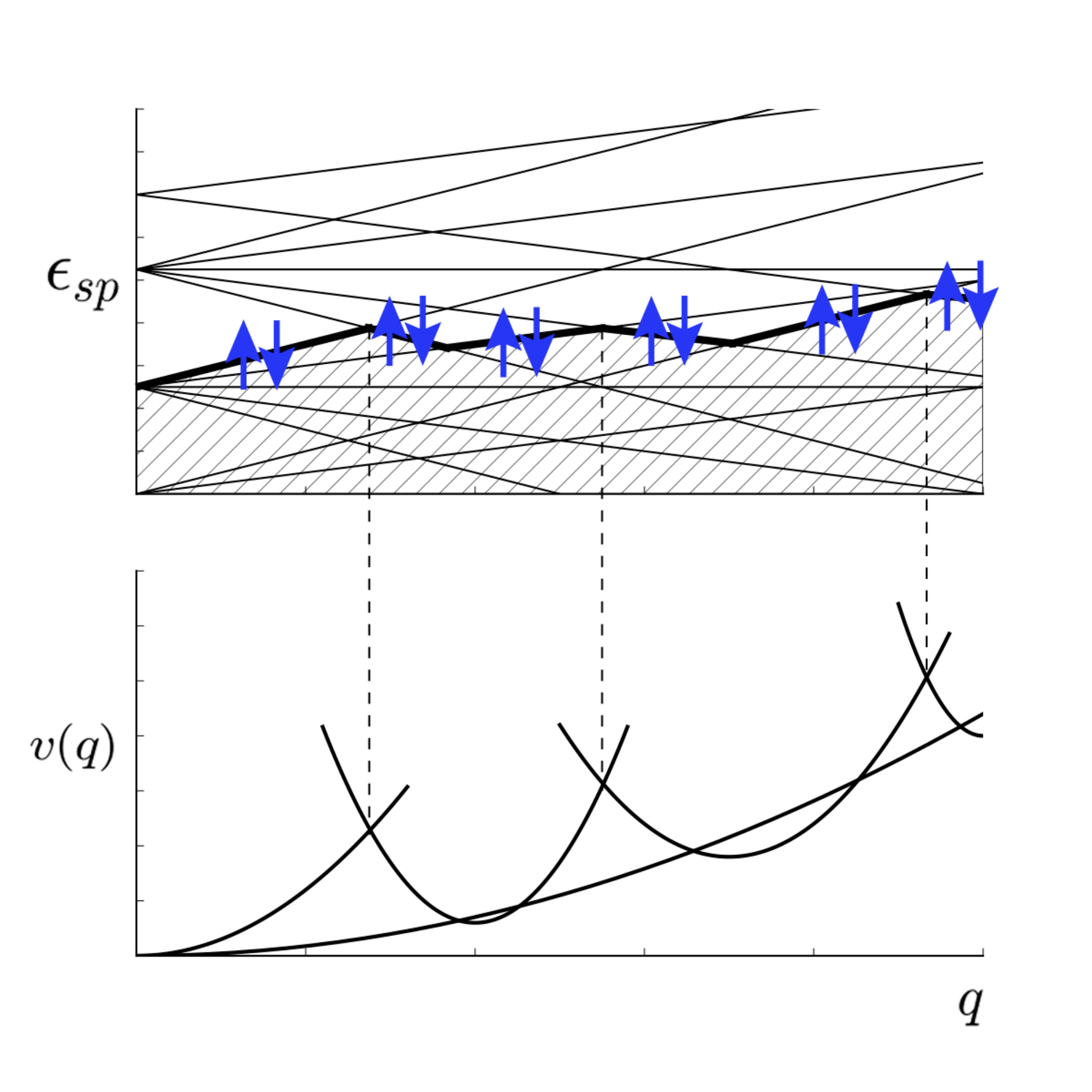}
\caption{Top: Schematic evolution of single-particle orbitals as a function of 
deformation. The thick line represent the last occupied state and the up and 
down arrows the Cooper pairs of nucleons on that level. Bottom: corresponding 
evolution of the total nuclear energy. Each crossing of the s.p. level result 
in the transition to a new intrinsic configuration and change of the potential 
energy. 
Figures reproduced with permission from \cite{bulgac2019fission} courtesy of 
Shi; copyright 2019 by The American Physical Society.
 }
\label{fig:crossings}
\end{figure}

Negele and collaborators reported the first application of the TDHF formalism 
to fission back in 1978 \cite{negele1978dynamics}. However, these initial 
simulations included a large number of rather strong simplifications; the 
earliest realistic application of TDHF to fission can probably be attributed to 
Simenel and Umar \cite{simenel2014formation}.
One important result of their 
work was the quantitative prediction of vibrational modes in the fission 
fragments and the determination of a mechanism to estimate the TKE of the 
reaction and the excitation energy of the fragments.
The implementation and release of efficient TDDFT codes without symmetry
assumptions~\cite{maruhn2014tdhf,schuetrumpf2018tdhf,jin2021lise}
together with the inclusion of superfluidity at the BCS~\cite{scamps2015superfluid}
and full HFB level~\cite{bulgac2016induced} were key to popularize this approach.
Among the most important 
early results was the confirmation that pairing degrees of freedom are 
indispensable to reach the point of scission: pure TDHF calculations initialized 
around the saddle point of the fission barrier in actinides failed to reach 
that point unless a significant boost in energy was imparted to the system 
\cite{goddard2015fission,goddard2016fission}. Including pairing degrees of 
freedom, even at the TDHF+BCS approximation with BCS occupation frozen after a 
given time, completely removed the need to introduce such boosts 
\cite{scamps2015superfluid}. This mechanism naturally results from the ability 
of pairing correlations to pass through single-particle level crossings, as 
schematically depicted in Figure \ref{fig:crossings}. For this reason, pairing 
has been dubbed the ``fission lubricant''.
The TDDFT approach successfully predicts many properties of the most probable
primary fission fragments such as their kinetic energy, with a precision of a few percents
~\cite{bulgac2016induced}, their deformation close to scission~\cite{scamps2018impact},
and more recently the spin of the fragments~\cite{bulgac2021fission}.

The TDHF and TDHFB formalisms are often called semi-classical, in the sense that 
they do not allow collective quantum tunneling: if a TDHF(B) evolution was initialized in 
a configuration $\rho(t=0)$ ($\mathcal{R}(t=0)$) inside a potential well determined
from constrained HF(B) calculations, the system would remain inside that well
\cite{ring2004nuclear}. Various 
extensions of the TDDFT methods to imaginary time have been proposed to handle 
quantum tunneling but have rarely been applied in practical calculations  
\cite{levit1980timedependenta,levit1980barrier,negele1982meanfield,
puddu1987solution,skalski2008nuclear,brodzinski2020instantonmotivated}; see 
also Section~\ref{subsec:sf}. 

In spite of its advantages, current implementations of TDDFT are also not well
adapted to describing large fluctuations of observables, i.e. the quantity 
$\braket{\Delta \hat{O}^{2}(t)} = \braket{\hat{O}^{2}(t)} - 
\braket{\hat{O}(t)}^{2}$ \cite{blaizot1985quantum,balian1984fluctuations,
balian1992correlations}. This limitation stems from the restriction of the 
many-body wave function to a Slater or Bogoliubov state that do not encode 
enough correlations. As a result
TDDFT predictions capture well the features of the {\it average}
fragmentation but systematically underestimate the associated quantum fluctuations.
For this reason, they are not the most efficient tools 
to generate, e.g., fission fragment distributions.
There have been several attempts to go 
beyond ``simple'' TDHF(B). In the stochastic mean field theory, random 
fluctuations of the density matrix are used to generate an ensemble of 
trajectories (in the space of density matrices) that are 
similar to TDHF \cite{ayik2008stochastic,lacroix2014stochastic,tanimura2017microscopic}. This trick allows extracting reasonable estimates of 
kinetic energy distributions but the quantum coherence of the system is lost.
Other on-going investigations along this line simply introduce various 
phenomenological terms simulating dissipation and fluctuations in the 
TDHFB equation \cite{bulgac2019unitary}.
The Time-dependent RPA as well as the Balian-Veneroni
variational principle are other approaches that could 
be used, in principle, to get additional fluctuation while retaining
quantum coherence \cite{balian1992correlations,balian1981timedependent},
but it has only been applied in heavy-ion collisions so far 
\cite{simenel2020timescales,sekizawa2019tdhf}.
Another possibility consists in mixing TDDFT trajectories in the quantum-mechanical
sense: early attempts were based on TDHF trajectories \cite{reinhard1983time} 
while more recent ones rely on full TDHFB solutions (albeit in a simplified 
system) \cite{regnier2019microscopic}.

As discussed in the previous section, the strongly dissipative nature of 
fission is another result that had been invoked as a reasonable hypothesis by
many authors and was observed in the context of TDHFB simulations 
\cite{tanimura2015collective,bulgac2016induced,bulgac2019fission}. 
However, whether this strong dissipation would remain when exploring a larger variational
space than in current implementations of TDDFT, e.g., a space that also contains collective 
correlations, is yet to be determined.
Finally, note that 
dissipation is also largely caused by pairing correlations: the same 
mechanism that allows the system to go across single-particle crossings 
explains both that the system can deform sufficiently up to scission {\it and} 
that the collective energy is dissipated into changes of intrinsic 
configurations \cite{cassing1983role}.


\subsection{Time-Dependent Generator Coordinate Method}
\label{subsubsec:TDGCM_GOA}

Because of the many successes of collective approaches, from the seminal paper 
of Bohr and Wheeler to the predictions of fission isomers, spontaneous fission 
half-lives or fission fragment distributions,
there is a rather large consensus that fission can be well described 
by invoking a few collective variables such as shape deformations.
This idea motivates the search for approaches reducing the complete quantum 
many-body dynamics to an evolution in a small-dimensional collective space.
The ATDHF theory mentioned in Section \ref{subsec:sf} in the context of 
spontaneous fission \cite{baranger1978adiabatic,holzwarth1973four,
brink1976derivation,villars1977adiabatic}, and its ATDHFB extension with pairing 
\cite{krieger1974application}, are the poster child of this philosophy.
They reduce the complexity of TDDFT, which involves all the one-body degrees of 
freedom, to a classical collective dynamics on a manifold parameterized 
by a few collective variables~\cite{ring2004nuclear}.
Another approach leading to a quantum collective dynamics is the time-dependent
generator coordinate method (TDGCM) and especially its Gaussian overlap
approximation (GOA).
Note that re-quantizing the ATDHFB theory leads in fact to an equation of motion
similar to the TDGCM+GOA.
The theoretical framework of the TDGCM is presented in great details in 
\cite{verriere2020timedependent}.
In this section, we will thus discuss only the case of the TDGCM under the 
Gaussian overlap approximation which is, as of today, the only
quantum microscopic approach that has been tested in predictions 
of actual primary fragments distributions.

%

The first step of the TDGCM approach consists in building an ensemble $S$ 
of generator states (sometimes referred to as a collective path) parameterized by a 
few collective variables $S = \{ \ket{\Phi(\gras{q})} | \gras{q} \in \mathbb{R}^N \}$. 
In the context of fission the generator states are most often time-even Bogoliubov states
obtained by solving the constrained HFB equations as discussed in Section \ref{subsec:EDF}. The collective variables
are typically related to the first multipole moments of the one-body
density in order to encode the physics of deformation.
With this family of generator states, the TDGCM proceeds with a variational
principle applied to the Hill-Wheeler ansatz for the wave function
\begin{equation}
\label{eq:tdgcm_ansatz}
|\psi(t)\rangle = \int_{\gras{q}}  f(\gras{q},t) |\phi(\gras{q})\rangle d\gras{q}.
\end{equation}
The time-dependent mixing function $f(\gras{q},t)$ is the direct variational 
unknown but it is often more convenient to deal with a transformed function
$g(\gras{q},t)$ that could be interpreted as a wave packet in the collective
space.
All operators associated with physical 
observables can also be mapped onto operators acting on the collective 
space (a subspace of functions of the $\gras{q}$ variable)~\cite{reinhard1987generator}. 
Under the Gaussian overlap approximation in its simplest form, the equation of 
evolution for the collective wave function reduces to
\begin{equation}
i \hbar\frac{\partial}{\partial t} g(\gras{q},t) 
= 
\left[ 
- \frac{\hbar^2}{2} \sum_{\alpha\beta} \frac{\partial }{\partial q_{\alpha}} B_{\alpha\beta}(\gras{q}) \frac{\partial}{\partial q_{\beta}} 
+
V(\gras{q})
\right] g(\gras{q},t),
\label{eq:evolution0}
\end{equation}
where $V(\gras{q})$ is the potential energy and $\newtensor{B}(\gras{q}) \equiv 
B_{\alpha\beta}$ is the collective inertia tensor. The potential energy and 
collective inertia have the same meaning as in Section 
\ref{subsubsec:t_classical} and Section \ref{subsec:sf}. 
An important point is that these quantities derive from the variational principle
and are, formally, the zero- and second-order derivatives of the reduced Hamiltonian
kernels between the generator states.

\begin{figure}[!ht]
\centering
\includegraphics[width=0.48\textwidth]{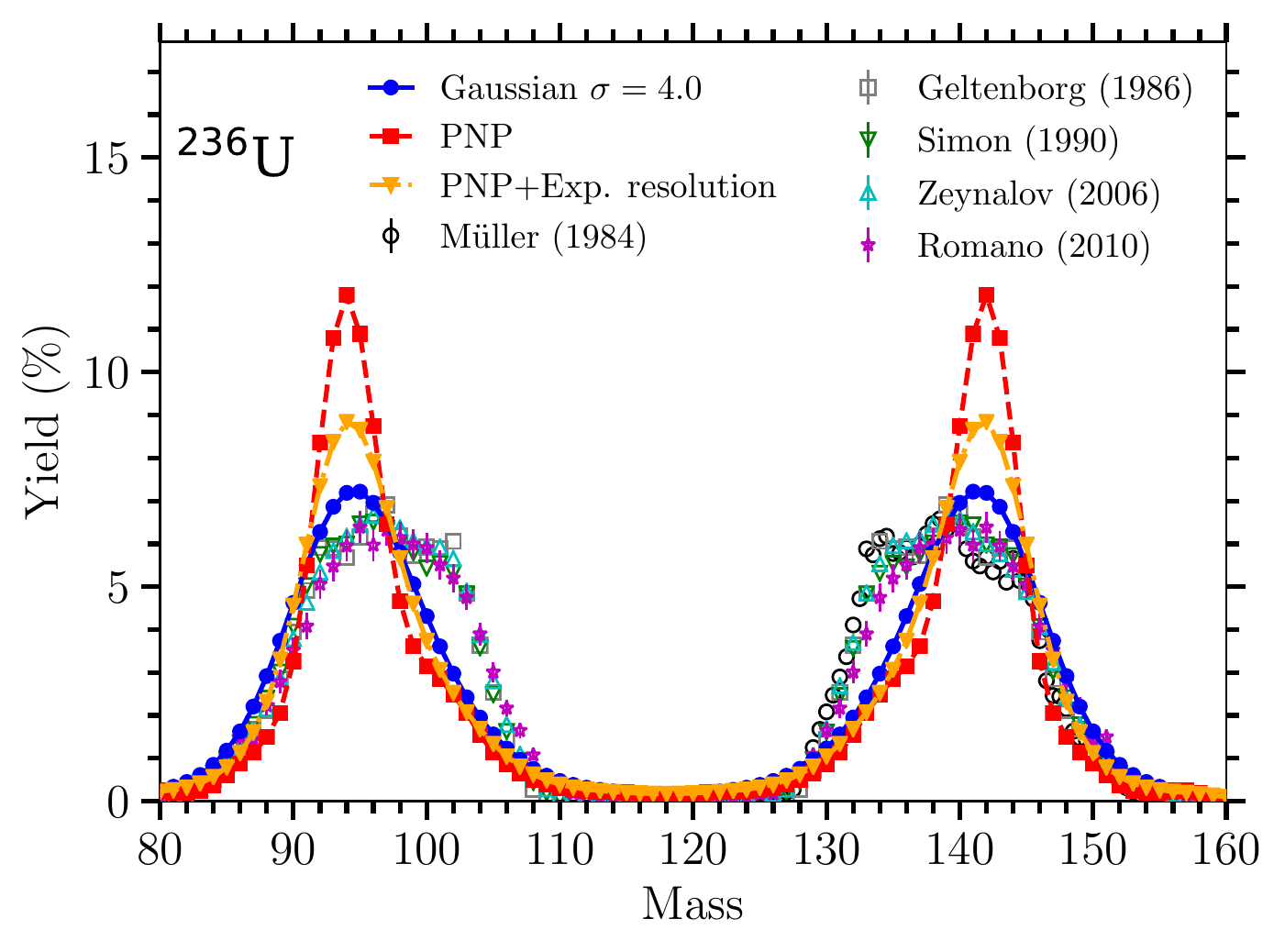}
\includegraphics[width=0.48\textwidth]{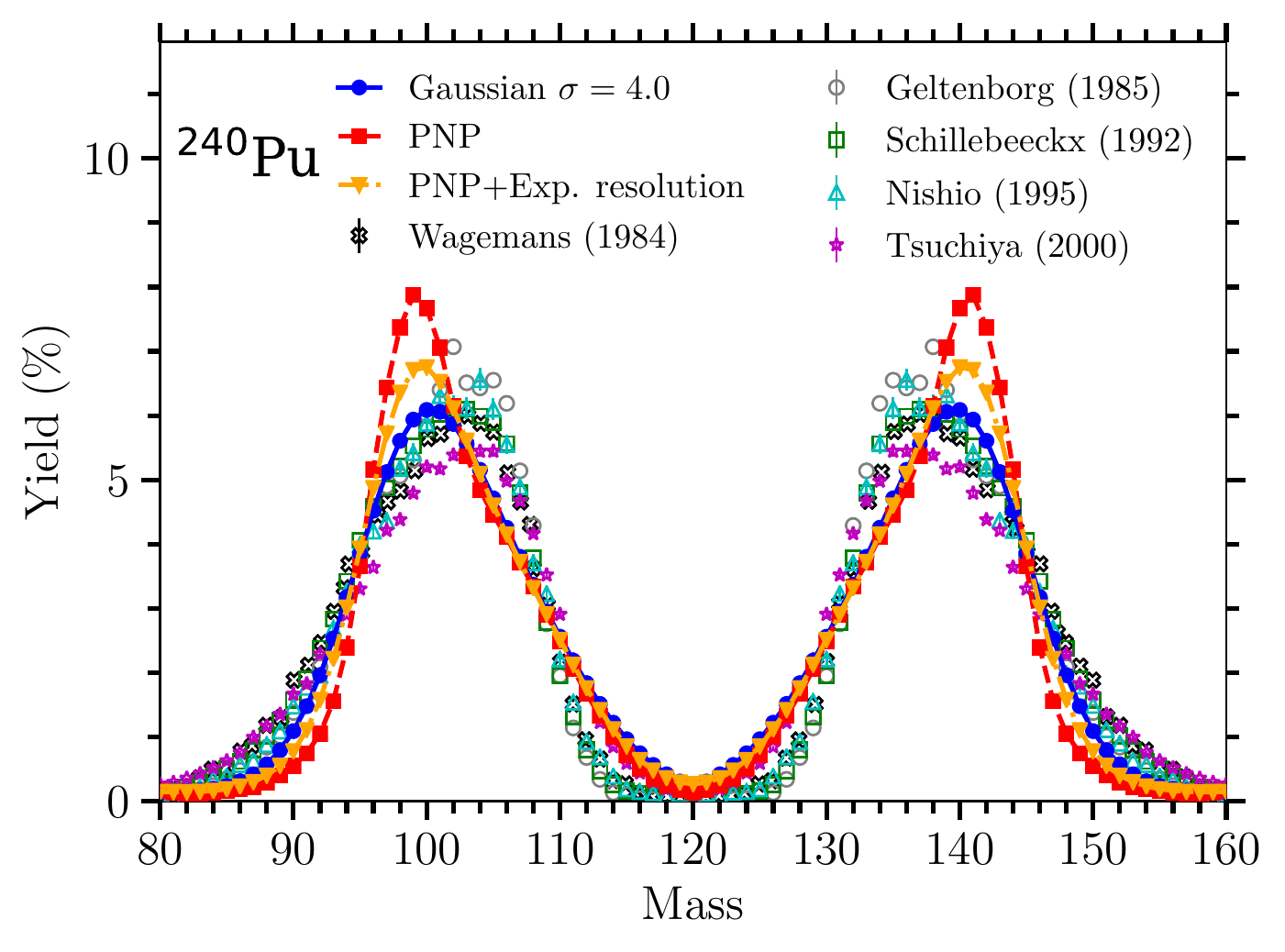}
\caption{Left: Fission fragment mass distribution of $^{236}$U for thermal 
neutrons obtained by solving the TDGCM+GOA equation with the SkM* 
parameterization of the Skyrme EDF, compared with experimental data taken from 
\cite{wagemans1984comparison,geltenbort1986precision,
schillebeeckx1992comparative,nishio1995measurement,tsuchiya2000simultaneous,
muller1984fragment,simon1990pulse,zeynalov2006investigation,romano2010fission}. 
The raw TDGCM+GOA results are convoluted following three different methods
that yields the blue circles, the red squares and the orange triangles.
Right: Same quantity for $^{240}$Pu.
Figures reproduced with permission from \cite{verriere2021microscopic} 
courtesy of Verriere; copyright 2021 by The American Physical Society.
 }
\label{fig:YA}
\end{figure}

The adaptation of the TDGCM formalism to fission dynamics was performed in the 
1980's by the group at CEA,DAM,DIF~\cite{berger1984microscopic,
berger1991timedependent}. These initial developments led to the first 
prediction of fission fragment mass distributions in a completely microscopic 
theory \cite{goutte2005microscopic}. Advances in computing capabilities and new 
algorithmic developments \cite{regnier2016felix1,regnier2018felix2} renewed 
interest in this approach in the 2010ies. Since then, the theory has been 
applied largely to compute the charge, mass and isotopic primary mass 
distributions of fission fragments \cite{younes2012fragment,regnier2016fission,
zdeb2017fission,tao2017microscopic,zhao2019microscopic,regnier2019asymmetric,
zhao2019timedependent,verriere2021microscopic}. Figure \ref{fig:YA} shows an 
example of such calculations for the primary mass distribution $Y(A)$ in 
$^{236}$U and $^{240}$Pu extracted from the solution to the TDGCM+GOA evolution 
equation.

Obviously, results of TDGCM+GOA calculations will depend on the choice and 
number of collective variables. We already mentioned the recent introduction of 
pairing collective variables \cite{sadhukhan2014pairinginduced,bernard2019role,
zhao2016multidimensionallyconstrained} for spontaneous fission; very few TDGCM 
calculations have been performed in such extended spaces \cite{zhao2021microscopic}.
Let us highlight 
an interesting attempt to use the collective variables 
$D = |z_{\rm CM}^{(1)} - z_{\rm CM}^{(2)}|$ and $\xi = |A_{1} - A_{2}|/A$, 
with $z_{\rm CM}^{(i)}$ the coordinate of center of mass of fragment $i$ and 
$A_i$ is total number of particles  \cite{younes2012fragment,
younes2019microscopic}. Such coordinates, which were inspired by classical 
shape parameterizations, are much better adapted to describing fission fragments
than the standard multipole moments and cover most of the set of all possible 
fragmentations. 
The role 
and importance of the collective inertia tensor also remains an open problem, and it 
is not clear if it has as large an impact for induced fission as it has for 
spontaneous fission. Similarly, the role of time-odd collective momenta, which 
have been neglected so far, is worth investigating since they provide the 
proper asymptotic values of the collective inertia (at least in the case of 
translations) \cite{goeke1980generatorcoordinatemethod,reinhard1987generator}. 
A preliminary study of these terms in the static case of particle number 
projection suggests a large effect \cite{hizawa2021generator}. 
An attempt to even remove the need for inertia by solving the complete
TDGCM without approximation was also reported in Ref.~\cite{verriere2017fission}.

Finally, a major limitation of the current implementations of the TDGCM is to rely only
on constrained HFB states with the lowest energy for a given constraint. 
This is the cause for discontinuities in the manifold of generator
states that break down the GOA~\cite{dubray2012numerical,lau2021smoothing}.
In addition, TDDFT calculations have shown that the most probable 
path toward fragmentation
involves HFB states with significant intrinsic excitation energy. 
The very fact that fission fragments emit light particles (cf. Section~\ref{sec:deexcitation})
proves that they are produced in excited states.
With the standard family of generator states, the ansatz~\eqref{eq:tdgcm_ansatz}
can not model such effects, which prevents the TDGCM from
describing properly the dynamics in the neighborhood of scission. 
In a pragmatic approach, the description of fission dynamics with the TDGCM 
should therefore always be stopped with di-nuclear 
configurations before scission, where the two prefragments still significantly 
interact by nuclear forces. A possible way to overcome
this issue consists in expanding the collective space to 
include intrinsic excitations, e.g., sets of two-quasiparticle states built on 
top of constrained HFB vacuua \cite{bernard2011microscopic}. Such an extension 
would allow incorporating both some amount of collective dissipation in the theory 
and the possibility to describe excited nascent fragments.
Let us also mention an alternative approach that consists in deriving a 
quantum theory of transport that 
combines a GCM description of collective dynamics with quantum 
statistics to account for thermal excitations and dissipation 
\cite{dietrich2010microscopic}.


\subsection{Comments on Adiabaticity and Dissipation}
\label{subsubsec:adiabatic}

The TDGCM, ATDHFB and adiabatic self-consistent collective (ASCC) methods are often 
referred to as adiabatic theories 
of large-amplitude collective motion. However, while this adjective has a clear 
definition in classical thermodynamics, it is often employed with a different 
(or several different) meaning(s) in fission theory and the theory of 
large-amplitude collective motion in general. In the literature, the concept of 
adiabaticity has been used as a proxy for three very different hypotheses, 
which are sometimes 
combined: (i) the collective motion, i.e., the change of shape leading to 
fission, is very slow \cite{hill1953nuclear,cassing1983role}, (ii) the 
coupling between such collective degrees of freedom and single-particle, or 
intrinsic degrees of freedom is so small that it can be neglected 
\cite{reinhard1978concept,goeke1980generatorcoordinatemethod,
goeke1983threedimensional}
and (iii) the intrinsic DoFs are associated with the lowest-energy configuration that
evolves with the
collective variable~\cite{bernard2011microscopic,tanimura2015collective}.
The variety of these definitions along with the rather arbitrary definitions of
collective and intrinsic excitations may lead to confusions.
We discuss below some common pitfalls related to this topic.

As a first example, theoretical evidence from TDDFT clearly suggests that
the collective motion is strongly dissipative, i.e. the shape of the system
evolves slowly and toward configurations which minimize the collective 
potential energy while the system populates single-particle states with 
high energies~\cite{scamps2015superfluid,bulgac2016induced,bulgac2019fission}.
Overall, TDDFT predicts time scales in the range of $10^{-20}-10^{-19}$ s to 
go from the saddle point to scission. 
On the one hand, this very slow collective motion justifies the 
reduction of TDDFT dynamics to its adiabatic counterparts, specifically, the 
hypothesis of small collective momenta that underpin, e.g., the ATDHF theory
\cite{baranger1978adiabatic}. On the other hand, TDDFT dynamics is often qualified
of diabatic in the sense of point (iii) above, because excited single- or quasi-particles states
are populated. One should thus be very careful when employing the adjective of adiabaticity to 
refer to slow collective motion.

The definition (ii) of adiabaticity especially focuses on the dissipation of 
energy from the collective DoFs to the intrinsic ones. Such dissipation is 
known to be a crucial ingredient in the Langevin dynamics as mentioned in 
Section~\ref{subsubsec:t_classical}. It appears in the form of the friction and random force 
terms in the collective equation of motion. The TDDFT picture
does not give direct insight into such dissipation since it does not rely on
an explicit splitting of the DoFs between collective and intrinsic.
It is possible to estimate \textit{a posteriori} the kinetic energy
associated with any collective variable represented by a local one-body 
operator~\cite{tanimura2015collective}. Such an approach suggested a 
strong dissipation of the quadrupole moment kinetic energy during the fission 
dynamics.
With this in mind, it might seem puzzling that the ATDHF approach based only on 
the assumption of slow collective motion (which seems to be valid) yields a 
collective equation of motion with no friction or dissipative terms~\cite{ring2004nuclear}.
In addition, the TDGCM produces primary fragments distributions
similar to the one obtained by Langevin approaches while also introducing no explicit 
friction term.
This suggest that primary fragments mass/charge distributions
are simply not very sensitive to dissipation. In contrast, predictions of the 
excitation energy of the primary fragments, which are not well described by 
TDGCM, are a much better probe into the dissipation mechanism.

In the case of ATDHF, there is another deeper reason for the absence of a
collective dissipation mechanism. Contrary to the current implementations
of TDGCM, which rely on a set of constrained HFB generator states, the original 
formulation of the ATDHF(B) approach  builds its collective path out of the 
time-even component of the states in a TDDFT trajectory. 
These states can already contain 
significant single- or quasi-particle excitations and only the small
energy component related to the residual time-odd part needs to be taken
into account by the collective dynamics. Following this idea, 
Reinhard \textit{et al.} proposed to build the collective path from
a variational principle~\cite{reinhard1978concept} that decouples by design
the collective dynamics from other DoFs.
This philosophy was formalized in the development 
of the ASCC model, which generalizes 
ATDHF theory while solving some of its inconsistencies 
\cite{matsuo2000adiabatic,hinohara2008microscopic,nakatsukasa2012density,
nakatsukasa2016timedependent,schunck2019energy}.
To summarize, the collective phase space constructed formally in these approaches
is in fact completely different from the one used either in current TDGCM implementations 
or in Langevin approaches. 
In other words, it is very possible that some theories  
need an explicit treatment of collective dissipation while some others do not.
The presence and strength of dissipative terms in the collective equation of motion
strongly depends on the arbitrary choice of the collective phase space.
One should keep this in mind while discussing adiabaticity in the sense of point (ii) above
or comparing the dissipative features of various theories.

\subsection{Ternary Fission}

Ternary fission is the rare process where the fission of a heavy element is 
accompanied by the emission of an $\alpha$ particle or, much less frequently, a 
light nucleus ranging from Lithium ($Z=3$) to Oxygen ($Z=8$) isotopes 
\cite{halpern1971three,theobald1989lowenergy}. More precisely, it is the 
(quasi)-simultaneous divide of the fissioning nucleus in three fragments -- one 
of them much lighter than the others. This phenomenon is usually quantified by 
the $\alpha$-emission probability $P_{\alpha}$ or, equivalently, the relative 
number $N_{\alpha}$ of emitted $\alpha$ over the entire number of fission 
events $N_{f}$ \cite{carjan1976origine}. Ternary fission does not lend itself 
easily to a description in terms of geometric collective variables as discussed 
in Section \ref{subsec:deformation} unless one enforces some amount of 
clustering, e.g. via a three-center basis \cite{poenaru2001nuclear}. Neither 
can such a process be easily simulated by the simple ansatz of TDDFT, where the 
nuclear wave function is a ``standard'' product state of particles or 
quasiparticles. In fact, the proper quantum-mechanical treatment of ternary 
fission should probably include a mechanism to rigorously simulate nuclear 
clustering. For these reasons, all models of ternary fission 
tend to be rather phenomenological. Probably the most advanced one relies on 
estimating $P_{\alpha}$ by solving the time-dependent Schr\"odinger equation 
for the $\alpha$ particle moving in a time-dependent mean-field potential, 
which is obtained by simply parametrizing with $t$ specific trajectories across 
the potential energy surface \cite{tanimura1987dynamic,schafer1995quantum}. 
Alternative methods involve the random-rupture neck model, where it is assumed 
that the neck between the prefragments ruptures at two different locations 
nearly simultaneously, the part of the neck between these two locations forming 
the light particle \cite{rubchenya1988dynamic}. The rupture-neck model was 
also combined with statistical theory (see Section \ref{subsec:statistical}) to 
account more realistically for the available excitation energy at scission 
\cite{lestone2004combined}.


\section{Emission of Light Particles}
\label{sec:deexcitation}

After the scission point, nuclear forces between the two fragments quickly drop 
to zero while Coulomb repulsion drives the acceleration and kinematics of the 
fission fragments. This is the acceleration phase of the fragments. 
We can estimate an order of magnitude for the time scale of the acceleration from the
classical dynamics of two point-wise particles. Let us consider an initial 
situation where the surfaces of the two fragments are separated by a distance 
of a few fermis and neglect the relative velocity between the fragments at this 
stage. Integrating the classical equation of motion results in a fast 
acceleration of the two fragments that reach 90\% of their final speed (2-5\% 
of the speed of light) within 5 zs. Within this time range, the fragments 
typically cover a distance of several tens of fermis.

The large deformation of the prefragments at scission, the effect of the Coulomb 
repulsion on the inhomogeneous charge densities in the fragments, as well as the internal 
excitation created during the descent from the saddle point to scission produce 
two excited fragments far from their ground-state equilibrium. On average, 
about 30 MeV of excitation energy is shared between the two fragments in the 
fission of an actinide while several theoretical and experimental arguments indicate
the typical spin of the fragments lies in the range of 5-12 $\hbar$ units 
(cf. Section~\ref{sec:initial_fragments}). The primary fragments lose their energy mostly 
by emitting neutrons and $\gamma$ rays. The post-neutron pair of fragments resulting
from this neutron emission phase are called secondary fragments. Further $\beta$ decay and possible 
delayed neutron/$\gamma$ emission may follow this process leading to a final pair of nuclei,
the fission products; see Fig.~\ref{fig:schematic}.

The characteristics of the emitted fission neutrons and $\gamma$ rays, 
such as their multiplicity and energy spectrum, are key nuclear data for 
energy applications and give stringent constraints on our understanding of 
the primary fragments formation. State-of-the-art theoretical approaches predicting 
these observables generally rely on a statistical deexcitation model
to simulate cascades of particle emissions from the excited fragments. 
In the specific case of fission, such deexcitation simulations have to be repeated 
for every combination of primary fragments pairs and set of relevant initial
properties that may impact the evaporation process. These generally include the
kinetic energy, excitation energy, spin and parity of the primary fragments.
The final observables are then obtained by averaging fission events both over
the large set of primary fragments configurations as well as over the set of 
deexcitation cascades for each of these configurations. 
Each event is weighted by its associated probability of appearance. 
This averaging can be performed in a deterministic way~\cite{tudora2017comprehensive}
as long as the phase space considered for fission events is small enough.
For more detailed descriptions of deexcitation cascades including for instance
the spin of the emitted particles the phase space of fission events becomes
so large that fragments deexcitation models often turn to a Monte Carlo 
estimation of observables. In this case, the code sample a large but 
limited number
of fission events according to their probability distribution and the 
observables are deduced as averages over these events.

The success of such approaches relies on high-accuracy and high-precision estimates
of the primary fission fragment distributions, especially for the 
total kinetic energy, which directly 
determines the total energy available for neutron/$\gamma$ emission.
As a consequence, it is customary to rely mostly on experimental data or empirical models for
the primary fragments distributions. In this section we will give an overview of such empirical 
treatments. Recent efforts directed toward replacing progressively these inputs by
genuine predictions from microscopic theories will be discussed in Section~\ref{sec:initial_fragments}.

\begin{table*}[t]
\begin{tabular}{lccccc}
\hline
\hline
 Code                    & PbP~\cite{tudora2017comprehensive} 
                                   & GEF~\cite{schmidt2016general} 
                                                 & FREYA~\cite{verbeke2015fission}
                                                                & CGMF~\cite{talou2021fission}
                                                                            & FIFRELIN~\cite{litaize2015fission} \\ 
 Pre-fission emission    &  --
                               & n$_{\text{pe}}$,n$_{\text{stat}}$,$\gamma$ 
                                             & n$_{\text{pe}}$,n 
                                                              & n,$\gamma$ & --  \\
 Wide range of fissioning systems 
                             & no   & yes        & no             & no       & no \\                                                       
 Neutron-$\gamma$ competition   & no   & yes        & partial        & full     & full \\
 Gamma deexcitation          & no   & yes        & yes            & yes      & yes \\
 Deexcitation model          & Weisskopf & Weisskopf & Weisskopf  & Hauser-Feshbach & Hauser-Feshbach \\
 Statistical fluctuation     & no   & no         & no             & no       & yes \\
 Open-source                 & no   & yes        & yes            & yes      & no \\
\hline
\hline
\end{tabular}
\caption{Summary of some features of fission fragments deexcitation models.
         In the pre-fission emission row, n$_{pe}$ denotes pre-equilibrium neutrons and 
         n$_{\text{stat}}$ corresponds to statistical neutrons. 
         'Statistical fluctuation' denotes taking into account fluctuations of nuclear
         structure properties on top of average quantities such as the level density 
         and the averaged partial widths.}
\label{tab:deexcitation_models}
\end{table*}

The statistical description of nuclear deexcitation cascades is by itself an old problem
quite independent of fission studies. It started with the seminal paper of Weisskopf in 1937
concerned with the spectrum of neutrons emitted from highly
excited nuclei~\cite{weisskopf1937statistics}. Grover \textit{et al.} pioneered the use of the Hauser-Feshbach
model of deexcitation for the study of neutron, $\gamma$ and $\alpha$ 
evaporation~\cite{grover1967deexcitation,grover1967dissipation,grover1967emission} and
published the first study quantifying the impact of the spin of the fission fragments on such decays~\cite{thomas1967angular}.
Many of the developments of statistical reaction theory in the 1970ies and 
1980ies briefly mentioned in Section \ref{subsubsec:fast_neutrons} aiming at going 
beyond Hauser-Feshbach theory \cite{kawai1973modification,
agassi1975statistical} were not really tested in practical simulation of 
fission fragment decay.
Interest for this subject surged in the 2000's, in particular due to 
the growing issue of $\gamma$ heating in nuclear reactors~\cite{plompen2006summary,oecdnea2006nuclear,lemaire2006monte}
as well as new technological applications such as neutron/$\gamma$ interrogation 
or advanced Monte Carlo simulations of detection devices.

Today, several codes, implementing different models, 
are dedicated to the description of fission fragment deexcitation. 
In this section we will only review the most commonly-used and complete ones: 
FREYA~\cite{verbeke2015fission}, GEF~\cite{schmidt2016general}, CGMF~\cite{talou2021fission}, 
FIFRELIN~\cite{litaize2015fission} and the Point-by Point-(PbP)~\cite{tudora2017comprehensive} model.
We summarize in Table \ref{tab:deexcitation_models} the main features of these codes.
The more recent code HF$^3$D~\cite{kawano2021influence,lovell2021extension} will not
be discussed since it implements a similar physics as CGMF with a different numerical
resolution (deterministic for HF$^3$D versus Monte Carlo for CGMF).
Other codes implement nuclear deexcitation cascades for specific applications, for example GEMINI++ for the 
deexcitation of fragments produced in heavy-ion 
reactions~\cite{sekizawa2017microscopic}. Conversely, general-purpose codes 
such as TALYS \cite{koning2008talys1} or EMPIRE \cite{herman2007empire} 
have a much broader scope of applications and aim at 
describing any type of nuclear reactions. For the sake of completeness, let us also mention the work of 
Lestone~\cite{lestone2016neutronfragment} focused on a very precise reproduction 
of the prompt fission neutron multiplicities, spectrum and angular correlations.

\subsection{Distribution of Primary Fragments Properties}

The properties of the primary fission fragments could in principle be obtained 
as the result of a statistical or dynamical approach to the large-amplitude collective motion 
as outlined in Section~\ref{sec:lacm}.
However, one should keep in mind that predictions of prompt particle
emission are very sensitive to the characteristics of the primary fragments.
For instance, microscopic theories can predict the kinetic energy of the fragments
at best within a few percent precision~\cite{bulgac2019fission}. 
In the fission of an actinide, this would typically lead to an uncertainty of a few MeV on the 
total excitation available for deexcitation, which would translate into a $0.5-1$ 
uncertainty on the average prompt neutron multiplicity. For comparison,  
measurements of the prompt neutron multiplicity can typically reach precision
below one percent ($\simeq 0.015$ neutrons for $^{252}$Cf spontaneous fission in Ref.~\cite{boldeman1967prompt}).

Adopting a pragmatic strategy, most fragments deexcitation models thus start by estimating
the primary fragments distributions either by directly using some available experimental data or 
by relying on empirical models previously calibrated to a collection of experimental data.
These distributions of probability are then used as an input to the statistical deexcitation
model. Note that such an approach discards any quantum correlations between the
various primary fragments configurations.
We now highlight the different ingredients and models used to build the empirical distribution
of the primary fragments configurations. 

\paragraph{Proton and neutron numbers}

The principal characteristics of the primary fragments configurations are the number of protons and neutrons in each fragment.
In the standard approach the probability distribution that fission produces a fragment with a number of mass $A_{\rm f}$ and a number of charge $Z_{\rm f}$
given the excitation energy $E^*_{\text{CN}}$ of the compound system is factorized into
\begin{equation}
 P(A_{\rm f},Z_{\rm f}| E^*_{\text{CN}}) = Y(A_{\rm f}| E^*_{\text{CN}}) P(Z_{\rm f}|A_{\rm f}).
\end{equation}
The factor $Y(A_{\rm f}| E^*_{\text{CN}})$ stands for the primary mass distribution whereas $P(Z_{\rm f}|A_{\rm f})$ accounts for the charge probability distribution given the mass number $A_{\rm f}$.
The complementary fragment is always determined through mass and charge conservation
(i.e. ternary fission is neglected)
\begin{equation}
A_{f2} = A_{\rm CN} - A_{f1}, \qquad Z_{f2} = Z_{\rm CN} - Z_{f1}.
\end{equation}
Fragments deexcitation models assume a normal distribution of charge with a mass-dependent mean $\bar{Z}_f(A_{\rm f})$ and standard deviation $\sigma_Z(A_{\rm f})$
\begin{equation}
P(Z_{\rm f}|A_{\rm f}) \propto \text{exp}
\left[ 
-\frac{1}{2}\left( \frac{Z_{\rm f} - \bar{Z}_f(A_{\rm f})}{\sigma_Z(A_{\rm f})} \right)^2
\right].
\end{equation}
The mean of this distribution is either directly determined from the Unchanged Charge 
Distribution (UCD) approximation (as in FREYA) or includes an additional polarization
charge effect.
In the later case, the polarization of charge is parameterized as a function of $A_{\rm f}$ 
(PbP model) or estimated from the Wahl's systematics~\cite{wahl1988nuclearcharge}
(FIRELIN, CGMF).

As for the mass distributions, a possible strategy relies on using experimental
data, either in tabulated form, or described in terms of a few fission modes~\cite{brosa1990nuclear}.
The codes FIFRELIN and PbP model implement such an approach that is limited
to fissioning systems and energies for which experimental data is available.
To overcome this issue, several deexcitation models favor a description of the primary 
mass yields as a sum of a three or five Gaussians. For a given fissioning system,
 the position (average mass), the width as well as the weight of each Gaussian 
 (also known as fission modes) follow empirical laws fitted to reproduce a set of experimental data spanning a range of excitation energies.
For example, the code CGMF implements a 
parameterization of the mass yields with 3 Gaussians whose means and widths linearly depend on the excitation energy~\cite{lovell2021extension,talou2021fission}.
A similar technique is adopted in FREYA based on  
five Gaussians with energy-independent means and a quadratic dependency of the 
widths~\cite{vogt2012eventbyevent}.
Such an approach enables systematic studies of the prompt particles evolution as a function of the input channel energy.
Because of the remaining dependency on experimental data, this approach only enables  calculations on a few well known fissioning systems.

Going one step further, the philosophy of the GEF code is to provide an empirical yet systematic
estimation of the mass yields for a wide range of energies and fissioning systems.
In this model, the fissioning system is described by an elongation degree of freedom
coupled to several transverse collective DoFs such as the mass of the fragments, 
and to a thermal bath describing the remaining degrees of freedom.
The distribution of the neutron and charge 
number of the fragments is determined from a two-step statistical process guided by 
the physics of the collective dynamics,
\begin{equation}
P(A_{\rm f},Z_{\rm f}) = P(\text{mode}) P(A_{\rm f},Z_{\rm f}|\text{mode}).
\end{equation}
In the first step, a statistical population of 
the available states in the neighborhood of the fission outer saddle point yields the probability for the system to choose a particular fission mode.
The probability associated with a fission mode $P(\text{mode})$ is proportional to
\begin{equation}
P(\text{mode}) \propto \text{exp} \left( -\frac{E^0_{\text{mode}}}{kT} \right),
\end{equation}
where $E^0_{\text{mode}}$ is the minimum of the potential energy of 
the mode at an elongation corresponding to the outer barrier, relatively to the energy of the lowest mode, and $T$ is the 
temperature of the system. In practice, five to seven modes are considered.
A second step determines the distribution of the particle 
numbers in the fragments $P(A_{\rm f},Z_{\rm f}|\text{mode})$ from the 
properties of the system close to scission. The potential energy 
surface in the transverse direction associated with the charge and mass numbers
are described as harmonic potentials functions of $Z$ and $A$ with a minimum
value at $Z_{\text{mode}}$ and $A_{\text{mode}}$.
A second statistical population of states in the neighborhood of scission configuration
then leads to the following distribution in each mode:
\begin{equation}
P(A_{\rm f},Z_{\rm f}|\text{mode}) = P(A_{\rm f}|\text{mode})P(Z_{\rm f}|A_{\rm f},\text{mode}),
\end{equation}
with
\begin{equation}
P(A_{\rm f}|\text{mode}) \propto \text{exp}\left( -\frac{(A_{\rm f}-A_{\text{mode}})^2}{2\sigma^2_{A,\text{mode}}} \right) ,
\quad
P(Z_{\rm f}|A_{\rm f},\text{mode}) \propto \text{exp}\left( -\frac{(Z_{\rm f}-Z_{\text{mode}})^2}{2\sigma^2_{Z,\text{mode}}} \right).
\end{equation}
The widths $\sigma_{A,\text{mode}}$ and $\sigma_{Z,\text{mode}}$ depend in particular on the energy balance estimated for each mode. 
Note that a special treatment is applied to the standard 2 mode ($S_2$) with a non
harmonic potential.
The GEF model finally includes the odd-even staggering of the fission yields as an
additive correction based on an equal-filling assumption of the intrinsic states.
Overall, the determination of the primary fission yields relies on a set of free 
parameters that are calibrated to a large body of experimental data, 
which includes high-resolution data from inverse kinematic
experiments~\cite{schmidt2000relativistic}. This approach is then capable of estimating 
systematically and with accuracy the primary fragments yields as a function of the mass,
charge and energy of the fissioning nucleus. 
This feature is quite unique, and other fragment deexcitation models
sometimes directly take the yields predictions from GEF as an input to their own 
deexcitation engine~\cite{vassh2019using,vassh2020probing,vogt2020employing}.

\paragraph{Energy sharing}

The excitation energy available in each primary fragments before particle 
evaporation is a crucial ingredient, since it is the primary driver of the
prompt neutron emission.
To estimate this quantity for any given mass and charge split, most of the 
deexcitation codes rely on an energy balance of the system before the reaction
and after the complete acceleration phase of the fragments. 
For neutron-induced fission, the energy balance reads~\cite{lemaire2006monte}
\begin{equation}
\label{eq:energy_balance}
{\rm TXE} + {\rm TKE} + M(Z_{\rm H},N_{\rm H}) + M(Z_{\rm L},N_{\rm L})
= 
M(Z,N)  + S_{n}(Z,N) + E_n,
\end{equation}
where $M(Z,N)$ is the mass of the nucleus $(Z,N)$, $S_{n}$ the neutron 
separation energy in the nucleus $(Z,N)$, $E_n$ the energy of the incident 
neutron in the center of mass frame, TXE the total excitation energy available to both fragments, 
and TKE the total kinetic energy of the fragments. 
In the case of 
spontaneous fission, this formula can be applied with $S_{n} = E_n = 0$; for 
photofission, $S_{n} = 0$ and $E_n \equiv E_{\gamma}$. 
The evaluation of masses and neutron separation energy is performed on a 
regular basis (see \cite{huang2021ame,wang2021ame} for the latest one) and data are available in nuclear structure
databases such as ENSDF~\cite{evaluated}, leaving only the total kinetic
energy and the total excitation energy to determine.
Because of the quantum fluctuations in fission, we can treat TKE as a random variable
with a large variance. 
In fragments deexcitation codes, its probability distribution is typically 
assumed Gaussian~\cite{talou2021fission} with a mean and variance that depend 
on the mass split
\begin{equation}
\label{eq:tke_distribution}
 P( \text{TKE} | A_{\rm f},Z_{\rm f},E^{*}_{\text{CN}}) 
 \propto \operatorname{exp}\left[ -\frac{1}{2}\left(\frac{\text{TKE} -\overline{\text{TKE}}(A_{\rm f})}{\sigma(A_{\rm f})} \right)^{2} \right].
\end{equation}
These moments of the probability distribution are in general extracted from experimental
measurements of the fragments kinetic energies (e.g. Ref.~\cite{varapai2005proceedings}) 
and may also include a dependency on the energy of the entrance 
channel~\cite{talou2021fission}.
Combining equations \eqref{eq:tke_distribution} and \eqref{eq:energy_balance} yield a distribution 
of probability for the total excitation energy TXE available for
the emission of particles.

Note that such a scheme neglects the possible emission of particles
(especially neutrons) before the complete acceleration of the fragments.
Although this hypothesis is strongly supported by the angular
distribution of the prompt neutrons~\cite{wagemans1984comparison}, which is peaked at 
forward angles, it remains
an open question whether, and to which extent, some of the neutrons are emitted before the complete
acceleration.
Numerous papers by Carjan and Rizea attempt the challenging prediction of 
scission neutrons and/or pre-acceleration neutrons based on a time-dependent 
average nuclear potential 
approach~\cite{carjan2019timedependent,carjan2019structures,carjan2015similarities}. 
Other studies were performed based on static potentials~\cite{capote2016scission}
and even on a full TDHFB calculation~\cite{bulgac2016induced}. As of today, no clear
consensus exists between the results of these theoretical approaches and experimentally verifying
their predictions remains difficult. In practice, most deexcitation codes 
are neglecting this kind of neutron emission.

After the determination of the total excitation energy available, one still needs to split
this energy between the two fragments. This energy sharing results from the complex competition 
between various energy reservoirs (intrinsic energy, deformation energy, etc) that 
interact with each other during the fission dynamics. The repartition of the energy strongly depends
on the mass of the fragments and it is expected to be highly correlated with the 
prompt neutron multiplicity $\bar{\nu}(A)$. Fragments deexcitation codes therefore rely on 
empirical laws for the total excitation energy sharing, which are guided by, or directly fitted on,
this neutron multiplicity.
The code CGMF assumes for instance that the excitation energy is of thermal nature and build
an empirical law for the temperature of the fragments as a function of its mass number. 
The code FREYA follows a similar approach with the introduction of a parameter that governs the
ratio of temperatures of the two fragments. In FIFRELIN, the excitation energy in each fragment
is recast into the sum of a rotational and a thermal contribution, with again a mass-dependent temperature ratio.
The point by point model (PbP) describes the excitation energy in terms of a thermal excitation
and a deformation component that depends on the mass. Finally, the GEF model shares the major
part of TXE according to a maximum entropy principle at scission and the remaining collective
kinetic energy is split evenly between the fragments.
Despite the apparent diversity of assumptions in all the models, we note that the shape relaxation
and the mass split dependency are the key ingredients that enable all these models to 
reproduce a sawtooth shape for the neutron multiplicity $\bar{\nu}(A)$.

\paragraph{Spin-Parity Distribution}

As will be shown in Section \ref{subsec:prompt}, the spin distribution of the 
fission fragments plays a very important role in setting the photon 
multiplicities, i.e., the average number of photons emitted during the 
deexcitation. The statistical theory assumes that the probability for a nucleus 
to have spin $J$ is approximately given by \cite{bloch1954theory}
\begin{equation}
p(J) \propto (2J+1)\exp \left[ -\frac{1}{2}\frac{(J+\tfrac{1}{2})^2}{\sigma^2} \right] ,
\label{eq:spin_distribution}
\end{equation}
where $\sigma$ is called the spin-cutoff parameter. Note that the spin-cutoff 
parameter is proportional to the expectation value $\braket{\hat{J}^{2}}$ of the 
total angular momentum in the system -- the mean of the actual distribution 
$p(J)$ \cite{madland1977influence,bertsch2019angular}. In the Fermi gas model, 
it is proportional to the moment of inertia of the system as well as to its 
temperature (or excitation energy) \cite{capote2009ripl}. Given this 
background, we can now summarize the different features of each deexcitation 
code:
\begin{itemize}
\item CGMF -- The spin of each fragment is sampled from the distribution 
\eqref{eq:spin_distribution}. The spin-cutoff parameter is explicitly 
parametrized by the product of the nuclear moment of inertia $\mathcal{I}$ and 
temperature $T$. The $Z$- and $A$-dependence of $\mathcal{I}$ is taken into 
account but not the variations as a function of deformation, angular momentum 
or temperature \cite{talou2021fission}. There is no constraint on the relative 
angular momentum between the fragments.
\item FIFRELIN --  The treatment of the spin is very similar to CGMF, but with 
an additional constraint used to simulate the damping of shell effects with 
increasing temperature \cite{litaize2010investigation,tamagno2018impact}. 
\item FREYA -- This code uses a semi-classical treatment of angular momentum 
\cite{vogt2021angular}. It assumes that the system can be treated as a dinuclear 
(binary) system of a heavy and a light fragment, each characterized by their 
moments of inertia $\mathcal{I}_{H}$ and $\mathcal{I}_{L}$ and spins 
$\gras{J}_{H}$ and $\gras{J}_{L}$, as well as a relative angular momentum 
$\gras{L}$. Angular momentum is thus explicitly conserved in this treatment. 
The total rotational energy of the dinuclear system is then expressed in terms 
of specific rotational modes, which are sampled separately based on a 
statistical formula \cite{randrup2014refined}. The dependency on nuclear 
structure inputs is encoded in the nuclear moments of inertia which, in most 
applications of FREYA, was approximated by its $A$-dependent rigid-body 
value. Recently, this approximation was improved following guidance from 
microscopic calculations \cite{vogt2021angular}.
\item PbP -- The spin of the fragments is not taken into account.
\item GEF -- As for CGMF and FIFRELIN, the spin of the fragments is sampled 
from \eqref{eq:spin_distribution}, but this is only done for the post-statistical 
emission spin. The statistical neutron and $\gamma$ emission is spin-independent.
\end{itemize}
Without any strong physical arguments as for the parity of the primary fragments,
all these models assume equiprobable parities.

\subsection{Prompt Particle Emission}
\label{subsec:prompt}

\begin{figure}[!htb]
\includegraphics[width=0.33\textwidth]{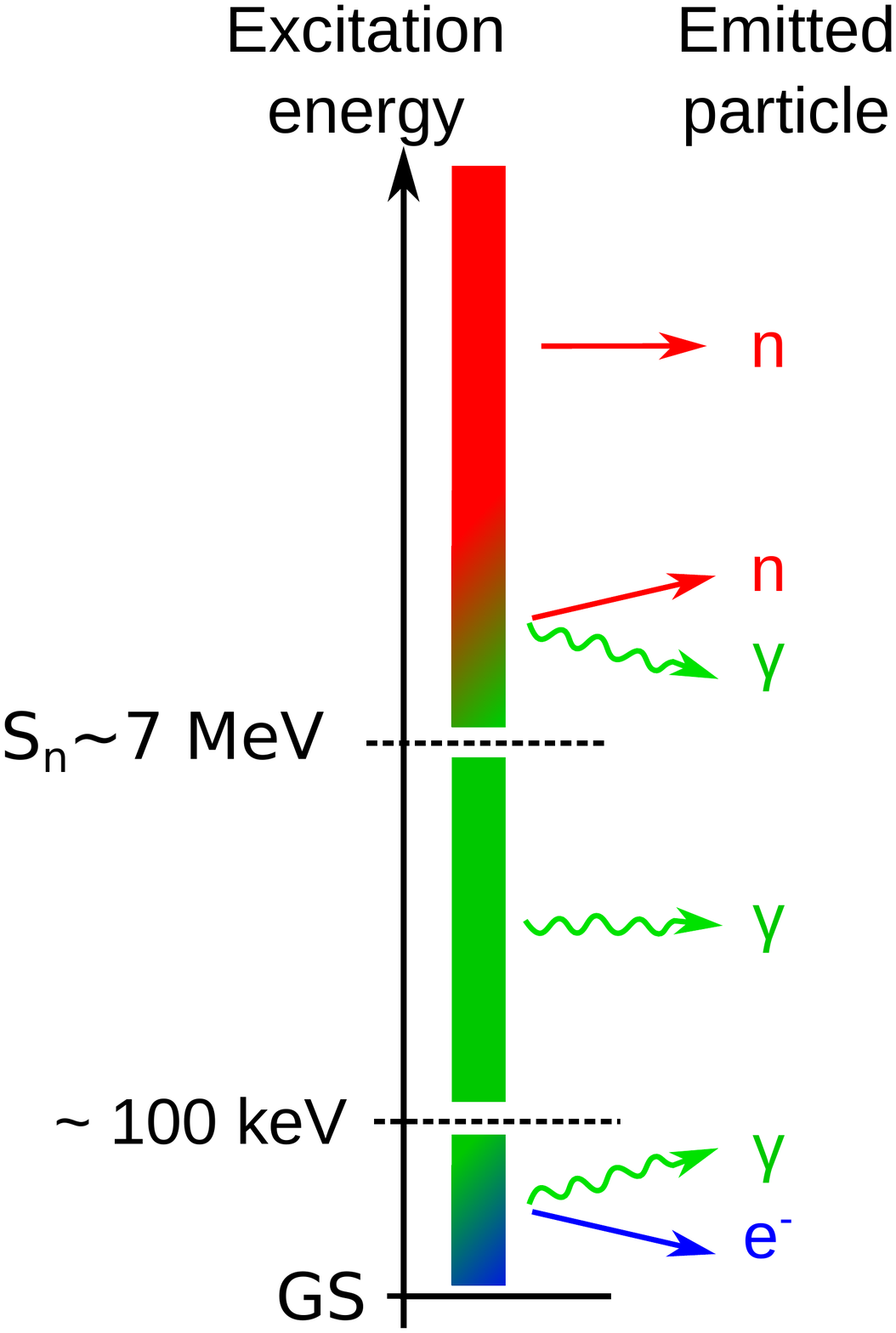}
\includegraphics[width=0.6\textwidth]{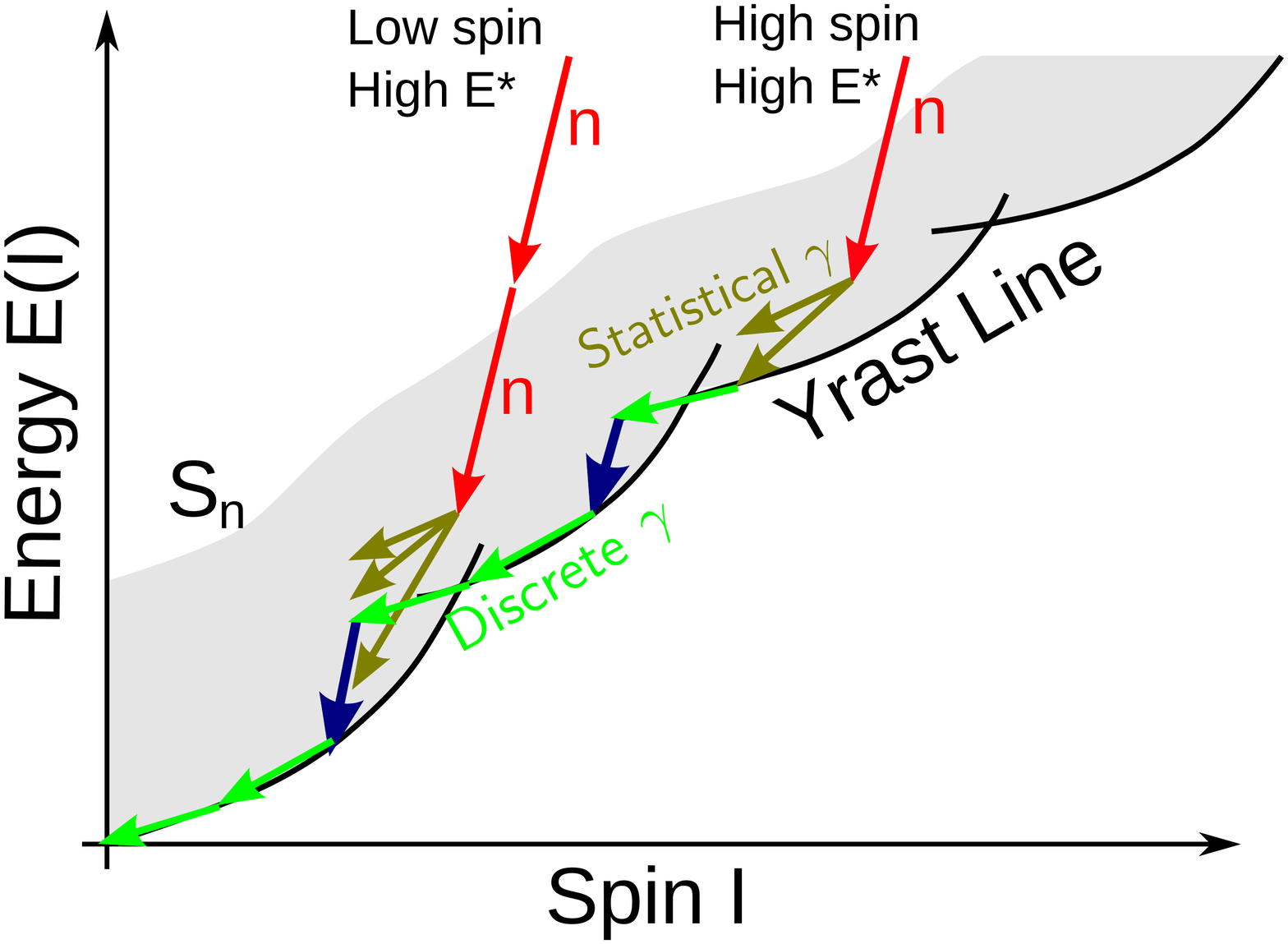}
\caption{Left: Schematic of particle emission as a function of the excitation 
energy. Right: Yrast plot showing decay mechanisms in the $E(I)$ plane. Neutron 
emission (in red) dominates at high excitation energy. Below the neutron 
separation energy $S_n$ (grey shaded area), statistical $\gamma$ emission (dark 
green) takes over until the Yrast line is reached, at which point discrete 
$\gamma$ transitions (green) occur until final deexcitation. Fast inter-band 
transitions, typically $E_1$ connect different rotational bands.}
\label{fig:fragmentDeexcitation}
\end{figure}

Starting from a highly exited state, the fission fragments first deexcite 
through successive emission of particles. As pictured in the left panel of 
Fig.~\ref{fig:fragmentDeexcitation}, the dominant type of particle emitted 
depends on the excitation energy and spin of the fragment. 
For fission fragments, the $\alpha$ emission channel is usually closed and 
the proton emission can be neglected.
At energies above 
10 MeV, the neutron emission channel is open and state-of-the-art theories agree on the fact that it is largely favored. Just above 
the neutron separation energy, the nucleus enters an area of competition
between the neutron and $\gamma$ emission. As we go lower in energy, the $\gamma$ 
emission channel becomes dominant up to energies of the order of 100 keV. 
At first, this $\gamma$ emission is mostly of $E2$ character and is statistical 
in nature and heavily fragmented \cite{dossing1996fluctuation,oberstedt2018prompt}, i.e., one 
cannot recognize individual rotational bands. When the nucleus has reached the 
lowest possible energy at a given spin $I$, what is known as the Yrast line 
\cite{bengtsson1985rotational}, discrete $\gamma$ transitions take place all the 
way down to the ground state. Several rotational bands possibly connected by 
fast $E1$ transitions may be populated. At energies of the order of $\sim$ 100 
keV, the internal conversion phenomenon becomes important: the electromagnetic 
quanta emitted by the nucleus has a non negligible probability to interact with 
the electrons of the atom, leading to the emission of an internal electron 
accompanied by X-rays.

This prompt emission of particles takes place over a wide range of time scales.
Arguments related to the statistical evaporation of the neutrons
lead to typical emission times ranging from 10$^{-18}$ to 10$^{-13}$ s~\cite{wagemans1991nuclear}. 
This upper bound is actually the official definition of prompt neutrons 
for the United State Department of Energy~\cite{doe}.
The prompt $\gamma$ emission mostly takes place within a few nanoseconds.
During this cascade, isomeric states of the fission fragments can be populated leading 
to $\gamma$ transition at 
post-scission times of  up to several microseconds. The relative population of these isomeric states, 
which is quantified by the branching ratios, has a strong impact on various 
fission properties \cite{madland1977influence,stetcu2013isomer,
stetcu2014angular,talou2016latetime}.

State-of-the-art theoretical approaches aiming at predicting the deexcitation 
process of the compound nucleus and/or the fission fragments are based on a 
statistical deexcitation approach. In such a model, the dynamics of the 
deexcitation process is not explicitly described. Starting from an initial 
excited state $|i\rangle$ of the nucleus, the statistical model predicts the 
probability for the emission of a series of particles leading to a final state 
$|f\rangle$ for the residual nucleus.

In the statistical model picture, a deexcitation cascade is seen as a 
multi-step process in which the system jumps from one intermediate state to  
another while emitting each time one particle. A cascade of deexcitation is 
therefore defined by (i) a series of transitions $|i_{k-1}\rangle \rightarrow |i_{k}\rangle$ 
between nuclear states or resonances, (ii) the particles $p_k$ emitted at each 
transition along with their physical properties (kinetic energy, spin/multipolarity, etc).
Owing to the quantum nature of the deexcitation dynamics, any possible 
deexcitation cascade or channel, generically denoted by the letter $c$ below, may happen with a given probability.
In standard approaches this probability is factorized as a product of
individual transition probabilities
\begin{equation}
P(c) =
P(i=i_0\rightarrow i_1, p_1) \times \cdots 
\times P(i_{k-1}\rightarrow i_k,p_k) \times \cdots
\times P(i_{n-1}\rightarrow i_n=f,p_n).
\end{equation}
The probability $P(i_{k-1}\rightarrow i_{k},p_k)$ represents the transition 
probability from the nuclear state $|i_{k-1}\rangle$ to the nuclear state 
$|i_{k}\rangle$ with the emission of a particle $p_k$. 
Such a factorization translates the Bohr hypothesis, which is assumed to be valid at each step
of the cascade. In particular, it neglects the possible quantum interferences
between the cascades.
After determining the probabilities associated with all possible deexcitation
cascades, one computes a target observable as a statistical average over all 
these events. For example, the average $\gamma$ multiplicity 
of the process is determined as
\begin{equation}
 M_{\gamma} = \sum_{c\in\mathcal{C}} P(c) M_{\gamma}(c) ,
\end{equation}
where $P(c)$ is the probability of occurrence of the cascade $c$ and 
$M_{\gamma}(c)$ the number of $\gamma$ rays emitted during this cascade.
A key aspect of the method is the proper determination of the set of
all possible cascades, which is related both to the structure of the nuclei
involved and to the probabilities $P(i_{k-1}\rightarrow i_{k},p_k)$
driven by the strength of the particle emissions.

At low excitation energies (below a few MeV), the structure
of most of the fission fragments is partially known from $\gamma$ spectroscopy. 
The ENSDF database records the first discrete energy levels, their spin and parity,
electronic conversion coefficient and possibly their transition rates to other discrete
levels. Deexcitation models such as FIFRELIN, CGMF or FREYA include this experimental structure
information in their deexcitation engine through tabulated databases such as the RIPL library \cite{capote2009ripl}.
At higher excitation energies the structure becomes unknown and ultimately neighboring neutron
resonances are overlapping. To tackle this energy range, deexcitation models rely on a
statistical description of nuclear structure~\cite{weidenmuller1984statistical,hansen1990applications,bohigas1988aspects}
identical to the ideas developed for the compound nucleus in the context of neutron-induced reactions (see Section~\ref{sec:proba}).
The level scheme is expressed in terms of a level density that encodes the 
average number of states/resonances in the neighborhood of a given excitation energy.
An additional description of the fluctuation of the the level spacing built from the 
Gaussian orthogonal ensemble (GOE) may complete the level density information.
A statistical description also applies to the probabilities of transfer from one level to the other.
The transmission coefficient $T_{a}(\epsilon)$ encodes the average probability 
to emit a particle characterized by the output channel $a$ and the energy $\epsilon$.
On top of this mean value, one can take into account the so-called Porter-Thomas 
fluctuations~\cite{porter1956fluctuations} or Ericson fluctuations~\cite{ericson1966fluctuations}
at higher energies.
These concepts form the backbone of fragment deexcitation models. Their detailed implementation
varies significantly from one model to another with two major trends: models based on a Weisskopf-Ewing
deexcitation spectrum and models based on the Hauser-Feshbach framework.

\paragraph{Weisskopf-Ewing evaporation}

Starting from an excited state of the nucleus, the Weisskopf-Ewing approach expresses
the energy spectrum of a particle evaporation. It originates from the seminal paper of
Weisskopf focused on neutron emission far above the neutron separation 
energy~\cite{weisskopf1937statistics}. 
The Weiskopf evaporation spectrum $\chi(\epsilon)$ in the center of mass frame reads
\begin{equation}
\label{eq:weisskopf_spectrum}
\chi(\epsilon) \propto \sigma_{\text{CN}}(\epsilon) \, \epsilon\,  \text{exp} \left(-\frac{\epsilon}{T} \right) ,
\end{equation}
where $\epsilon$ is the kinetic energy of the emitted particle (e.g. neutrons), 
$\sigma_{\text{CN}}(\epsilon)$ is the inverse cross section for the formation of the excited 
compound nucleus
by the process of the particle absorption and $T$
is the temperature of the residual nucleus (after neutron emission). Such a spectrum results from the summation over
all possible transitions for the particle energy $\epsilon$~\cite{hodgson1987compound}.
This implies that the detailed structure of the nuclei involved in the deexcitation
cascade does not need to be computed within this approach.
Also, the spin/multipolarity of the particle and the spin-parity of the 
residual state are not explicit taken into account. 

The deexcitation engine of the codes FREYA, PbP and GEF relies on the Weiskopf-Ewing evaporation
spectrum. For the neutron emission, the probability $P(i_{k-1}\rightarrow i_{k},p_k)$ of a transition
is proportional to Eq.~\eqref{eq:weisskopf_spectrum}. The PbP model relies on an optical model
calculation for the inverse cross section $\sigma_{\text{CN}}(\epsilon)$ whereas FREYA assumes it is
constant and GEF takes its energy dependency as $1 / \sqrt{\epsilon}$.
The Weisskopf-Ewing approach does not provide the relative 
probability of the neutron emission versus $\gamma$ emission. These deexcitation models therefore assume
that successive neutron emissions take place first, followed by the emission of $\gamma$ rays 
(see Fig.\ref{fig:fragmentDeexcitation}). The excitation energy at which the nucleus switches to $\gamma$
emission is set to $S_n$ in the case of the PbP model. 
To better account for the neutron/$\gamma$ competition, FREYA implements an empirical shift of this
energy threshold toward higher energies.
Another option explored by GEF consists in using an empirical formula for the relative neutron versus $\gamma$
probability that only depends on the excitation energy of the fragment.
Once the neutron evaporation is completed, the $\gamma$ emission can be treated in a similar manner.
For instance, FREYA uses the following Weiskopf-Ewing spectrum for $\gamma$ emission
\begin{equation}
 \chi_\gamma(\epsilon) \propto \epsilon^2 \, \text{exp} \left(-\frac{\epsilon}{T} \right).
\end{equation}
Contrary to a complete Hauser-Feschbach approach, such Weisskopf-Ewing evaporation
is expected to fail when the level density of the residual nucleus
starts to be discrete. The work of Marcath~\cite{marcath2018measured} compares the predictions
obtained from FREYA and the Hauser-Feshbach code CGMF to experimal measurement on $^{252}$Cf spontaneous
fission. It highlights quantitative differences between the models predictions but an overall
agreement of both models to experimental data.

\paragraph{Hauser-Feshbach Formalism}

As briefly descried in Section~\ref{subsubsec:fast_neutrons}, the 
Hauser-Feshbach approach associates with each open channel 
(neutrons, $\gamma$, fission, etc.) a partial width $\Gamma$ that is inversely proportional
to the characteristic time of the corresponding transition. The probability of a transition is
directly proportional to its associated partial width
\begin{equation}
\label{eq:hauser_proba}
 P(i_{k-1} \rightarrow i_{k}, p_k) = 
 \frac{\Gamma(i_{k-1}\rightarrow i_{k}, p_k)}{\sum_{i'p'} \Gamma(i_{k-1}\rightarrow i', p')}.
\end{equation}
Contrary to the Weisskopf-Ewing approach, the partial widths depend here on all
characteristics of the emission channel, including the spin of the emitted particle
as well as the spin-parity of the residual nucleus.
In addition, this approach naturally accounts for the competition between different kinds of emissions.
Computing the probability distribution of one emission requires the knowledge of the 
structure (discrete levels or statistical level scheme) of all possible residual nuclei.
The implementation of one decay step amounts to 
(i) listing all possible emission channels starting from the initial state,
(ii) determining the structure of all potential residual nuclei,
(iii) computing the partial widths of all possible transitions,
(iv) computing the sum of these partial widths to normalize 
 the distribution of probability.
In the context of fission, the codes CGMF an FIFRELIN implement such an Hauser-Feschbach scheme.

At low excitation energy, the partial widths may be known from $\gamma$ spectroscopy and directly
incorporated into Eq.~\eqref{eq:hauser_proba}.
For higher excitation energies, one relies on a statistical description of these quantities.
For neutron emission, an optical model calculation determines the average partial width
from 
\begin{equation}
\bar{\Gamma}_n(i\rightarrow f, l,j) = \frac{T_{l,j}(\epsilon)}{2\pi \rho(E_i, J_i, \pi_i)}.
\end{equation}
The transmission coefficient $T_{l,j}(\epsilon)$ corresponds to a neutron emission with
an orbital moment $l$, a total spin $j$ and a kinetic energy $\epsilon$.
In the denominator, $ \rho(E_i, J_i, \pi_i)$ stands for the spin-parity-dependent 
level density of the mother nucleus.
The average $\gamma$ partial width associated with the emission of a $\gamma$ ray of 
multipolarity $XL$ ($X=E,M$) and energy $\epsilon$ is given by
\begin{equation}
\label{eq:DefFonctionDeForce}
\bar{\Gamma}_{\gamma}(i\rightarrow f, XL) =
\frac{\epsilon_{\gamma}^{2L+1} f_{XL}(\epsilon)}
{\rho(E_i, J_i, \pi_i)} .
\end{equation}
The $\gamma$-strength function $f_{XL}$ contains the information related to the inverse 
cross section reaction. Although it can be inferred from the results of QRPA 
calculations \cite{goriely2002largescale,goriely2004microscopic,
hilaire2017quasiparticle}, it is most often parameterized as a Lorentzian shape 
corrected for various physical effects (Pigmy resonance, low energy shift). 
Such a parameterization is rooted in the Brink-Axel hypothesis that establishes 
a simple link between the $\gamma$-strength function depending only on the energy 
and multipolarity, and the photoabsorption cross section from any excited 
state \cite{axel1962electric}. More elaborate models such as the Enhanced Generalized
LOrentzian (EGLO) overcome this assumption by including a temperature dependency
in the gamma strength~\cite{capote2009ripl}.

These partial widths correspond to probabilities of transition averaged 
over the detailed nuclear structure near the initial and final states $i_{k-1}$ and $i_{k}$.
Significant 
variations from these averages can rise due to the Porter-Thomas fluctuations 
\cite{porter1956fluctuations} (fluctuation in the strength of the decay) or 
Ericson fluctuations at higher energies~\cite{ericson1966fluctuations}.
Different methods are possible to
take into account these fluctuations~\cite{regnier2016improved}. Although their 
impact on the deexcitation observables is weak in general, it proved to be of 
significant importance in the area where the number of open neutron channels 
comes close to unity.

\subsection{Pre-fission Particle Emission}
\label{sec:prefission}

If the entrance channel of the reaction supplies sufficient energy to the 
system, the emission of particles before fission happens is likely. For 
example, in the neutron-induced fission of actinides with fast neutrons, the 
emission of neutrons before fission leads to the well-known phenomenon of 
second- and third-chance fission: not only does the original compound nucleus 
with $Z$ and $N$ neutrons undergoes fission, but so does the isotope with 
$N-1$ neutrons (second-chance), $N-2$ neutrons (third-chance), etc. Since such 
processes carry away part of the energy and spin of the initial system, it is 
essential to take them into account to obtain realistic estimates of the prompt 
emission characteristics. The fission models discussed here may take into 
account this phenomenon with different degrees of refinement.

As already mentioned in Section~\ref{sec:proba}, fission competes with other
light particles channels and the ratio of the total widths of each 
channel gives their respective probability. For instance the fission
probability reads
\begin{equation}
P_f = \frac{\Gamma_f} {\Gamma_f + \Gamma_n + \Gamma_{\gamma}}.
\end{equation}
Codes like FIFRELIN or PbP do not implement this competition mechanism
and assume a probability one for fission. This restricts their usage at energies below
the threshold of emission before fission.
Other codes like GEF and FREYA account for pre-fission emission of neutrons and/or $\gamma$
rays. To do so they parameterize the total widths of pre-fission emissions channels 
with analytical formulas. GEF evaluates for instance the neutron and $\gamma$ channels from
the compound nucleus mass, temperature of the intrinsic DoFs at different steps of
the fission process as well as the fission barrier~\cite{schmidt2016general}.
In FREYA, only neutron emission is considered {\it before} the fission 
process (no $\gamma$). Both statistical and pre-equilibrium neutrons emission are 
possible. The ratio $\Gamma_n/\Gamma_f$ is determined based on the level 
density of the residual nucleus. For the case of fission, the density 
considered is at the top of the fission barrier. (Eq. 2 of \cite{verbeke2015fission});
see \cite{vogt2017improved} for more detailed information.
Finally, the code HF$^3$D, a deterministic equivalent of CGMF, computes the 
fission widths coming from the Hill-Wheeler formula \eqref{eq:Tf} like in TALYS 
\cite{koning2008talys1} and EMPIRE \cite{herman2007empire}. In these 
calculations, the fission width competes with possible neutron- and 
$\gamma$-emission widths depending on the energies and possibly multipolarities of 
the emission: this is a fully fledged Hauser-Feschbach calculation. 
Additional details are 
given in \cite{lovell2021extension}.

\subsection{Applications of Fragments Deexcitations Models}

The deexcitation codes developed specifically to describe the decay of fission 
fragments are obviously important tools to evaluate the prompt fission 
spectrum or provide constraints on the theoretical models aiming to describe 
the physics of scission. In recent years, their range of application has also 
grown to encompass more specific applications in nuclear technologies or 
transport codes. In this section, we summarize some of these developments.

\subsubsection{Benchmarks of Prompt Observables}

The fission fragments (FF) deexcitation models have been extensively compared to the available 
experimental data on prompt neutron and $\gamma$ rays characteristics during the 
last decade. Most of the relevant literature is focused on the spontaneous 
fission of $^{252}$Cf, because of the quality of the experimental data, and 
on the thermal neutron-induced fission of $^{235}$U and $^{239}$Pu because of 
their importance in nuclear technology. Global benchmarks of the models with 
experimental data can be found in Refs.~\cite{litaize2015fission} for FIFRELIN, 
~\cite{vogt2017improved} for FREYA, ~\cite{tudora2017comprehensive} for PbP, 
~\cite{schmidt2016general} for GEF and ~\cite{stetcu2014properties,
talou2014prompt} for CGMF. A few benchmark comparisons between some of these 
models have also been published \cite{talou2018correlated}.

\begin{figure}[!ht]
\centering
\includegraphics[width=0.49\textwidth]{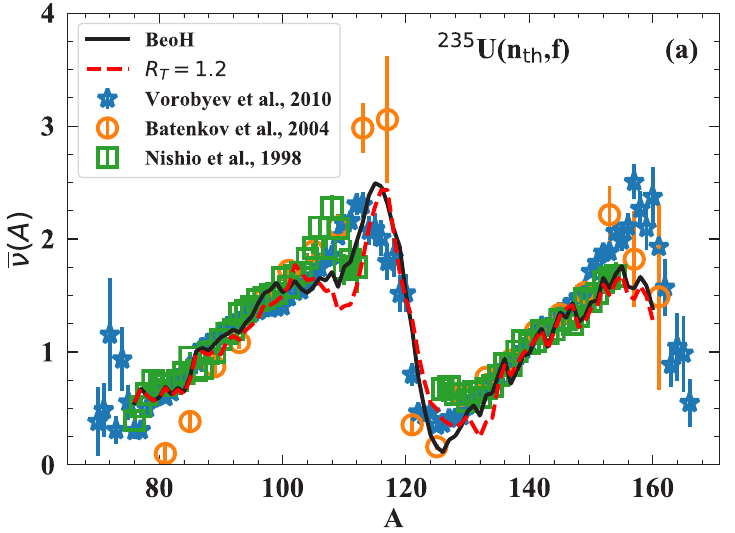}
\includegraphics[width=0.50\textwidth]{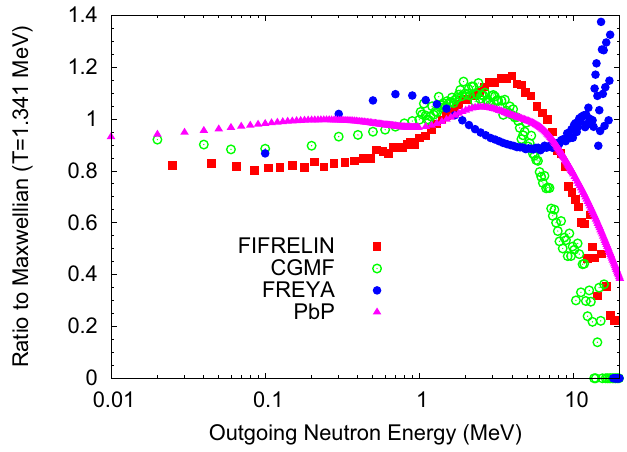}
\caption{Left: Average prompt neutron multiplicity of $^{235}$U(n$_{th}$,f) as a 
function of the primary fission fragment mass. Results predicted by the BeoH 
code compared to experimental data. Figure reproduced with permission from
~\cite{lovell2021extension} courtesy of A.~Lovell; copyright 2013 by The Amercian 
Physical Society. 
Right: Prompt neutron spectrum from the thermal induced fission on $^{235}$U. 
The spectrum is represented as its ratio to a Maxwellian with temperature 
$T=1.341$ MeV. The predictions obtained with the FF deexcitation models 
FIFRELIN, CGMF, FREYA and PbP are compared using the same primary FF mass and 
charge yields. Figure reproduced with permission from~\cite{capote2016prompt} 
courtesy of R.~Capote; copyright by Elsevier.
 }
 \label{fig:lovell_prc_2021_fig6}
\end{figure}

One of the key observables in FF deexcitation is the average neutron 
multiplicity as a function of the mass/charge of the primary fragment. The 
number of emitted neutrons is indeed closely related to the available 
excitation energy in the primary FFs and presents a so-called sawtooth shape as 
shown in the left panel of Fig.~\ref{fig:lovell_prc_2021_fig6}. This sawtooth 
behavior is mostly understood as coming from the shape relaxation of the 
fragments from their configuration close to scission to their ground-state 
deformation. Typically, heavy fragments close to $^{132}$Sn are assumed to be 
produced only with a low deformation at scission, which translates into a low 
excitation energy after shape relaxation. This reasoning explains the dip 
in the observed neutron multiplicity $\bar{\nu}(A)$ near $A=132$. Reproducing 
the sawtooth shape in the neutron multiplicity has led to the current 
implementations of excitation energy sharing between the fragments enumerated 
in Section~\ref{subsec:initial_e}. All deexcitation models discussed in this 
section reproduce qualitatively this sawtooth behavior. A widespread 
discrepancy compared to experiment was the prediction of a minimum in the 
neutron multiplicity at $A=132$ instead of the the measured A=$130$ in the 
fission of actinides. This deviation was recently attributed to the assumption 
of shell closure at A=$132$ which is actually shifted by a few mass units if 
one also considers octupole deformations in the fragments 
\cite{scamps2018impact}.

Along with the neutron multiplicity, the prompt neutron spectrum is of 
particular interest for nuclear technology applications. On this specific 
topic, Ref.~\cite{capote2016prompt} reviews in details predictions obtained 
from the available deexcitation models with regard to measurements and needs 
for evaluated data. Despite the efforts of the community, deexcitation models 
do not yet predict the prompt fission neutron spectrum (PFNS) with the accuracy required for 
such applications. This is illustrated in the right panel of 
Fig.~\ref{fig:lovell_prc_2021_fig6}, which shows the large discrepancy between 
the PFNS predicted by four state-of-the-art deexcitation models. Recent studies 
suggested that modifications both of the width of the FF distribution and of 
the spin assignment of a few low-lying states in the FF structure could improve 
significantly the PFNS predictions \cite{kawano2021influence}.

\begin{figure}[!ht]
\centering
\includegraphics[width=0.46\textwidth]{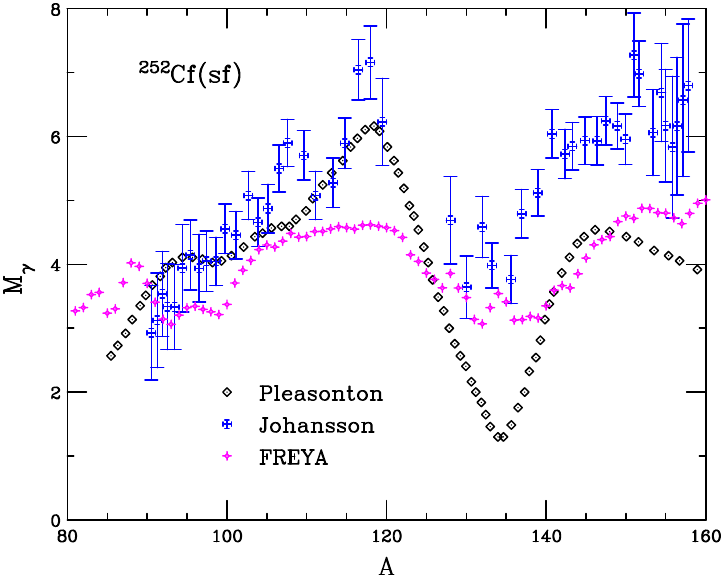}
\includegraphics[width=0.49\textwidth]{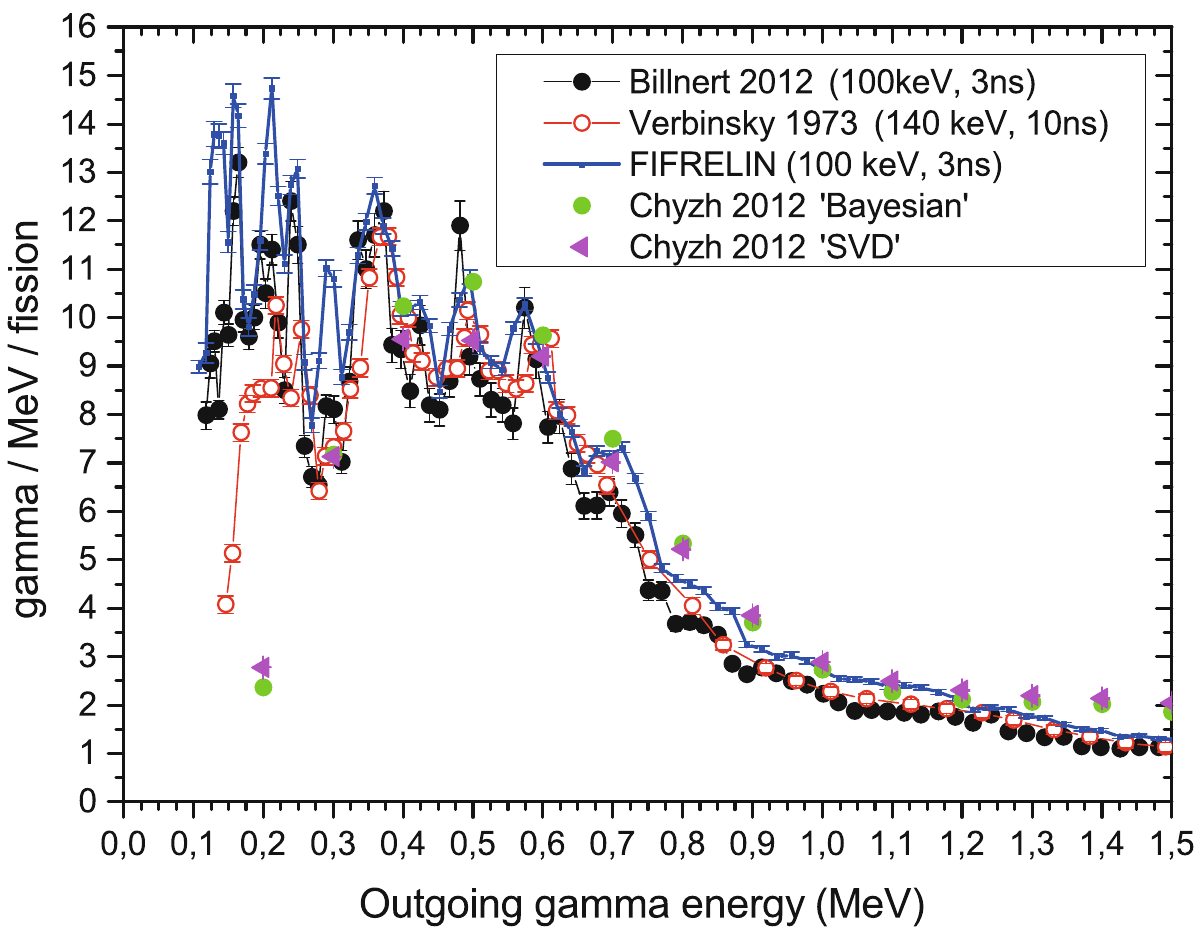}
\caption{Left: Prompt $\gamma$ multiplicity of $^{252}$Cf(sf) predicted by 
FREYA and compared with two experimental results. 
Figure reproduced with permission from~\cite{vogt2017improved} courtesy of R. Vogt;
copyright 2013 by the American Physical Society.
Right: Comparison of the prompt $\gamma$ spectrum of $^{252}$Cf predicted by with 
FIFRELIN to experimental data. Figure reproduced with permission from~\cite{litaize2015fission}
courtesy of O.~Litaize; copyright by Springer.}
\label{fig:vogt_prc_2017_fig21}
\end{figure}

Figure \ref{fig:vogt_prc_2017_fig21} is representative of typical predictions 
for the prompt $\gamma$ multiplicity and spectrum. In general, the total prompt 
$\gamma$ ray multiplicity is less known than the neutron multiplicity and 
differences in theoretical predictions are more pronounced. The total $\gamma$ 
multiplicity as a function of the primary fragment mass is particularly 
sensitive to the initial distribution of spins of the fragments 
\cite{stetcu2014properties,thulliez2019neutron}. This makes deexcitation models 
a good tool to constrain our knowledge of the primary fragments spins from 
$\gamma$ multiplicity measurements as discussed in Section~\ref{subsec:spin}.
The right hand side of Fig.~\ref{fig:vogt_prc_2017_fig21} compares the 
low-energy part of the prompt $\gamma$ spectrum of $^{252}$Cf spectrum obtained 
from FIFRELIN with three sets of experimental data. The overall magnitude of 
the spectrum reproduces the experimental spectrum within $\pm$1 $\gamma$ per 
fission. It is also very noticeable that the deexcitation codes based on the 
Hauser-Feshbach method reproduce very well the low-energy structures in the 
spectrum. This is directly related to the fact that the low-energy structure of 
all possible FFs is taken into account as an input of these codes, usually by 
reading it from existing data libraries \cite{capote2009ripl}.

\subsubsection{Probes into the Scission Mechanism}

\begin{figure}[!ht]
\centering
\includegraphics[width=0.6\textwidth]{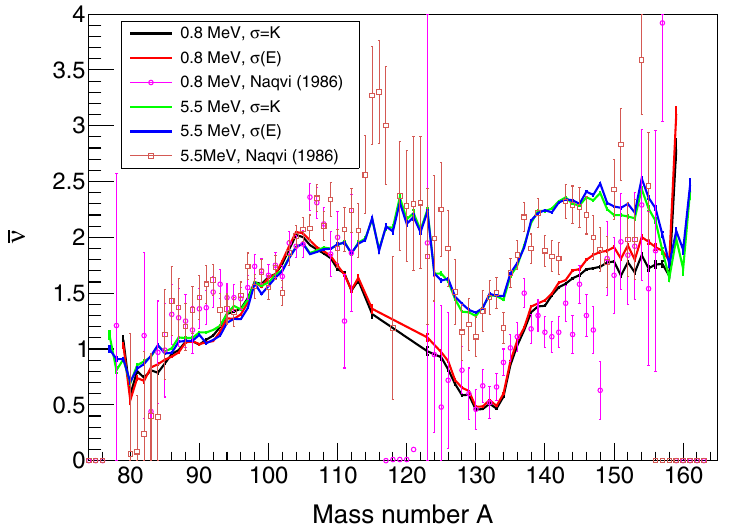}
\caption{Neutron multiplicity as a function of the primary fragment mass for 
two excitation energies of the fission system $^{238}$Np. Increasing the 
excitation energy of the system mostly increases the neutron multiplicity of 
the heavy fragment. This behavior is correctly reproduced when assuming 
an energy and mass dependent temperature ratio between the two fragments.
Figure reproduced with permission from \cite{thulliez2019neutron} courtesy
of O.~Litaize; copyright 2013 by the American Physical Society.}
\label{fig:thulliez_prc_2019_fig5}
\end{figure}

Beyond their ability to predict the prompt particles multiplicities and 
spectra, most of the deexcitation models provide information on other 
observables such as the multipolarity of the $\gamma$ transitions as well as their 
correlations. Comparing these results with a large body of experimental data 
puts stringent constraints on the empirical modeling of the properties of the 
primary fragments and enables to infer information on the physics of scission 
itself. A recent example is the study of the $^{237}$Np(n,f) reaction as
a function of incident neutron energy~\cite{thulliez2019neutron}.
This paper shows that different models with incompatible energy sharing mechanism and primary
fragment spin distributions 
reproduce equally well the experimental neutron multiplicity $\bar{\nu}(A)$ 
(cf. Fig.~\ref{fig:thulliez_prc_2019_fig5}) but differ significantly
on their predictions on the neutron-$\gamma$ correlation $M_\gamma(\bar{\nu})$.

Many studies have also attempted to unfold the primary fragments spin distribution
by matching particle emission predictions to known measurements. 
Work along this line have focused on the $\gamma$ multiplicity
~\cite{talou2014prompt,stetcu2014properties,stetcu2021angular,randrup2014refined} as well 
as well as on the isomeric ratios in the fragments~\cite{stetcu2013isomer,okumura2018235u,stetcu2021angular}.
We discuss in more details theoretical efforts to predict from microscopic 
theory the primary fragments spins
in Section~\ref{subsec:spin}.

\subsubsection{Evaluated Data for Nuclear Technology}

One possible application for deexcitation codes consists in contributing to the 
evaluation of nuclear data. The fact that a model predicts the neutron 
and $\gamma$-rays spectra and multiplicities in a consistent way is a unique opportunity
to augment 
the nuclear databases with correlated evaluations as well as their associated 
uncertainties.
The joint effort of Ref.~\cite{capote2016prompt} is a first step in this direction, 
focusing on the evaluation of the PFNS based on various models including the 
PbP model and FREYA. Covariance matrices of the uncertainty of the PFNS at 
various outgoing neutron energies were computed within FREYA. In practice, 

Whereas alternative models exist to evaluate the prompt neutron spectrum, 
deexcitation models are basically the only existing tools capable of predicting 
prompt $\gamma$ observables. In 2000's, the needs for better nuclear data 
associated with the prompt $\gamma$ production were officially formulated in the 
Nuclear Data High Priority Request List of the Nuclear Energy Agency of the 
OECD~\cite{oecdnea2006nuclear}. This surge of interest was mainly due 
to the topic of $\gamma$-ray heating in the structural materials of GEN-IV 
reactors, especially in large reflectors~\cite{privas2018reflector}. For this 
reason, prompt $\gamma$ multiplicity distributions predicted with the CGMF code 
have been incorporated in the ENDF-BIII.0~\cite{stetcu2020evaluation}. In this 
work the deexcitation code also supported the evaluation process of the prompt 
$\gamma$ spectrum. Similarly, in the JEFF-3.3 libraries, the code FIFRELIN was
used to complete the experimental prompt fission $\gamma$ spectrum data below and above 
experimental energy thresholds~\cite{plompen2020joint}. These efforts illustrate 
a recent trend of including deexcitation codes into the evaluation process. 
Further extensions of the scope of such codes suggested by the work of 
Ref.~\cite{okumura2018235u} would be the evaluation of the independent yields 
and of the branching ratios toward metastable states in the deexcitation 
cascade.

\subsubsection{Fission Event Generators in Neutron Transport Codes}

In general-purpose neutron transport codes such as MCNP or TRIPOLI, fission 
events and their associated emission of prompt particles are taken into account 
in an average way. The number of prompt neutrons generated is typically 
sampled between the two integers closest to the average neutron multiplicity 
instead of a complete distribution ranging from 0 up to 10 neutrons. The 
correlations in energy or emission angles between the emitted particles are 
neglected. This simple modeling of the prompt particles emission turns out to 
be sufficient for criticality problems in large systems. On the other hand, 
neutron transport codes are nowadays targeting new applications requiring more 
precise inputs related to fission. This includes simulation of detectors for 
experimental nuclear physics and neutron interrogation technologies
~\cite{hall2007nuclear,hausladen2007portable}.
Such applications often involve smaller geometries for which the anisotropy of 
transported particles becomes important. 
The development of these applications together with the capability of
FF deexcitations codes to predict angular correlations (see for instance
~\cite{lovell2020correlations}) 
is an incentive to use these models as fission events 
generators in neutron transport codes.

As of today, the codes FREYA and CGMF can be used as fission event generators 
in the MCNP6.2~\cite{werner2018mcnp} code and were also interfaced with the 
MCNPX-PoliMi transport code~\cite{talou2018correlated}. On the other 
hand, the codes FREYA and FIFRELIN can also be used within the TRIPOLI-4 
transport code~\cite{petit2016fifrelin,verbeke2018correlated}. Only a few 
applications have been published so far. These include the simulation of a prompt 
neutron-neutron angular correlation detection apparatus 
\cite{verbeke2018correlated}, which would not have been possible without a 
fission event generator, and the study of the neutron/photon multiplicities 
correlations measurement~\cite{marcath2018measured}.

\subsubsection{Experimental Data Analysis} 
\label{sec:experimental_interpretation}

Fission fragments deexcitation models often provide important insights into the 
interpretation of experiments thanks to their ability to estimate the 
sensitivity of various observables to an isolated physical effect. For example 
the effect of the Doppler shift of the prompt $\gamma$-ray emission due to the 
velocity of the FFs on the low energy $\gamma$ spectrum can be explicitly 
calculated \cite{regnier2013contribution}. Taking into account this effect 
reproduces nicely the experimental prompt $\gamma$-ray spectrum of $^{252}$Cf 
measured by Verbinsky \text{et al.}~\cite{verbinski1973prompt}. 

Fission events generators are capable of estimating the impact of experimental 
parameters such as the coincidence time window or the detection energy 
threshold on the measured $\gamma$ multiplicity. These two parameters turn out to be 
critical for the precise determination of the $\gamma$ multiplicity. Typically, 
the late-time $\gamma$ emission from a few nanoseconds to a few microseconds after 
scission represents a significant part of 3\% to 7\% of the total number $\gamma$ 
rays~\cite{talou2016latetime,chyzh2018dependence}. 

Finally, some specific features of the overall prompt particles spectra may 
sometimes be assigned to a specific FF by a deexcitation code analysis. 
Interpreting an experiment measuring the high-energy $\gamma$ spectrum 
between 0.8 and 20 MeV of $^{235}$U(n,f) with the code CoH, the authors of 
Ref.~\cite{makii2019effects} showed that the high-energy $\gamma$ ($>$ 10 MeV) mostly 
come from a few FFs. They were also able to assign a bump in the spectrum near 
4 MeV to emission from the heavy fragments near the 
shell closure $Z=50$, $N=82$.

Beyond these applications in the context of fission, fragments deexcitation
codes contains general deexcitation models and have thus been used in other 
fields such as the physics of neutrino detection~\cite{almazan2019improved,thulliez2021calibration}.

\subsection{Delayed Emission}
\label{subsec:delayed}

After the prompt emission, $\beta$ decay in the fission fragments can 
still populate daughter nuclei with a significant amount of energy.
Reaching the ground state of the fission products requires the emission
of delayed neutrons and/or $\gamma$ rays. 
A typical fission of actinide will emit approximately one delayed neutron 
every 100 fissions~\cite{wahl1988nuclearcharge}.
Most of the deexcitation codes reviewed in this section do not simulate
this delayed emission yet. There have been a few studies performed with the HF$^3$D code 
to simulate the delayed neutron emission~\cite{okumura2018235u,okumura2019prompt}.
They especially focus on the prediction of the cumulative fission yields
as well as of isomeric ratios.


\section{Toward a Consistent Description of Fission}
\label{sec:initial_fragments}

The deexcitation of fission fragments is largely dictated by the properties 
of the primary fragments after their complete acceleration.
We have shown in Section~\ref{sec:deexcitation} that state-of-the-art fragment
deexcitation models rely on empirical estimates of these properties, which are calibrated
on experimental measurements. Yet, both technological and astrophysical applications require 
robust predictions of the light-particle emission resulting from the fission of many
short-lived isotopes over a broad range of energies. In many cases, experimental data for 
such systems either is very incomplete or simply does not exist. 
This provides a strong incentive to use theoretical models of 
the formation of primary fragments to provide inputs to fragments deexcitation codes.
The past decade has witnessed tremendous progress in this direction especially
in the context of the EDF framework. In this section, we report on
such recent achievements, which bring us closer to a consistent description of both the formation 
and deexcitation of fission fragments.

The EDF framework, whether explicitly time-dependent 
or not, provides potentially powerful tools to characterize the primary 
fission fragments starting from (effective) nuclear forces among nucleons.
The most important problem is to identify 
nearly-isolated fragments with good quantum numbers and well-defined excitation 
energy. Until very recently, most work focused on estimating the properties of 
the fragments close to scission were based on average properties only 
(i.e. expectation value of one-body operators). For instance 
the expectation value of any one-body, spatially local operator $\hat{O}$ 
such as, e.g., particle number, angular 
momentum, multipole moments, etc., in the left fragment can be 
computed simply from
\begin{equation}
\braket{\hat{O}}_{\rm L} = 
\int_{-\infty}^{\infty} dx
\int_{-\infty}^{\infty} dy
\int_{-\infty}^{z_{\rm N}} dz\, \hat{O}(\gras{r})\rho(\gras{r}) ,
\label{eq:obs_classical}
\end{equation}
where $z_{\rm N}$ refers to the position of the neck between the two 
prefragments along the axis of elongation. Several pioneering studies of 
neutron-induced fission relied precisely on such average estimates in the 
neighborhood of scission~\cite{goutte2004mass,goutte2005microscopic,
dubray2008structure,younes2009microscopica,younes2011nuclear,
schunck2014description,schunck2015description}. The principal deficiency of 
this procedure is that some results are not realistic. For example, the 
particle numbers in the fragments typically take non-integer values. In 
addition, static calculations do not deal with isolated fragments but with 
entangled prefragments \cite{younes2011nuclear}. Although special procedures 
can be designed to mitigate and somewhat quantify this effect, they do not 
completely solve the problem \cite{younes2011nuclear,schunck2014description}. 
In this respect, TDDFT provides much cleaner estimates since one can wait until 
the fragments have fully separated and only interact with one another through 
the Coulomb repulsion \cite{simenel2014formation}. 
In the following, we are 
reviewing the recent improvements on the determination of these primary fragments
properties crucial for the simulation of light particles emission.


\subsection{Number of Particles}
\label{subsec:particles}

The number of particles in the fragments is the most basic property one would 
expect a fission model to provide. Yet, the naive application of 
\eqref{eq:obs_classical} almost always gives fragments with non-integer numbers 
of particles -- a problem made worse in static theories where 
results depend on the definition of scission; see Section~\ref{subsec:scission}. 
The solution is to extend well-known particle number projection (PNP) techniques to the 
case of a fission fragment following an idea originally proposed by Simenel 
for heavy-ion collisions \cite{simenel2010particle}. In this case, the 
``only'' difference with standard projection is the definition of the particle 
number operator, which now reads, e.g., for the left fragment,
\begin{equation}
\hat{N}_{L} = \sum_{\sigma}
\int_{-\infty}^{\infty} dx
\int_{-\infty}^{\infty} dy
\int_{-\infty}^{z_{\rm N}} dz\, c^{\dagger}(\gras{r},\sigma) c(\gras{r},\sigma),
\end{equation}
where $(c^{\dagger}(\gras{r},\sigma),c(\gras{r},\sigma))$ are the usual 
particle creation and annihilation operators. The first application of this 
technique to fission was reported in \cite{scamps2015superfluid}. It was later 
extended and formalized in \cite{verriere2019number} and applied to estimate 
particle numbers for a large set of static scission configurations of $^{240}$Pu and $^{236}$U 
\cite{verriere2021improvements,verriere2021microscopic}. 

\begin{figure}[!htb]
\centering
\includegraphics[width=0.51\textwidth]{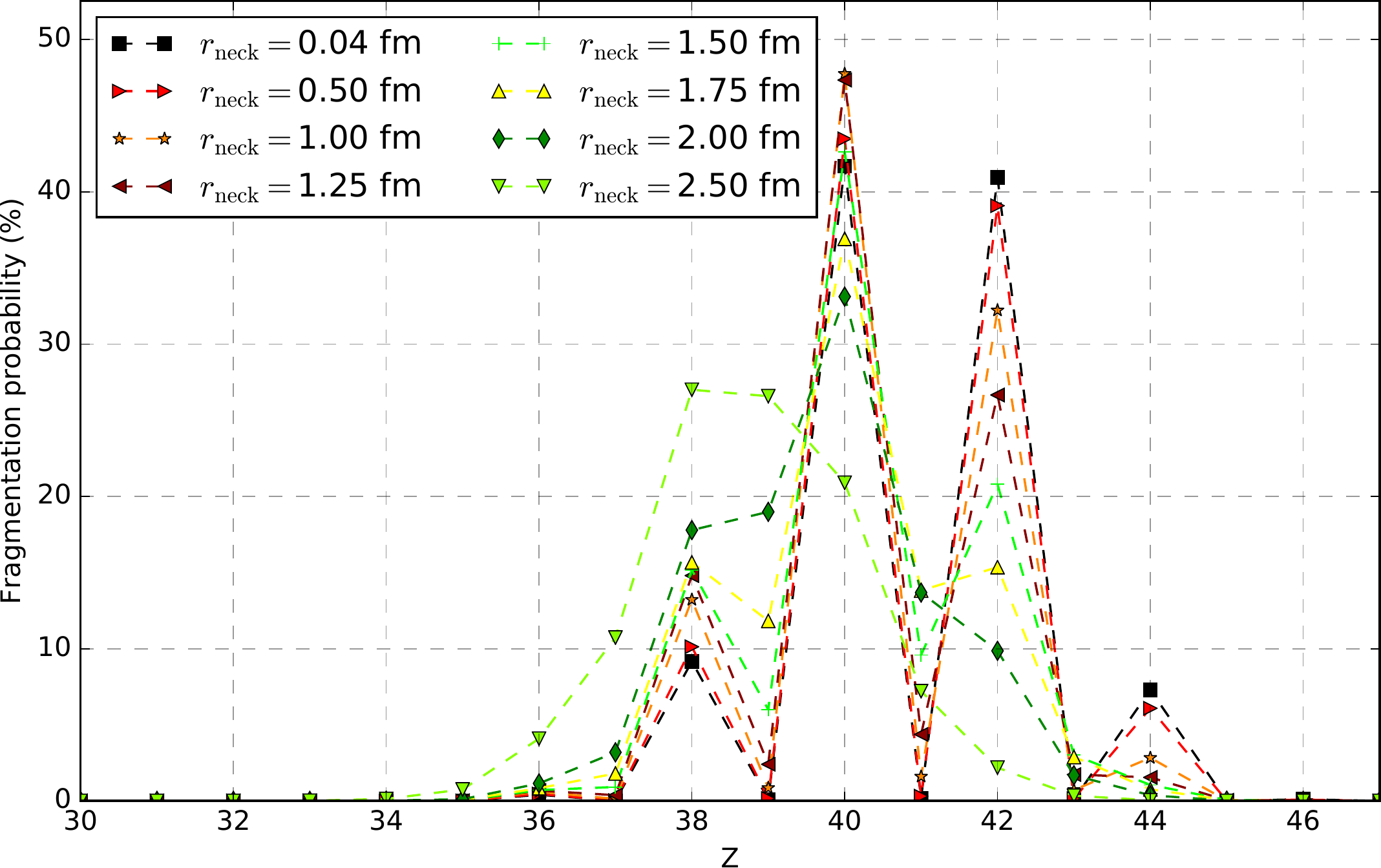}
\includegraphics[width=0.44\textwidth]{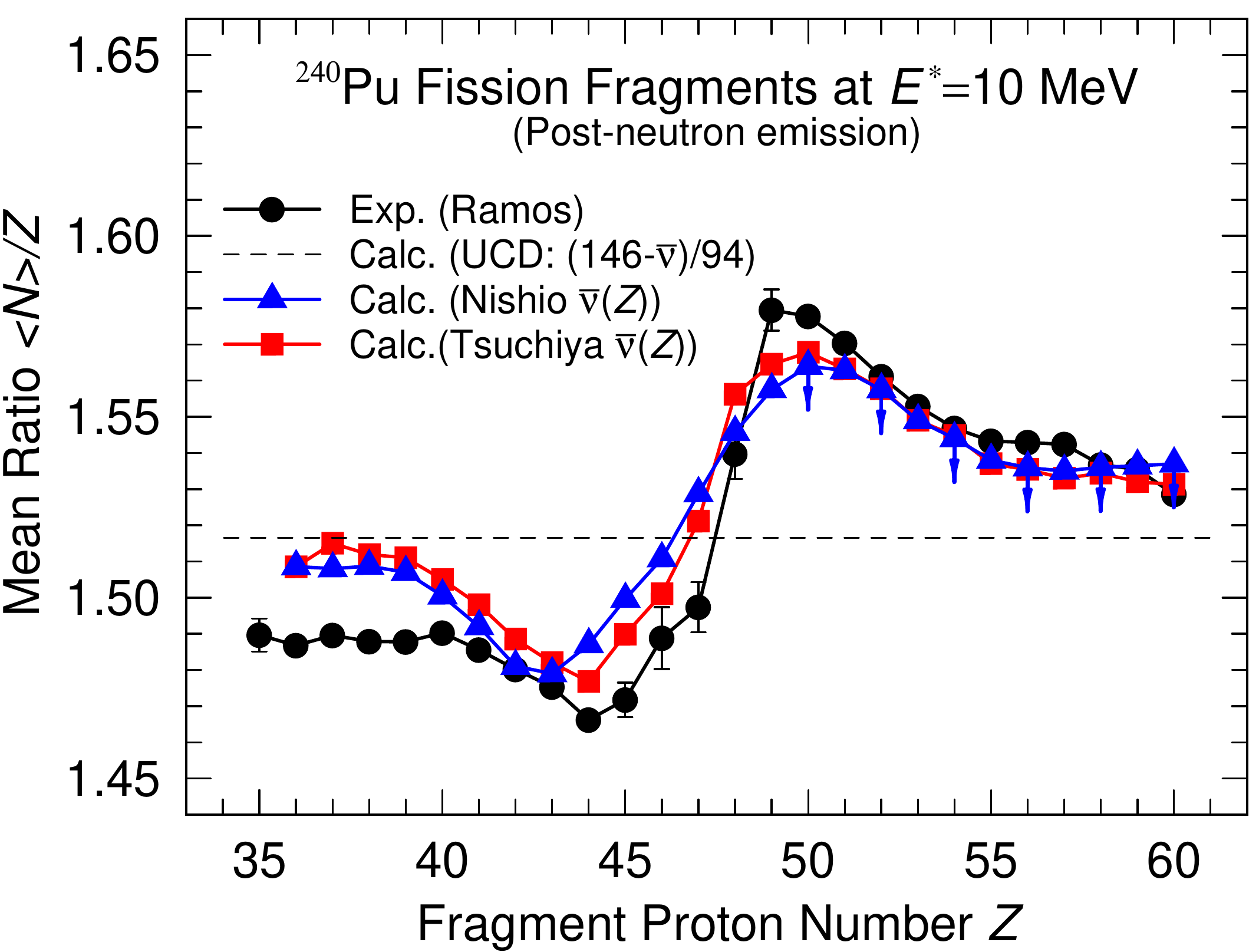}
\caption{Left: Charge fragmentation probabilities for the light fragment for a 
scission configuration in $^{240}$Pu. Each curve corresponds to a different 
size of the neck between the prefragments. 
Figure reproduced with permission from \cite{verriere2019number} 
courtesy of Verriere; copyright 2019 by The American Physical Society.
Right: Calculated post-neutron fission-fragment average neutron-to-proton ratio 
$\braket{N}/Z$ versus fragment charge compared with experimental data and the 
unchanged charge distribution (dashed line). 
Curves with blue and black 
symbols rely on different estimates of the average number of 
emitted neutrons.
Figure reproduced with permission from \cite{schmitt2021isotopic} 
courtesy of Schmitt; copyright 2021 by The American Physical Society.
}
\label{fig:fragments_particles}
\end{figure}

The left panel of Fig. \ref{fig:fragments_particles} illustrates that every 
single scission configuration, which is characterized by an average number of 
particles as estimated from \eqref{eq:obs_classical}, contains in fact of 
superposition of states with different, integer-valued particle numbers. For 
the smaller values of the neck radius, the method predicts the appearance of an 
odd-even staggering in the charge distribution. This technique can be applied 
to both proton and neutron numbers in every scission configuration. When 
combined with the time evolution of the collective packet in the TDGCM to 
populate these scission configurations (see Section \ref{subsubsec:TDGCM_GOA}), 
this method enabled the first prediction of two-dimensional $Y(A,Z)$ isotopic 
primary fragments distributions within the EDF 
framework~\cite{verriere2021microscopic}.

Another method to obtain integer-valued 
particle numbers in the fragments has been explored in the context of the microscopic-macroscopic models.
The idea is to 
rely on collective variables that specifically represent the 
average number of particles in each fragment on a discrete mesh \cite{moller2015method}. In 
addition to giving integer numbers, this technique was used to probe different 
neutron/proton ratios in the fragments and therefore computing isotopic yields
\cite{schmitt2021isotopic,moller2015method}. 
The right panel of Fig. \ref{fig:fragments_particles} shows an application of this technique to compute the 
ratio of the mean number of neutrons per element in the fragments, as a 
function of the fragment charge number. Results clearly indicate that the 
popular Unchanged Charge Distribution (UCD) \cite{wahl2002systematics}, which 
assumes that this ratio does not depend on the fragment charge number and is the 
same as in the fissioning nucleus, is not valid. This result, well known from
experiments, was also reproduced by microscopic calculations with PNP \cite{verriere2021microscopic}.


\subsection{Deformations of the Fragments}
\label{subsec:frag_deform}

The deformation of fission fragments is another important property, which has a 
direct impact on the number of emitted neutrons \cite{ruben1990fission} and on 
the spin distribution of the fragments; see also next section. One of the major 
advantages of EDF over macroscopic-microscopic approaches is that these 
deformations are {\it predicted} at scission, simply owing to the variational 
nature of the HFB equation and the fact that deformations are outputs of the 
calculations \cite{dubray2008structure,younes2009microscopica}. By contrast, 
models based on parametrizing the nuclear shape treat deformations as inputs: 
to obtain deformed fission fragments, one has to explicitly set these 
deformations. 

\begin{figure}[!htb]
\centering
\includegraphics[width=0.75\textwidth]{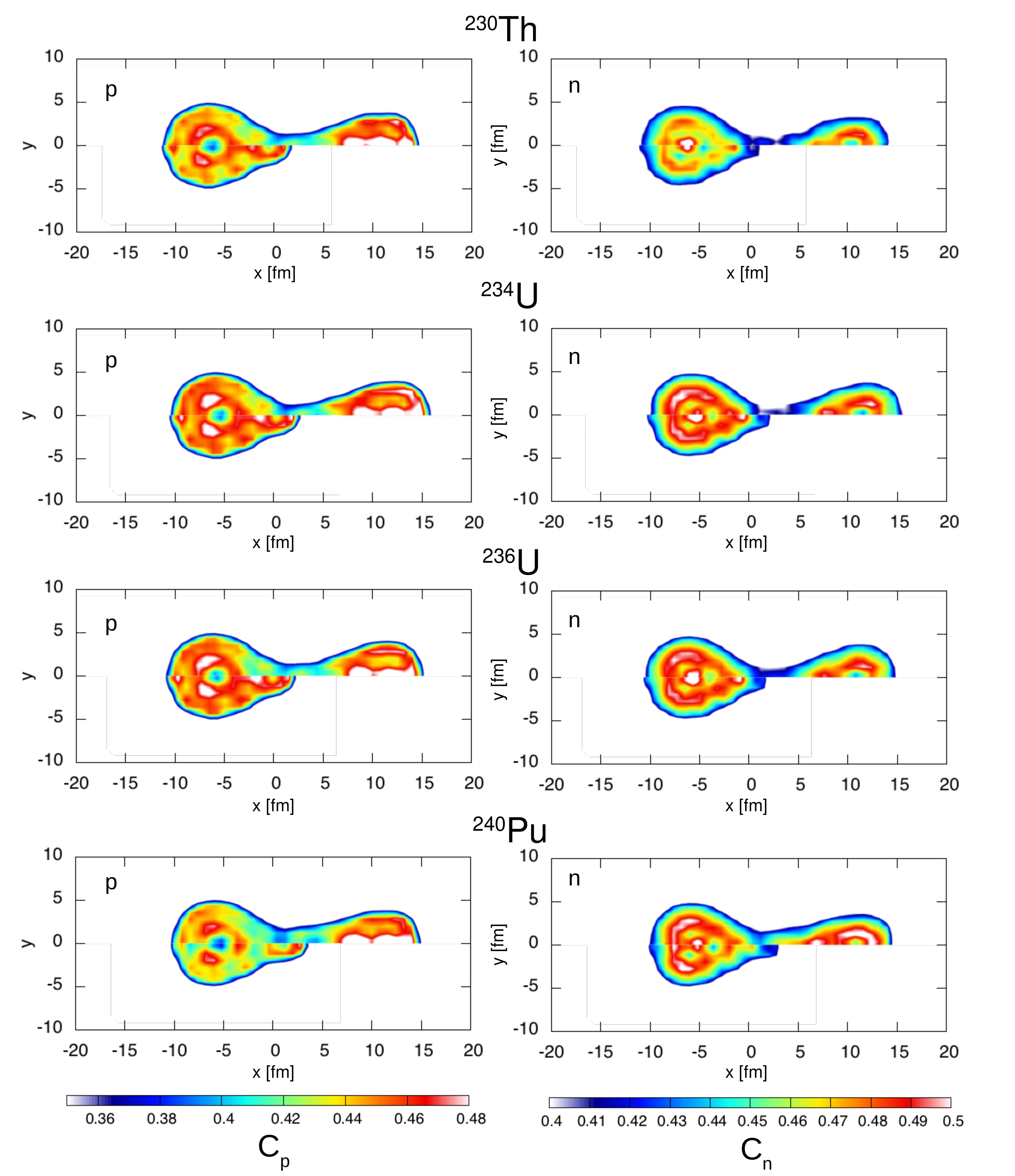}
\caption{Identification of the heavy pre-fragment in asymmetric fission of 
actinides. Localization functions of the fissioning nucleus at scission (upper 
half of the contour plot in each frame) and localization function of the 
octupole deformed $^{144}$Ba (lower-half of the contour plots).
Figures reproduced with permission from \cite{scamps2018impact} 
courtesy of Scamps; copyright 2021 by Nature.
}
\label{fig:fragments_deformations}
\end{figure}

Figure \ref{fig:fragments_deformations} is a spectacular confirmation of this 
property. Systematic TDDFT simulations of the fission of actinide nuclei showed 
that the heavy fragment has a sometimes substantial octupole deformation 
\cite{scamps2018impact}. This explains why the charge of the most likely heavy fragment in 
actinide fission is not centered around $Z=50$, as would naively be expected 
from considerations of spherical shell effects in the fragments, but rather 
around $Z=52-54$, depending on the fissioning system. This small shift is 
caused by the well-known octupole shell closure at $Z=56$ 
\cite{butler1996intrinsic}. 


\subsection{Spin Distributions}
\label{subsec:spin}

\begin{figure}[!htb]
\centering
\includegraphics[width=0.51\textwidth]{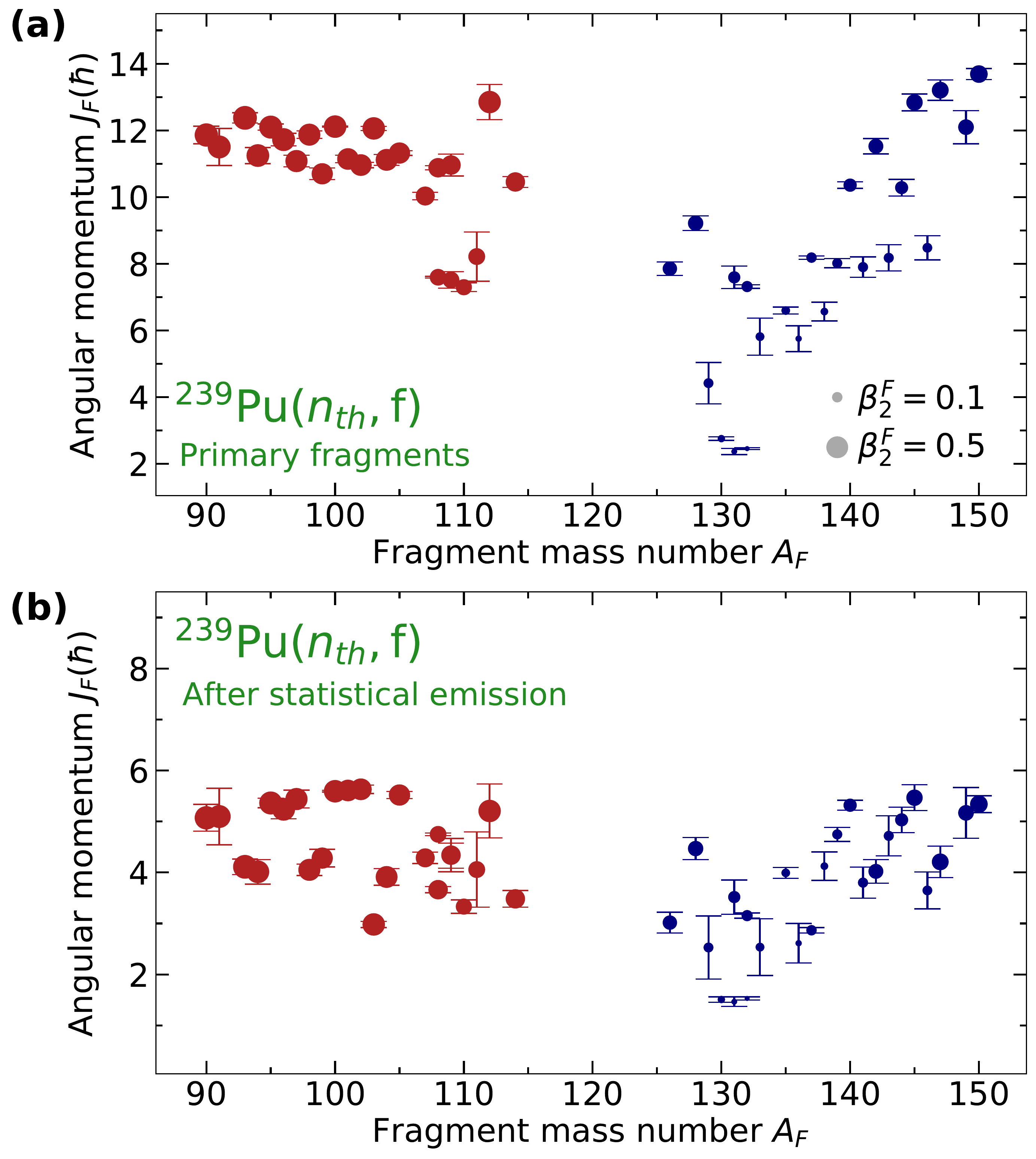}
\includegraphics[width=0.44\textwidth]{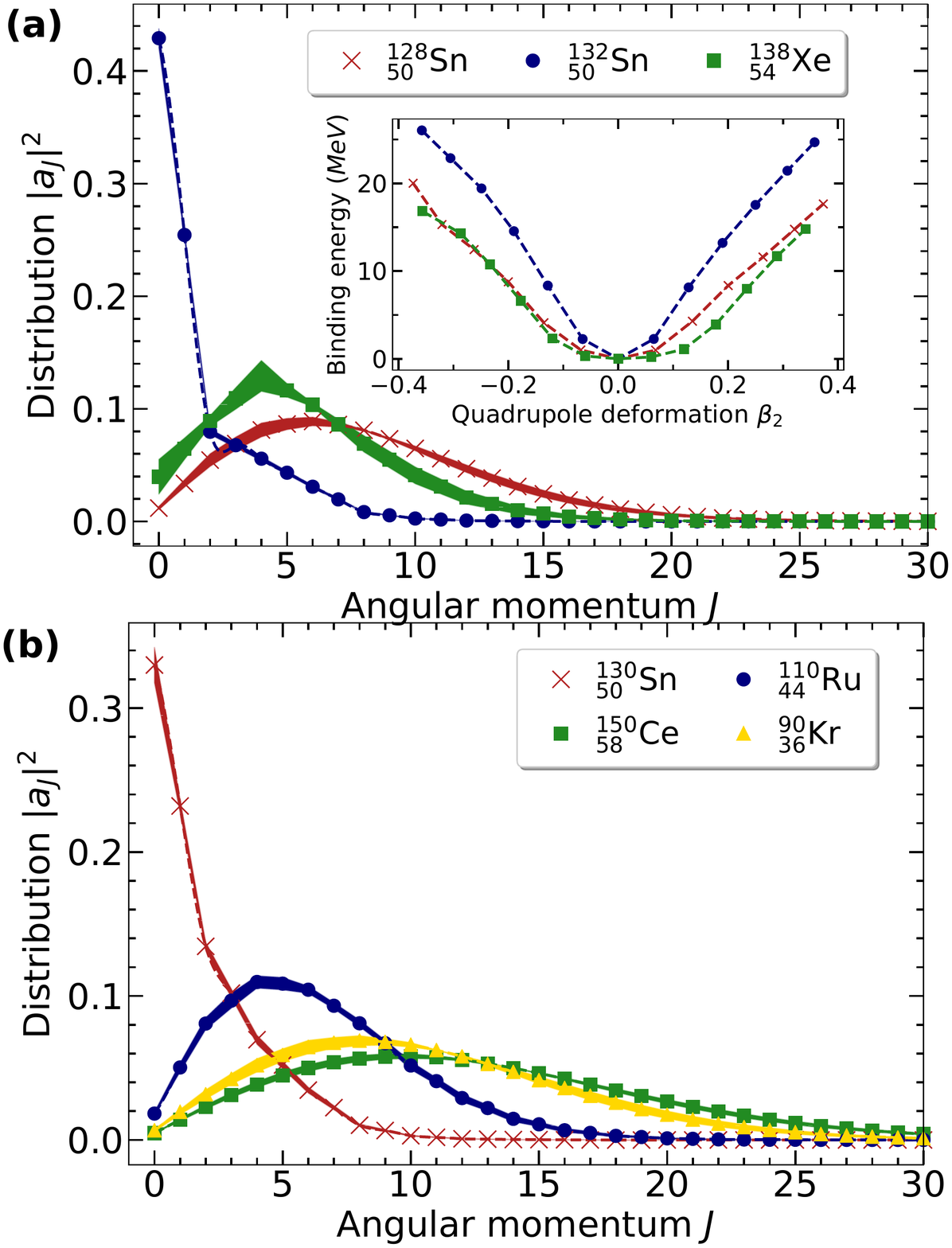}
\caption{Left: Average angular momentum of the fission fragments of $^{240}$Pu 
before (top) and after (bottom) statistical emission. Right: spin distributions 
of select isotopes near the doubly-close shell $^{132}$Sn (top).
Figures reproduced with permission from \cite{marevic2021angular} 
courtesy of Marevic; copyright 2021 by The American Physical Society.
}
\label{fig:fragments_spins}
\end{figure}

In addition to particle number, projection techniques have also been recently 
proposed to estimate the spin of fission fragments using angular momentum 
projection (AMP), following again an idea borrowed from theoretical models of 
heavy ion collisions \cite{sekizawa2017microscopic}. The extension of AMP to 
fission fragments is very similar to particle number and involves redefining 
the angular momentum operator so that it acts only in the subspace containing a 
single fragment. the method was tested both in static HFB calculations for 
nearly all mass fragmentations of $^{240}$Pu \cite{marevic2021angular} and in 
TDDFT for the most likely fission \cite{bulgac2021fission}. 
Note that in this context the projection is always performed after the variational 
determination of the (TD)HFB many-body state.
Results of static 
calculations suggest that the average spin of the light fragment is higher than 
the one of the heavy fragment and that shell effects at scission are very 
important to set the initial amount of angular momentum in the fragment. This 
is shown in Fig.~\ref{fig:fragments_spins}. The left panel shows the average 
spin of the fragments before and after statistical emission of photons, while 
the right panel shows that the deformations of the fragments {\it at scission} 
can be very different from the equilibrium deformations of the corresponding 
nucleus. Surprisingly, static calculations (constrained HFB) and full TDHFB 
simulations give very similar results -- even if the fragments are much more 
excited in TDHFB than they are in static HFB calculations. The additional 
insight of the time-dependent calculations is to show that the intrinsic spins 
of the two FFs are essentially perpendicular to each other.

These theoretical results partly confirm recent measurements of the spin  
distribution after the emission of statistical neutrons and $\gamma$ in three 
actinide nuclei \cite{wilson2021angular}. Specifically, the overall dependency 
of the spin on the fragment mass is very similar, and both theory and 
measurements suggest that the spin of the light fragment in the most probable 
fragmentation is larger than the one of the heavy fragment. 
Such conclusions were also obtained independently from 
studies of the fission spectrum with the code 
FIFRELIN~\cite{thulliez2019neutron}, where a light fragment spin higher than 
the heavy one naturally emerges from the fit of the spin-cutoff parameter to 
the light and heavy neutron multiplicities; see Section \ref{subsec:prompt} for
more details. Although several parameterizations of this code may lead to 
similar results, it shows at least that $\langle J_L \rangle> \langle J_H \rangle$
is not incompatible with a 
qualitative reproduction of both prompt-neutron and prompt-$\gamma$ observables. Contrary 
to the argument put forward in \cite{wilson2021angular}, the most advanced 
theoretical studies of the deexcitation of these fragments did not show 
evidence that statistical photon emission leads to a constant shift of the 
average angular momentum \cite{randrup2021generation,stetcu2021angular}. All 
three different theoretical approaches explain this result by the fact that the 
spin of the fragments is largely dependent on their deformation at
scission~\cite{randrup2021generation,marevic2021angular,bulgac2021fission}.


\subsection{Excitation Energy}
\label{subsec:initial_e}

The determination of the excitation energy of fission fragments is more 
difficult. 
Two approaches are possible: (i) 
estimate the total kinetic energy TKE (or rely on existing measurements), which 
sets the value of TXE, and model how this total excitation energy is shared 
among the fragments, i.e., with a statistical model; (ii) try to directly 
estimate the individual excitation energy of each fragment $E_{\rm H}^{*}$ and 
$E_{\rm L}^{*}$. In this case, the total excitation energy is simply 
${\rm TXE} = E_{\rm H}^{*} + E_{\rm L}^{*}$ and can be used to reconstruct the 
value of the TKE.

Adiabatic models based on static PES give by definition cold 
fragments where the excitation energy comes entirely from deformation effects
and the fact that a fragment $(Z,N)$ may have a different deformation from its 
ground-state value. An estimate of the value of the excitation energy can still
be set ``by hand'' based on the total energy balance of the reaction at scission 
similar to \eqref{eq:energy_balance}.
In this case, the first step consists in determining the total kinetic energy
of the static configurations in the neighborhood of scission.
Since most of the kinetic energy 
carried away by the fragments derives from the Coulomb repulsion between them, 
the most simple estimate of TKE is given by 
\begin{equation}
{\rm TKE} \approx \frac{e^2}{4\pi\epsilon_0}\frac{Z_{\rm H}Z_{\rm L}}{R} ,
\end{equation}
with $R$ the distance between the two fragments, which can estimated from the 
position of the two centers of mass \cite{brosa1983exit}. Refinements of this 
formula involve adding the pre-scission kinetic energy of the fragments, which 
can be estimated as the collective energy term $\frac{1}{2} \sum_{\alpha\beta} 
\dot{q}_{\alpha}\dot{q}_{\beta}$ in the context of the Langevin equation 
\eqref{eq:langevin}; see, e.g., \cite{davies1976effect,usang2016effects,
usang2017analysis}. The remaining total excitation energy of the system
then needs to be shared between the fragments.
This mechanism of energy sharing, also called energy 
sorting, is often described based on a statistical argument 
\cite{madland1982new}: in the Fermi gas model, the excitation energy of the 
system $E^{*}$ is related to the temperature through $E^{*} = aT^2$, where $a$ 
is the level-density parameter \cite{bohr1998nuclear}. If we assume statistical 
equilibrium between the two fragments at scission, then the temperatures satisfy 
$T_{\rm H} = T_{\rm L}$, and one can easily estimate $E_{\rm H}^{*}$ and 
$E_{\rm L}^{*}$. Simulations of the fission spectrum prompted a discussion of 
the validity of the equality between the two temperatures and the need to 
phenomenologically parametrize the ratio $R_{T} = T_{\rm H} / T_{\rm L}$ 
\cite{takaakiohsawa.1991representation,litaize2010investigation,
schmidt2010entropy,schmidt2011final,talou2011advanced}. Recently, yet another 
energy sharing mechanism was proposed, based on more realistic 
calculations of level densities in the fragments 
\cite{albertsson2020excitation,albertsson2021correlation}.

The second approach to estimating directly the excitation energy of each 
fragment is the direct result of progress in microscopic fission theory, 
especially real-time TDDFT simulations of fission events. In such cases, the 
total energy of the system is conserved, and simulations can be run until the 
two fragments are well separated (up to $\simeq 30$ fm between the fragments in
Ref.~\cite{bulgac2021fission}). At that point, the excitation energy of the 
fragment is simply the difference between the computed value and the 
ground-state binding energy. In addition, the TKE can be computed directly as 
the relative kinetic energy of the two moving fragments, rather than through 
the Coulomb energy proxy. This approach was first outlined in 
\cite{simenel2014formation} and applied in \cite{goddard2015fission} in the 
context of TDHF theory. With the inclusion of pairing in full TDHFB 
calculations, estimates of TKE and fragment excitation energy in the 
low-energy fission of nuclei such as $^{240}$Pu became possible 
\cite{bulgac2019fission} up to percent precision. Results for the most likely
fission suggest the two fragments do not have the same temperature. 


\section{Conclusion}

In spite of its applications in nuclear engineering and its essential role in 
answering fundamental science questions about the stability of superheavy 
elements or the formation of heavy elements in the Cosmos, the phenomenon of 
fission remains shrouded in mysteries. In terms of sheer complexity, it has few 
rivals as it compounds the challenges of any quantum many-body theory with 
those of open quantum systems, nuclear forces and out-of-equilibrium processes. 
From a nuclear theory perspective, fission is a deep and unforgiving probe into 
many properties of atomic nuclei: small errors in computing nuclear deformation 
properties can lead to orders of magnitude discrepancies on actual observables. 

In this article, we attempted to give as thorough a review on fission theories 
as possible. We highlighted that there are, broadly speaking, three main 
phases in the process: the probability for a nucleus to 
fission, the large-amplitude collective motion leading to the production of
primary fragments and the 
deexcitation of these fragments after they are formed. The first phase 
requires the tools of reaction theory to account for the competition between 
fission {\it per se} and every other decay channel. Even though most of the 
theoretical formalism was developed between the 1950ies and the early 1980ies, 
lack of proper nuclear structure inputs has been the major bottleneck and 
progress, while real, has been relatively slow. In contrast, it is fair to say 
that the description of large-amplitude nuclear collective dynamics has seen 
the most spectacular progress in the last two decades. This was largely caused 
by the possibility to perform large-scale, precise energy density functional 
calculations with realistic energy functionals on supercomputers. The last 
phase of the fission process, the prompt and delayed emission of particles from 
the fragments, is described by the same methods of nuclear reaction theory as 
the entrance channel. In recent years, several codes have been published to 
perform complete simulations of fission events and thereby connect with 
transport codes used in technological applications.

Looking ahead, we see many opportunities to bridge these three phases of 
fission into a single, consistent theoretical framework. For example, the 
recent studies on the number of particles, excitation energy or spin 
distribution of the fission fragments that were discussed in 
Section~\ref{sec:initial_fragments} have shown that one can start replacing 
phenomenological inputs of deexcitation models with microscopic predictions 
based on actual simulations of fission dynamics. Similarly, both the 
statistical and R-matrix formalisms for fission cross sections summarized in 
Secs.~\ref{subsubsec:fast_neutrons} and \ref{subsubsec:resonances} rely, most 
generally, on nuclear structure theory inputs that could be computed with more 
advanced models than used until now. This global theoretical framework that we 
envision is bound to be based on the EDF approach, at least in the foreseeable 
future: large-amplitude collective motion, $\gamma$ emission or absorption and 
$\beta$ decay are naturally described within this framework. It is reasonable 
to think that formal developments and computational advances should soon enable 
the calculation of the complete sets of nuclear wave functions with good 
quantum numbers needed to describe the entrance channel. Putting together all 
these ingredients will without a doubt require considerable effort, but 
is also the promise of a bright future for fission theory.


\section*{Acknowledgments}

The authors would like to thank O.~Litaize, N.~Pillet, J.~Randrup, P.~Romain, 
P.~Tamagno and I.~Thompson 
for reading parts of the manuscript and making relevant suggestions.
This work was supported in part by the NUCLEI SciDAC-4 collaboration 
DE-SC001822 and was performed under the auspices of the U.S.\ Department of 
Energy by Lawrence Livermore National Laboratory under Contract 
DE-AC52-07NA27344. Computing support came from the Lawrence Livermore National 
Laboratory (LLNL) Institutional Computing Grand Challenge program.



\bibliography{references_new,books}

\end{document}